\documentclass[12pt]{article}

\usepackage{amssymb,amsmath,amsfonts,amssymb}
\usepackage{graphics,graphicx,color}
\usepackage{empheq}
\usepackage{slashed}
\usepackage{tikz,tkz-fct}
\usepackage{caption,subcaption}

\usepackage{hyperref}

\usepackage{algorithm}
\usepackage{algpseudocode}
\usepackage{footnote}

\def\@abssec#1{\vspace{.05in}\footnotesize \parindent .2in
{\bf #1. }\ignorespaces}
\graphicspath{{/EPSF/}{Figures/}}
\DeclareGraphicsExtensions{.eps}
\newtheorem{theorem}{Theorem}[section]

\newtheorem{remark}[theorem]{Remark}

\def \Rm {\mathbb R}
\def \Nm {\mathbb N}
\def \Cm {\mathbb C}

\def \Mm {\mathbb M}
\newcommand{\eps}{\varepsilon}

\newcommand{\dsum}{\displaystyle\sum}
\newcommand{\dint}{\displaystyle\int}

\newcommand{\pdr}[2]{\dfrac{\partial{#1}}{\partial{#2}}}

\newcommand{\dr}[2]{\dfrac{d{#1}}{d{#2}}}

\newcommand{\aver}[1]{\langle {#1} \rangle}

\newcommand{\mF}{\mathcal F}
\newcommand{\mG}{\mathcal G}

\newcommand{\mM}{\mathcal M}

\newcommand{\mP}{\mathcal P}
 
\newcommand{\mV}{\mathcal V}

\newcommand{\fa}{{\mathfrak a}}

\newcommand{\cout}[1]{}

\newcommand{\sgn}[1]{\,{\rm sign}(#1)}

\newcommand{\epm}{\epsilon_m}
\newcommand{\psin}{\psi_{\rm in}}
\newcommand{\psiout}{\psi_{\rm out}}

 \renewcommand{\arraystretch}{1.5}

\makeatletter
\let\save@mathaccent\mathaccent
\newcommand*\if@single[3]{%
  \setbox0\hbox{${\mathaccent"0362{#1}}^H$}%
  \setbox2\hbox{${\mathaccent"0362{\kern0pt#1}}^H$}%
  \ifdim\ht0=\ht2 #3\else #2\fi
  }
\newcommand*\rel@kern[1]{\kern#1\dimexpr\macc@kerna}
\newcommand*\widebar[1]{\@ifnextchar^{{\wide@bar{#1}{0}}}{\wide@bar{#1}{1}}}
\newcommand*\wide@bar[2]{\if@single{#1}{\wide@bar@{#1}{#2}{1}}{\wide@bar@{#1}{#2}{2}}}
\newcommand*\wide@bar@[3]{%
  \begingroup
  \def\mathaccent##1##2{%
    \let\mathaccent\save@mathaccent
    \if#32 \let\macc@nucleus\first@char \fi
    \setbox\z@\hbox{$\macc@style{\macc@nucleus}_{}$}%
    \setbox\tw@\hbox{$\macc@style{\macc@nucleus}{}_{}$}%
    \dimen@\wd\tw@
    \advance\dimen@-\wd\z@
    \divide\dimen@ 3
    \@tempdima\wd\tw@
    \advance\@tempdima-\scriptspace
    \divide\@tempdima 10
    \advance\dimen@-\@tempdima
    \ifdim\dimen@>\z@ \dimen@0pt\fi
    \rel@kern{0.6}\kern-\dimen@
    \if#31
      \overline{\rel@kern{-0.6}\kern\dimen@\macc@nucleus\rel@kern{0.4}\kern\dimen@}%
      \advance\dimen@0.4\dimexpr\macc@kerna
      \let\final@kern#2%
      \ifdim\dimen@<\z@ \let\final@kern1\fi
      \if\final@kern1 \kern-\dimen@\fi
    \else
      \overline{\rel@kern{-0.6}\kern\dimen@#1}%
    \fi
  }%
  \macc@depth\@ne
  \let\math@bgroup\@empty \let\math@egroup\macc@set@skewchar
  \mathsurround\z@ \frozen@everymath{\mathgroup\macc@group\relax}%
  \macc@set@skewchar\relax
  \let\mathaccentV\macc@nested@a
  \if#31
    \macc@nested@a\relax111{#1}%
  \else
    \def\gobble@till@marker##1\endmarker{}%
    \futurelet\first@char\gobble@till@marker#1\endmarker
    \ifcat\noexpand\first@char A\else
      \def\first@char{}%
    \fi
    \macc@nested@a\relax111{\first@char}%
  \fi
  \endgroup
}
\makeatother
\title{Asymmetric transport computations in Dirac models of topological insulators}

\author{Guillaume Bal  \thanks{Departments of Statistics and Mathematics and CCAM, University of Chicago, Chicago, IL 60637; {\tt guillaumebal@uchicago.edu}} \and Jeremy G Hoskins \thanks{Department of Statistics and CCAM, University of Chicago, Chicago, IL 60637; {\tt jeremyhoskins@uchicago.edu}} \and   Zhongjian Wang \thanks{Department of Statistics and CCAM, University of Chicago, Chicago, IL 60637; {\tt zhongjian@uchicago.edu}}}
\begin{document}
 
\maketitle

%\tableofcontents

\begin{abstract}
  This paper presents a fast algorithm for computing transport properties of two-dimensional Dirac operators with linear domain walls, which model the macroscopic behavior of the robust and asymmetric transport observed at an interface separating two two-dimensional topological insulators. Our method is based on reformulating the partial differential equation as a corresponding volume integral equation, which we solve via a spectral discretization scheme. 
  
  We demonstrate the accuracy of our method by confirming the quantization of an appropriate interface conductivity modeling transport asymmetry along the interface, and moreover confirm that this quantity is immune to local perturbations. We also compute the far-field scattering matrix generated by such perturbations and verify that while asymmetric transport is topologically protected the absence of back-scattering is not.
\end{abstract}
 
%\begin{AMS}
%\end{AMS}

\renewcommand{\thefootnote}{\fnsymbol{footnote}}
\renewcommand{\thefootnote}{\arabic{footnote}}

\renewcommand{\arraystretch}{1.1}

%\begin{keywords}
%\end{keywords}

%\begin{AMS}
%\end{AMS}

%\pagestyle{myheadings}
%\thispagestyle{plain}

%%%%%%%%%%%%%%%%%%%%%%
%%% BEGINNING TEXT %%%
%%%%%%%%%%%%%%%%%%%%%%
%
%
\section{Introduction}
\label{sec:introduction}
Two-dimensional Dirac operators appear naturally in the analysis of topological phases of matter \cite{WI,moessner2021topological} and one-particle models of topological insulators and topological superconductors \cite{Be13,PSB16,sato,VO,FC,lee2017elliptic,B19b,drouot2020edge}.
This paper focuses on efficient numerical simulation of the transport properties of one such operator, which for concreteness we take as
\begin{align}
    \label{eq:Dirac}
    H_V = D_x \sigma_1 + D_y \sigma_2 + m(y) \sigma_3 + V = \begin{pmatrix}
    m(y)+V_{11} & D_x-iD_y + V_{12} \\ D_x+iD_y+V_{12}^* & -m(y)+V_{22}
    \end{pmatrix}.
\end{align}
Here, $\sigma_{1,2,3}$ are the standard Pauli matrices (which along with the identity matrix $\sigma_0=I_2$ form a basis of $2\times2$ Hermitian matrices), $D_a=-i\partial_a$ for $a=x,y$, $m(y)=y$ is a linear domain wall, and $V(x,y)$ is an arbitrary local Hermitian-valued perturbation with compact support. Physically, the scalar component of $V$ corresponds to an electric potential, the components of $V$ in front of $\sigma_{1,2}$ to a magnetic potential, and the component in front of $\sigma_3$ to a local perturbation of the domain wall $m(y)=y$. We also use the notation $H_V=H+V$ with $H$ the unperturbed Dirac operator. 

The domain wall $m(y)$ models a topological phase transition between two bulk phases \cite{Be13,FC,B19b}, one in the domain $y>0$ where $m(y)>0$ and one in the domain $y<0$ with $m(y)<0$. A striking feature of topologically non-trivial materials is that transport (say, electronic) along the interface $y\sim0$ is {\em asymmetric}, in the sense that we observe stronger transport towards negative values of $x$ than transport in the opposite direction. Moreover, this transport asymmetry is {\em quantized} and related to a difference of bulk topological invariants following the celebrated bulk-edge correspondence \cite{Be13,PSB16} (more precisely in fact a bulk-difference invariant \cite{B20} in the context of Dirac operators). Finally, the topological nature of the asymmetry ensures its robustness to perturbations. In our context, the transport asymmetry is independent of any perturbation $V(x,y)$ (no matter how large or how oscillatory) that vanishes at infinity sufficiently rapidly.

The robustness of this asymmetric transport is one of the main practical appeals of such materials. It stands in sharp contrast to topologically trivial materials, for instance modeled by a one-dimensional wave equation along the interface $y=0$, where transmission is exponentially suppressed by Anderson localization in the presence of strong perturbations \cite{PSB16,fouque2007wave,topological} (so that incoming signals may be fully back-scattered). However, robust asymmetric transport does not imply the absence of back-scattering; rather, merely that some quantized transmission is guaranteed.  See \cite{PSB16,LJS,delplace,souslov} for references on topologically-protected asymmetric transport in numerous applications. 

In order to understand the transport properties of models such as \eqref{eq:Dirac}, we develop an efficient algorithm to compute its generalized eigenvectors. This allows us to estimate a conductivity quantifying asymmetric transport and scattering matrices generated by the perturbation $V$. For $\psi_{\rm in}$ an incoming plane wave solution of $(H-E)\psi_{\rm in}=0$, we look for outgoing solutions of
\begin{align}\label{eq:psiout}
    (H+V-E)\psi_{\rm out} = -V \psi_{\rm in} .
\end{align}
We will construct explicitly the (unique) outgoing Green's function, i.e., the kernel of the operator $(H-E)^{-1}\equiv (H-(E-i0^+))^{-1}$. The above equation can then be reduced to a volume integral equation for a new unknown $\rho$ which we will refer to as the density corresponding to $\psi_{\rm out}.$ Specifically,
\begin{align}\label{eq:rho}
    \psi_{\rm out}=(H-E)^{-1} \rho\quad \implies \quad \rho + V (H-E)^{-1} \rho = -V \psi_{\rm in}.
\end{align}
A computational advantage of the latter integral equation is that by construction, $\rho(x,y;E)$ is supported on the same domain as $V$. Numerical methods to solve such Fredholm integral equations of the second kind are discussed, e.g., in \cite{atkinson2009numerical}.

The Green's function $G(x,y;x_0,y_0;E)$, the $2\times2$ matrix-valued kernel of $(H-(E-i0^+))^{-1}$, does not admit a simple, explicit expression. It may, however, be efficiently written in a basis of Hermite functions in the $y-$variable. The construction is detailed in section \ref{sec:GF}. The Green's function admits two main features. One in the singularity at the source location where the Green's function is inversely proportional to the distance between $(x,y)$ and $(x_0,y_0)$ as for any two-dimensional Dirac operator. Another feature is the behavior of the Green's function `at infinity' along the $x$ axis reflecting the wave-guide nature of the model generated by the domain wall as well as the transport asymmetry along the edge $y\sim0$. %\jh{I am not sure what these singularities are}.

For $V\in C^\infty_c(\Rm^2;\Mm_2)$ a smooth compactly supported matrix-valued perturbation, solving the volume integral equation for $\rho$ requires inverting an operator of the form $I+K$ with $K$ a compact operator (from $H^k(\Rm^2;\Cm^2)$ to itself for each integer $k\geq0$). When $-1$ is not an eigenvalue of $K$ (this is independent of $k$), we thus obtain that the unique solution $\rho$ of \eqref{eq:rho} is itself smooth in $C^\infty_c(\Rm^2;\Cm^2)$. The numerical simulation of \eqref{eq:rho} thus boils down to efficiently approximating the integral operator $V(x,y)\int_X G(x,y;x_0,y_0)\rho(x_0,y_0)dx_0dy_0$ for $X$ a domain including the support of $V$.

Due to the nature of the problem, it is natural to discretize $\rho$ using a finite Cartesian product basis of Legendre polynomials in the $x$ direction and Hermite functions in the $y$ direction. The details of the construction and demonstration of convergence are presented in section \ref{sec:single_discretization}.

We accelerate the computation of the density $\rho$ using a hierarchical merging approach in the $x$-direction. First, the $x-$axis is subdivided into adjacent small slabs which we refer to as {\em leaves}. The full transmission-reflection matrix (solution operator) for each small leaf is then computed explicitly. A standard algebraic merging procedure, described in section \ref{sec:meging_alg}, is then used to solve for $\rho(x,y)$ globally. This approach is a slight generalization of the method in \cite{greengard1991numerical} for solving two point boundary value problems. More broadly, it is closely related to {\it fast direct solver} methods for constructing compressed representations of Green's functions for linear PDEs (\cite{beams2020,hao2016,gillman2012,gillman2014,gillman2015,fortunato2021,martinsson2009,martinsson2013,corona2015,minden2017}). Additionally, similar techniques are frequently employed in electromagnetic waveguide problems to solve the PDE directly, often though not always with the additional assumption that $V$ is piecewise constant in the direction of propagation. See, for example, the {\it $S$-matrix propagation algorithm} \cite{li2003note}, the {\it reflection-transmission coefficient matrix (RTCM)} method \cite{kexiang1999modal}, and the {\it transfer-matrix method (TMM)} \cite{mackay2020transfer}, as well as the references therein. In particular, see \cite{phan2021electronic} for an example of TMM applied to the Dirac equation in a different context. {Finally, we also note that equation (\ref{eq:Dirac}) shares some similarities with the {\it gravity Helmholtz equation}, which arises in the study of the ionosphere, nano-scale optical devices, seismology, and acoustics (see \cite{barnett2015} and the references therein).} 

\medskip

We next describe applications of the above algorithm to compute accurate approximations of several relevant transport properties of \eqref{eq:Dirac}. The interface conductivity, a physical observable describing asymmetric transport may be defined as follows. Let $P_{x_0}(x,y)=P_{x_0}(x)$ be the Heaviside function equal to $1$ for $x>x_0\in\Rm$ and equal to $0$ for $x<x_0$. For notational convenience, in the following we will suppress the dependence of $P$ on $x_0.$ For $\psi$ a solution of the Schr\"odinger equation $i\partial_t \psi =H\psi$, we define $\aver{P}=\aver{\psi|P|\psi},$ which physically corresponds to the probability that the particle with wavefunction $\psi$ is on the right of the line $x=x_0$. The time derivative 
\begin{align} \nonumber
    \dr{}t \aver{P} = \aver{\psi|i[H,P]|\psi}
\end{align}
then describes the probability current crossing the line $x=x_0$. 

Let $(E_1,E_2)$ be an energy interval and $\chi(h)$ a smooth non-negative scalar function such that $\chi(h)=0$ for $h<E_1$ and $\chi(h)=1$ for $h>E_2$. Thus, $\chi'(h)$ may be interpreted as a density of states confined to the energy range $(E_1,E_2)$. The conductivity modeling asymmetric transport for this choice of density of states is then given by
\begin{align}\label{eq:sigmaI}
    2\pi \sigma_I = 2\pi {\rm Tr} \ i[H,P] \chi'(H).
\end{align}
For the Dirac operator, $2\pi\sigma_I$ is quantized and equals $-1$ \cite{B19b,bal2021topological}, indicating that the overall current integrated over the states present in the system always equals $-1$ independent of the choice of $V$ and of the energy interval $(E_1,E_2)$. See \cite{B19b,B20,Be13,Drouot,Elbau,elgart2005equality,SB-2000} for details and context on the above observable describing asymmetric transport. The theoretical computation of \eqref{eq:sigmaI} is often greatly simplified by appealing to the bulk-edge correspondence, relating it to the difference of topological invariants of the two insulating phases; for derivations and applications of such a correspondence see \cite{Be13,BM21,SB-2000}.

In section \ref{sec:conductivity}, we show how $\sigma_I$ may be related to the generalized eigenfunctions $\psi_{\rm out}(E)$ computed in \eqref{eq:psiout}; see \eqref{conductivity_formula} below. The explicit expression involves a number of propagating modes that depends on the energy range $(E_1,E_2)$. To simplify the presentation, we will assume $E_2=E_1$ there.

Another natural transport property of \eqref{eq:Dirac} is the scattering matrix generated by the perturbation $V$. The scattering matrix describes the far field behavior of $\psiout$ away from the support of $V$ associated in \eqref{eq:psiout} to any incoming plane wave $\psin$. The description of the $E-$dependent plane waves $\psin$ is described in section \ref{sec:SD} while scattering matrices are defined and computed numerically in section \ref{sec:scattering}.

\medskip

This paper focuses on the Dirac model \eqref{eq:Dirac} with a linear (unbounded) domain wall $m(y)=y$ so that the Green's function given by the kernel of $(H-E)^{-1}$ is explicitly constructed. We note that a method to compute $\sigma_I$ for a large class of partial differential operators was developed in \cite{QB22} along with a rate of convergence estimate. While the latter method does not require knowledge of an unperturbed Green's function, it is based on embedding a domain of interest in a larger box augmented with periodic boundary conditions. The method proposed in this paper is considerably faster numerically as the solutions $\rho$ in \eqref{eq:rho} are computed only on the support of $V$ and no artificial boundary conditions are necessary; see also \cite{colbrook2021computing} and references there for a computation of invariants for discrete (tight-binding) Hamiltonians.

Generalizations of the above integral formulation to bounded domain wall profiles \cite{QB22}, where high-energy bulk propagation is possible, and other partial differential models \cite{delplace,souslov} will be considered elsewhere. We also note that the above transport descriptions are purely spectral. For an analysis of localized wavepackets propagating along curved domain walls (the domain wall $y$ above is replaced by a more general function $m(x,y)$), we refer the reader to \cite{bal2021edge,bal2022magnetic,hu2022traveling}. 
\medskip

An outline for the rest of the paper is as follows. Section \ref{sec:IF} presents the integral equation \eqref{eq:rho} for the density $\rho$ following a spectral decomposition of the unperturbed Dirac operator in section \ref{sec:SD} and a computation of the associated Green's function in \ref{sec:GF}. Numerical approximations of the objects appearing in the integral equation \eqref{eq:rho} are given in section \ref{sec:prelim} and an efficient merging algorithm in section \ref{sec:meging_alg}. The algorithm to solve  \eqref{eq:rho} is then described in detail in section \ref{sec:alg}. Numerical results on accuracy and convergence are displayed in section \ref{sec:num}. Applications to transport calculations, including the computation of the quantized line conductivity and of scattering matrices of illustrative choices of the perturbation $V$ are given in section \ref{sec:appli}. Evidence for the presence of point spectrum (localized modes) embedded in the continuous spectrum is also presented in section \ref{sec:localizedmodes}. Section \ref{sec:conclu} offers some concluding remarks.

\section{Integral formulation of the perturbed system}
\label{sec:IF}

This section derives the integral equation \eqref{eq:rho} and presents several relevant properties of its solutions. We first provide a spectral decomposition of the unperturbed ($V=0$) operator in section \ref{sec:SD} and construct the associated Green's function in section \ref{sec:GF}. The integral equation for $\rho$ when $V\not=0$ and its properties are detailed in section \ref{sec:gal}.

To simplify calculations, we perform a unitary transformation replacing $H$ in \eqref{eq:Dirac} by $QHQ^*$, which, with some slight abuse of notation we still call $H$, and where $Q=\frac{1}{\sqrt2} (\sigma_1+\sigma_3)$. We observe that $Q^*=Q$ and that $Q(\sigma_1,\sigma_2,\sigma_3)Q=(\sigma_3,-\sigma_2,\sigma_1).$ In this new basis, we thus observe that
\begin{align} \label{eq:DiracNew}
     H = D_x \sigma_3 - D_y\sigma_2 + m(y) \sigma_1
\end{align}
We also still denote by $V$ the transformed perturbation $QVQ^*$ so that $H_V=H+V$ in the new basis as well.
\subsection{Spectral decomposition of the unperturbed operator}
\label{sec:SD}
The objective of this section is to find solutions in the kernel of $H-E$  for $E\in\Rm$ when the perturbation $V=0$. This provides a description of the propagating and evanescent modes we need later on and of the branches of absolutely continuous spectrum of the operator.

The unperturbed operator $H$ in \eqref{eq:DiracNew} is invariant under translations in the $x-$variable. Denoting by $\xi$ the dual Fourier variable to $x$ and $\mF$ the corresponding transform, we observe that $H=\mF^{-1}\hat H(\xi)\mF$ satisfies
\begin{align} \nonumber
   \hat H(\xi)-E = \xi\sigma_3 - D_y\sigma_2 + y \sigma_1 -E.
\end{align}

Let $\fa=\partial_y+y$ and define $\varphi_n(y)=a_n(\fa^*)^n\varphi_0(y)$, the normalized (in $L^2(\Rm_y)$, $a_n$ is normalizing factor) Hermite functions satisfying,
\begin{align*}
    \fa^*\fa \varphi_n=2n \varphi_n,\quad \fa \varphi_n = \sqrt{2n} \varphi_{n-1}, \quad \fa^*\varphi_n=\sqrt{2n+2} \varphi_{n+1},\quad \varphi_0(y)=\pi^{-\frac14} e^{-\frac12 y^2}.
\end{align*}
They form an orthonormal basis of $L^2(\Rm_y)$.

We define the countable set $M$ as the union of the following (pairs of) indices $m$. For $\Nm\ni n\geq1$, we define $m=(n,\epm)$ with $\epm=\pm1$, while for $n=0$, we define $m=(0,-1)$. In particular, $(0,1)\not\in M$.

While a description of the continuous spectrum of $H$ involves the branches $\xi\to E_m(\xi)$, we will need a characterization of the propagating and evanescent modes described by the `inverse' maps $E\to\xi_m(E)$. 

We thus fix energy $E\in\Rm$. The solutions to $E^2=\xi^2+2n$ are given explicitly by
\begin{align}\label{eq:xim}
    \xi_m=\epm(E^2-2n)^{\frac12}, \qquad m\in M,
\end{align}
with the choice $(E^2-2n)^{\frac12}=\sqrt{E^2-2n}$ when $E^2\geq2n$ and $(E^2-2n)^{\frac12}=i\sqrt{-E^2+2n}$ when $E^2\leq2n$. 

We then define $\phi_m$ to be the solution of
\begin{align} \nonumber
  (\hat H(\xi_m)-E) \phi_m =  \begin{pmatrix}  \xi_m-E & \fa \\  \fa^* & -(\xi_m+E) \end{pmatrix} \phi_m=0.
\end{align}
The normalized solutions are given by $\xi_0=-E$ and $\phi_0=(0,\varphi_0)^t$ when $m=(0,-1)$ and for $n\geq1$ otherwise by
\begin{align}\label{def:phi}
    \phi_m = c_m \begin{pmatrix} \fa \varphi_n \\ (E-\xi_m) \varphi_n \end{pmatrix} =  c_m \begin{pmatrix} \sqrt{2n} \varphi_{n-1} \\ (E-\xi_m) \varphi_n \end{pmatrix} ,\qquad c_m^{-2} = 2n + |E-\xi_m|^2. 
\end{align}
%\gb{I did not correct below.}
We observe that for $m=(n,\eps_m)$ and $q=(p,\eps_q)$ such that $n\not=p$,
\begin{align} \nonumber
   (\phi_m,\phi_q)_y = 0,
\end{align}
where $(\cdot,\cdot)_y$ is the usual inner product on $L^2(\Rm_y;\Cm^2)$. When $n=p$ while $\eps_m\not=\eps_q$, $\phi_m$ and $\phi_q$ are linearly independent vectors in
\begin{align}
    \nonumber \text{span}\Big\{\varphi_{n-1}\begin{pmatrix}1\\0\end{pmatrix},\varphi_{n}\begin{pmatrix}0\\1\end{pmatrix}\Big\}.
\end{align} The functions $\phi_m(y;E)$ may be shown to be complete and thus form 
%%%% an orthonormal  %%% PROBABLY FALSE THEN?
a basis of $L^2(\Rm_y;\Cm^2)$.

In particular, we find that if $\psi_m$ is defined via the following formula
\begin{align}\label{eq:psim}
    \psi_m(x,y;E)=e^{i\xi_m x}\phi_m(y;E)
\end{align}
then $(H-E)\psi_m(x,y;E)=0$. Moreover, when $\xi_m$ is real-valued (i.e., when $E^2\geq 2n$), $\psi_m(x,y;E)$ is a (bounded) generalized eigenvector (plane wave) of $H$ associated with energy $E$. These generalized eigenvectors form the set of incoming plane waves $\psin$ mentioned in the introduction, section \ref{sec:introduction}. We refer to them as propagating modes. 

When $\xi_m$ is purely imaginary, the corresponding $\psi_m(x,y;E)$ is an {\em evanescent} mode. Accounting for evanescent modes is necessary in the presence of perturbations $V$ and in the computation of the Green's function described in the next section.

\medskip

The above calculations described propagating and evanescent modes depending on the reality of $\xi_m$. The same calculations provide a complete spectral decomposition of $H$ as follows. Define
\begin{align}
    E_m(\xi)=\epm \sqrt{\xi^2+2n},\qquad \phi_m = c_m \begin{pmatrix} \fa \varphi_n \\ (E_m-\xi) \varphi_n \end{pmatrix},\quad c_m^{-2} = 2n + |E_m-\xi|^2. 
\end{align}
Each $\xi\to E_m(\xi)$ is a branch of absolutely continuous spectrum with generalized eigenvectors solving $(H-E_m(\xi))\psi_m=0$ and given by $\psi_m(x,y;\xi)=e^{i\xi x}\phi_m(y;\xi)$.

\begin{figure}[htbp]
    \centering
    \includegraphics{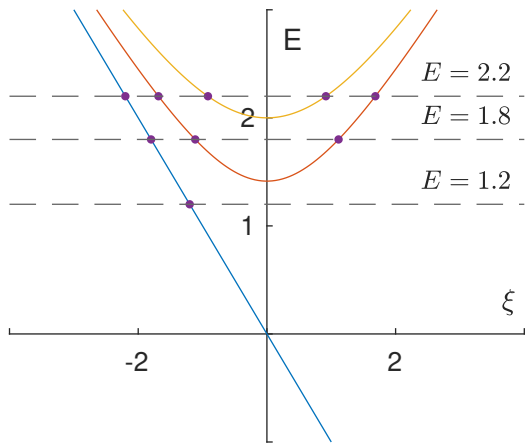}
    \caption{Three branches of absolutely continuous spectrum of the unperturbed operator $H$ with dots describing the propagating modes for different fixed energy levels $E$. While no backscattering is present for $|E|<\sqrt 2$, this is no longer true for $|E|>\sqrt 2$.}
    \label{fig:EvsXi}
\end{figure}

Thus, $\sigma(H)=\sigma_{\rm ac}(H)=\Rm,$ i.e., the whole real line is absolutely continuous spectrum, with the following energy-dependent structure; see also Fig.\ref{fig:EvsXi}. 

In the range $E^2<2$, we obtain a unique branch of absolutely continuous spectrum $m=(0,-1)$ parameterized by a simple branch of generalized eigenvectors $\psi_{(0,-1)}(x,y;\xi)$ with group velocity equal to $\partial E(\xi)/\partial\xi=-1$ (straight, blue curve in Fig.\ref{fig:EvsXi}).
In this energy range, no mode displays a positive group velocity. A perturbation $V$ of $H$ can only generate a phase shift that does not perturb the particle density.  Back-scattering is therefore entirely suppressed in this energy range, both for topological and energetic reasons. 

The spectrum is symmetric in $E\to-E$ and it therefore suffices to assume $E\geq0$. For $2<E^2<4$, the absolutely continuous spectrum has a degeneracy equal to $3$ corresponding to three propagating modes $\psi_{\rm in}^j$ for $1\leq j\leq 3$ and generating corresponding outgoing modes $\psi_{\rm out}^j$ solution of \eqref{eq:psiout} (see level set $E=1.8$ in Fig.\ref{fig:EvsXi}). When $V=0$, two modes propagate towards negative values of $x$ while one propagates in the opposite direction. In terms of asymmetric transport, $-2+1=2\pi\sigma_I$ is valid.
When $V\not=0$, the three propagating modes interact via the perturbation as we will demonstrate in section \ref{sec:scattering}. Back-scattering is then clearly present even though the `topology' of the Dirac operator is still non-trivial. We will present numerical simulations showing that the structure of the scattering matrix strongly depends on the choice of perturbation $V$ while $2\pi\sigma_I=-1$ holds up to arbitrary (i.e., at least $14$ digits here) accuracy even in the presence of strong perturbations $V$.

More generally, the degeneracy of the spectrum is equal to $2n+1$ for energies in the range $2n<E^2<2(n+1)$. The perturbation $V$ then generates $(2n+1)\times(2n+1)$ (far field) scattering matrices; see level set $E=2.2$ in Fig.\ref{fig:EvsXi} when $n=2$. We illustrate such computations for $n=0,1,2$ in section \ref{sec:scattering}.

\subsection{Green's Function of the unperturbed system}
\label{sec:GF}
We now construct the outgoing Green's function of the operator $(H-E)$ for $E\in\Rm$ (i.e., the Green's function of $H-(E-i0^+)$) when the perturbation $V=0$. We will in fact avoid a set of (Lebesgue) measure zero and assume that $E^2\not=2n$ for $n\in\Nm_*$. 

Consider the Green's function $G$ which is the solution of the following equation
\begin{align} \nonumber
   (H-E) G = \delta(x-x_0)\delta(y-y_0) I.
\end{align}
Without loss of generality, since $H$ is translationally invariant in $x$, we can assume that $x_0 = 0.$ Then,
\begin{align} \nonumber
   (H+E)(H-E)G&=(H^2-E^2)G= (H+E) \delta(x)\delta(y-y_0) I,\\ \implies G&= (H+E) (H^2-E^2)^{-1} \delta(x)\delta(y-y_0) I,\nonumber
\end{align}
{where we have used the fact that clearly $(H-E)$ and $(H+E)$ commute.} Additionally, we note that here 
\begin{align} \nonumber
H^2-E^2= \begin{pmatrix} D_x^2 -E^2 + \fa\fa^* & 0 \\ 0& D_x^2-E^2 + \fa^*\fa  \end{pmatrix}
\end{align}
is a diagonal matrix of scalar operators. We thus need to solve two similar equations 
\begin{align}\nonumber
   (-\partial_x^2 -\partial_y^2 + y^2 \pm 1-E^2)G_\pm =\delta(x) \delta(y-y_0),
\end{align}
and then observe that the complete Green's function is given by
\begin{align} \nonumber
    G = (H+E) {\rm Diag} (G_+,G_-).
\end{align}
We recall that $\fa^*\fa\varphi_n=2n\varphi_n$ and $\fa\fa^* \varphi_n=(2n+2)\varphi_n$ and assume that $E^2\not=2n$ for each $n$.

Expanding $G_-$ in the basis $\phi_n$, and applying the operator $D_x^2-E^2+2n$, we see that if
\begin{align} \nonumber
  G_-=\dsum_n G_{-,n}(x) \varphi_n(y),\quad{\rm then} \quad (D_x^2-E^2+2n) G_{-,n}(x) = \delta(x) \varphi_n(y_0).
\end{align}
When $2n>E^2$, we obtain from the jump condition at $x=0$ the evanescent modes
\begin{align} \nonumber
   G_{-,n}(x) = \frac{1}{2\sqrt{2n-E^2}} e^{-\sqrt{2n-E^2}|x|} \varphi_n(y_0) ,\qquad 2n>E^2.
\end{align}
When $2n<E^2$, two propagating modes satisfy the above scalar equation for $G_-$. Keeping only the outgoing mode since we are interested in computing the outgoing Green's function, (i.e., keeping $e^{ikx}$ for $k$ and $x$ positive while $e^{-ikx}$ would be considered incoming), 
we thus observe that 
\begin{align}\label{eq:outgoingGn}
   G_{-,n}(x) =   \frac{1}{-2i\sqrt{E^2-2n}} e^{i\sqrt{E^2-2n}|x|} \varphi_n(y_0) ,\qquad 2n<E^2.
\end{align}

To combine the above calculations into a single expression, we define
\begin{align} \label{eq:thetan}
 \theta_n=  i\sqrt{E^2-2n} =i\xi_{(n,1)}.
\end{align}
The complex numbers $\theta_n$ are thus related to $\xi_m$ introduced earlier in the analysis of the Dirac operator.

We thus find
\begin{align}\label{G_minus}
    G_-(x,y;y_0) = \dsum_{n\geq0} \frac{-1}{2\theta_n} e^{\theta_n|x|} \varphi_n(y) \varphi_n(y_0).
\end{align}

For $G_+$, we have the same expression except that $2n$ is replaced by $2n+2$:
\begin{align}\label{G_plus}
   G_+(x,y;y_0) = \dsum_{n\geq0} \frac{-1}{2\theta_{n+1}} e^{\theta_{n+1}|x|} \varphi_n(y) \varphi_n(y_0).
\end{align}
It remains to apply $H+E$ to get the final result
\begin{align}\label{green_function_matrix}
   G =  \begin{pmatrix} (D_x+E) G_+ & \fa G_- \\ \fa^* G_+ & (-D_x+E) G_- \end{pmatrix}.
\end{align}
Note that $G$ is a $2\times2$ matrix-valued function. We also use the notation $G(x,y;x_0,y_0)=G(x-x_0,y;y_0)$. See also \eqref{Gphi1} below for an explicit expression of each entry of the matrix.

\begin{figure}[htbp]
    \centering
		\begin{subfigure}{0.4\textwidth}
		\includegraphics[width = \linewidth]{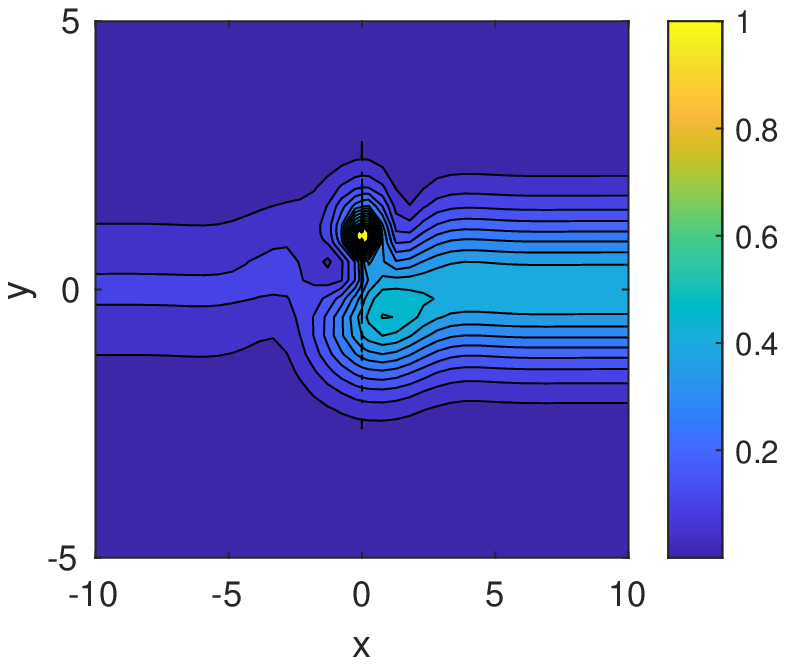}
		\caption{$|G_{11}|$}
	\end{subfigure}
	\begin{subfigure}{0.4\textwidth}
		\includegraphics[width = \linewidth]{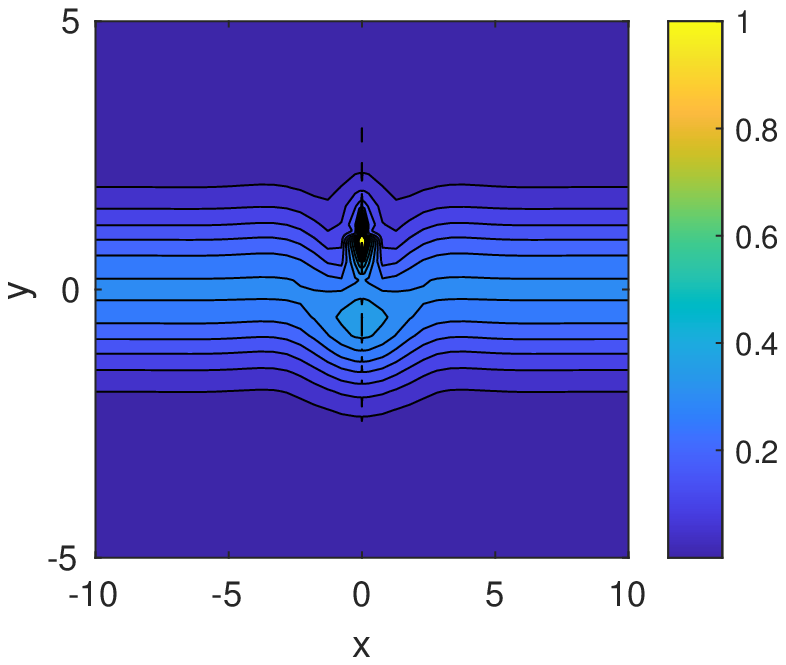}
		\caption{$|G_{12}|$}
	\end{subfigure}
	\begin{subfigure}{0.4\textwidth}
		\includegraphics[width = \linewidth]{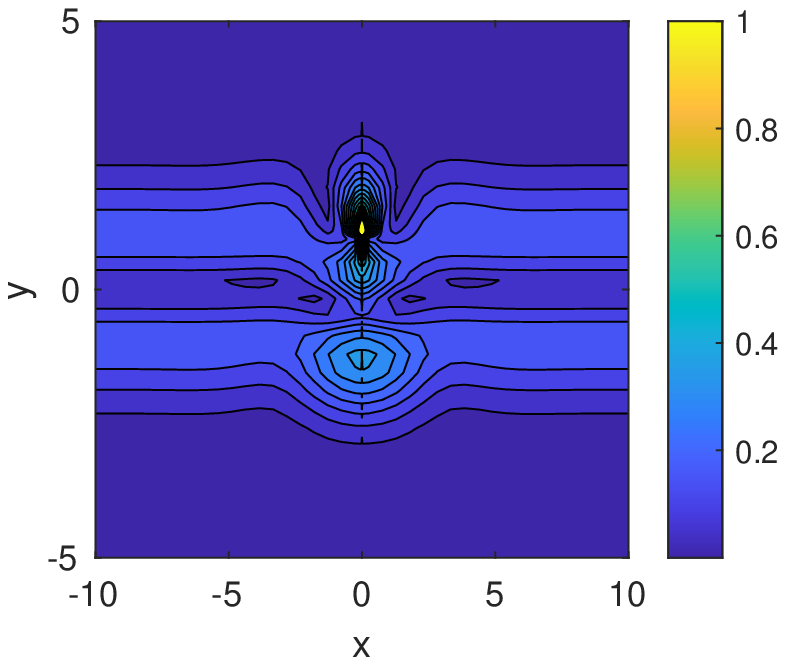}
		\caption{$|G_{21}|$}
	\end{subfigure}
	\begin{subfigure}{0.4\textwidth}
		\includegraphics[width = \linewidth]{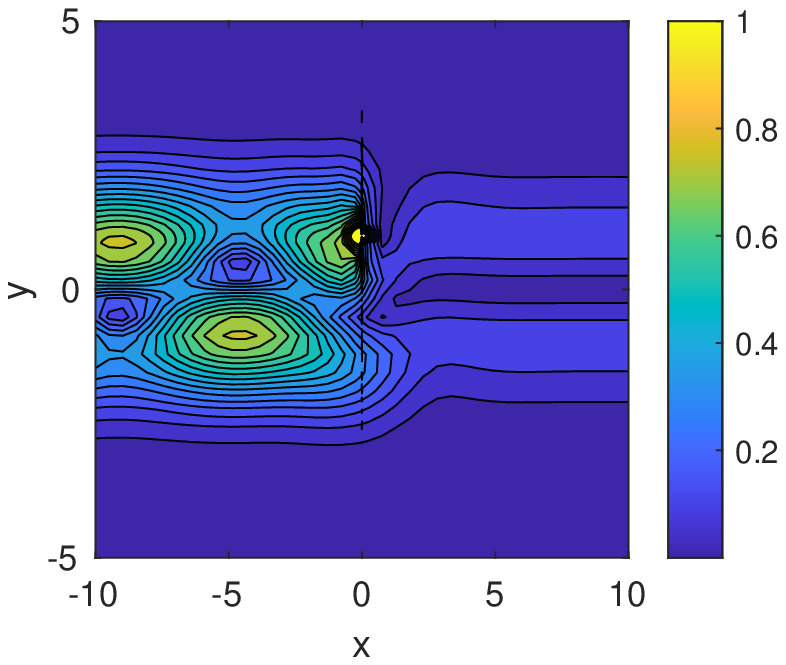}
		\caption{$|G_{22}|$}
	\end{subfigure}
    \caption{ Absolute value (truncated to $[0,1]$) of the components of the Green's function with $x_0=0$, $y_0=1$, and $n_y=100000$.}
    \label{fig:green_function_global}
\end{figure}

\begin{figure}[htbp]
    \centering
		\begin{subfigure}{0.4\textwidth}
		\includegraphics[width = \linewidth]{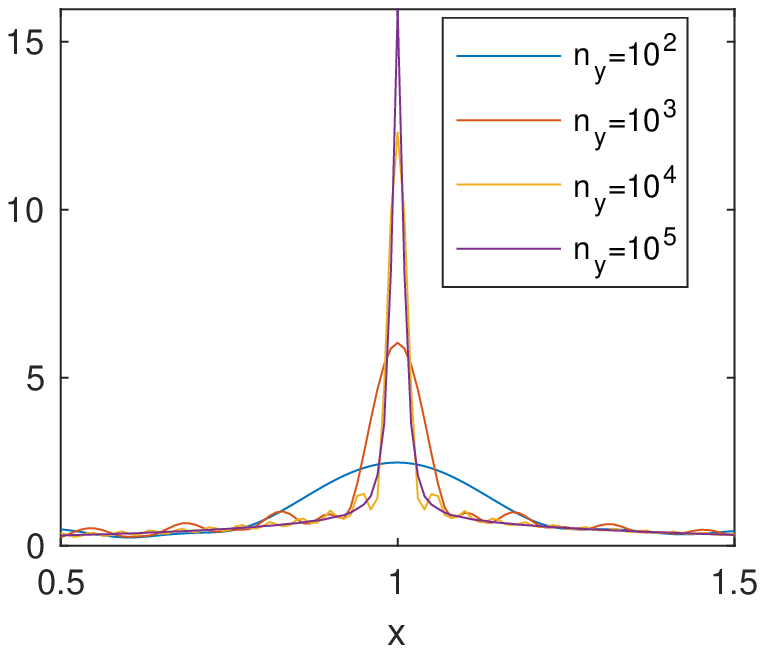}
		\caption{in $x$ direction, $y=y_0+10^{-2}$}
	\end{subfigure}
	\begin{subfigure}{0.4\textwidth}
		\includegraphics[width = \linewidth]{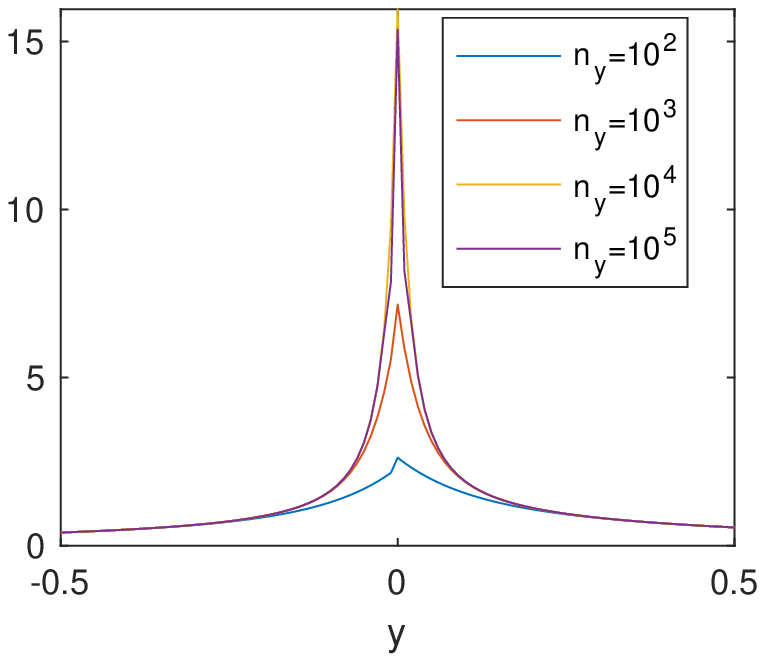}
		\caption{in $y$ direction, $x=x_0+10^{-2}$}
	\end{subfigure}
    \caption{ Local behaviour of approximations of $|G_{11}|$ near $(x_0,y_0)=(0,1)$ for several values of $n_y$.}
    \label{fig:green_function_local}
\end{figure}

\paragraph{Numerical Green's Function.} As a numerical illustration, consider the Green's function $G(x,y,x_0,y_0)$ with $E=1.8$. The source is located at $(x_0,y_0)=(0,1)$. In Fig.\ref{fig:green_function_global}, we plot contours of absolute values of the four entries of $G$. For the first plots, we truncate the series expansion for $G$ at $n_y=10^5,$ and then show how the $|x-x_0|^{-1}$ singularity  near the source \cite{evans2010partial} is captured as $n_y$ varies in Fig.\ref{fig:green_function_local}. We observe in Fig.\ref{fig:green_function_local} that the Green's function is asymmetric in the $x$ direction, which is a manifestation of the transport asymmetry of the unperturbed operator. In the $y$ direction, which is confined by the domain wall, the Green's function decreases exponentially as $|y|$ increases.

In Fig.\ref{fig:green_function_local}, we investigate the local behaviour of the Green's function. It is known that $G$ has a $\frac{1}{r}$ singularity. In \eqref{green_function_matrix}, the spectral series expansions of $G_+$, see \eqref{G_plus}, and $G_-$, see \eqref{G_minus}, converge slowly and have oscillations when $n$ is small near the source. 
\begin{remark}
The convergence of the spectral decomposition of $G$ is necessarily slow since $G$ is singular. However, in the algorithm described below to compute generalized eigenfunctions and densities $\rho$, only the integration of $G$ against smooth functions is involved.
\end{remark}
\subsection{Generalized eigenfunctions and integral formulation}
\label{sec:gal}
We now use the above unperturbed Green's function to solve for generalized eigenfunctions of the perturbed operator $H_V=H+V$. Let $\psin$ be an incoming plane wave, e.g. $\psi_m$ as described in \eqref{eq:psim} with $E^2>2n$ so that $\xi_m$ is real-valued. We then look for solutions of $(H+V-E)\psi=0$ of the form 
\begin{align} \nonumber
    \psi=\psin+\psiout,
\end{align} 
so that
\begin{align}\label{eqn:eigenproblem}
    (H+V-E)\psiout=-V\psin,
\end{align}
where we assume $\psiout$ to be {\em outgoing}. In other words, using the unperturbed outgoing Green's function constructed in the preceding section, we look for 
\begin{align}\label{representation_psiout}
   \psiout(x,y)= \int G(x,y;x_0,y_0)\rho(x_0,y_0)dx_0dy_0,
\end{align}
where $\rho(x,y)$ is an unknown density. Since $G$ is the kernel of the integral operator $(H-E)^{-1}$ (with outgoing conditions), we readily obtain that $\rho$ satisfies the integral equation
\begin{align}\label{IntegralFormulation}
\rho(x,y)+V(x,y)\int G(x,y;x_0,y_0)\rho(x_0,y_0)dx_0dy_0=-V(x,y)\psin(x,y).
\end{align}

In what follows, we assume that $V$ is a smooth Hermitian matrix-valued compactly supported function. As is apparent from the above formulation, the density $\rho$ vanishes outside of the support of $V$. The above formulation is therefore convenient numerically as only the support of $V$ needs to be discretized in practice. 

%\jh{
\begin{remark}
For ease of exposition, here we only consider the case in which the right-hand side of \eqref{IntegralFormulation} is generated by an incoming propagating mode. It is relatively straightforward to modify the approach to include more general (compactly-supported) right-hand sides.
\end{remark}
%}
%
\paragraph{Properties of the density $\rho(x,y)$.}
The above equation for $\rho$ can be re-written in the form 
\begin{align}\label{eq:rho1}
    \rho + V \mG \rho = -V\psin 
\end{align}
with $\mG=(H-E)^{-1}$ the linear operator with kernel $G(x,y;x_0,y_0)$. Since the Dirac operator is elliptic, the inverse $\mG$ maps $H^k(\Rm^2;\Cm^2)$ to $H^{k+1}(\Rm^2;\Cm^2)$, where $H^k(\Rm^2;\Cm^2)$ is the space of $\Cm^2$ valued functions with square integrable derivatives of order up to $k$. Since $V$ is smooth and compactly supported, we obtain that $V\mG$ is a compact operator from $H^k(\Rm^2;\Cm^2)$ to itself for any $k\geq0$. 

The above equation thus admits a unique solution provided $-1$ is not an eigenvalue of $V\mG$ (in any $H^k(\Rm^2;\Cm^2)$). The complement of the latter condition is then equivalent to the existence of a normalized (in $L^2(\Rm^2;\Cm^2)$) solution of the problem
\begin{align}\label{eq:pointspectrum}
    (H+V-E)\psi=0.
\end{align}

While solutions of \eqref{eq:pointspectrum} for Schr\"odinger and Dirac (with constant mass term) operators $H$ are known not to exist for $E$ in the continuous spectrum of $H$ when $V$ is compactly supported, see for instance \cite{berthier1987point} and \cite[Theorem XIII.58]{RS4}, the confinement afforded by the domain may in fact generate such point spectrum. We revisit the question in section \ref{sec:localizedmodes}, where we show the presence of localized modes for certain choices of $V$ which are compactly supported in $x$.

When $-1$ is not in the spectrum of the compact operator $V\mG$, we apply the Fredholm alternative to \eqref{eq:rho1} and formally define
\begin{align} 
    \rho = -(I+V\mG)^{-1}  V\psin.
\end{align}
Since $V\psin$ is smooth and supported on the support of $V$, we obtain that $\rho$ is itself smooth and compactly supported and therefore an element in $C^\infty_c(\Rm^2;\Cm^2)$.

We now show that $-1$ may be in the spectrum of $V\mG$ only for discrete values of $E$ when $V$ is compactly supported (exponential decay being sufficient). Indeed, following the proof in, e.g., \cite[Theorem XI.45]{RS3}, $E\to V\mG$ is analytic in the vicinity $\mathcal V$ of any interval $[E_1,E_2]\subset\Rm$ such that $\{2n;\ n\in\Nm\}\cap [E_1^2,E_2^2]=\emptyset$. Since $H+V$ is self-adjoint, $-1$ cannot be an eigenvalue of $V\mG$ as soon as $E\in\mV$ is not purely real. The analytic Fredholm theory \cite[Theorem VI.14]{RS} then states that $-1$ is an eigenvalue of $V\mG$ only for discrete (real) values of $E$. For the rest of the paper, we assume that $E$ does not belong to any of these discrete sets or to $\{\sqrt{2n};\ n\in\Nm\}$.

\paragraph{Basis expansion of the density $\rho$.}
 In the following derivation, we denote
\begin{align}\label{rho_coeff}
    \rho(x,y)=\sum_{i,n} \rho_{i,n}
P_i(x)\varphi_n(y),
\end{align}
where $P_i$ denotes the i-th Legendre polynomial and $\varphi_n$ denotes $n$-th order Hermite functions, while $\rho_{i,n}\in \Cm^2$ are the corresponding generalized Fourier coefficients. 
The generalized eigenfunction solution of $(H+V-E)\psi=0$ is then given by $\psi=\psin+\mG\rho$.

From the expression \eqref{green_function_matrix} for the Dirac Green's function, we observe that the integral operator in \eqref{IntegralFormulation} involves explicit integrals given by
\begin{align}
   &\int G(x,y;x_0,y_0)\begin{pmatrix}
    \varphi_n(y_0)\\0
    \end{pmatrix}dy_0=
    \begin{pmatrix}
    \frac{1}{2}(i \sgn{x-x_0}-\frac{E}{\theta_{n+1}})e^{\theta_{n+1}|x-x_0|}\varphi_{n}(y)\\
     -\frac{\sqrt{2n+2}}{2\theta_{n+1}} e^{\theta_{n+1}|x-x_0|}\varphi_{n+1}(y)
    \end{pmatrix}\label{Gphi1}
\\ 
   &\int G(x,y;x_0,y_0)\begin{pmatrix}
    0\\\varphi_n(y_0)
    \end{pmatrix}dy_0=
    \begin{pmatrix}
    -\frac{\sqrt{2n}}{2\theta_n} e^{\theta_n|x-x_0|}\varphi_{n-1}(y)\\
    \frac{1}{2}(-i \sgn{x-x_0}-\frac{E}{\theta_n})e^{\theta_n|x-x_0|}\varphi_{n}(y)
    \end{pmatrix}. \label{Gphi2}
\end{align}
We may then compute $\psiout$ from $\rho$ using \eqref{representation_psiout}. Outside of the support of $\rho$, we observe that the approximation of $\psiout$ decays exponentially as $|y|$ increases and is a sum of propagating modes that oscillate in the $x-$variable along the interface $y\equiv0$ and of evanescent modes that decay exponentially away from the support of $\rho$.

\section{Numerical preliminaries}
\label{sec:prelim}
\subsection{Global discretization}
\label{sec:single_discretization}
In order to obtain a numerical approximation of the density $\rho,$ we truncate the expansion of $\rho$ in \eqref{rho_coeff}, keeping a finite number of coefficients. In the following, we denote the solution of this truncated system by $\hat{\rho}.$ Additionally, we denote the finite product space consisting of products of linear combinations of the  first $n_x-1$ Legendre polynomials in $x$ with linear combinations of the first $n_y$ Hermite functions in $y$ by $\mV_{n_x,n_y}.$ In particular,
$\mV_{n_x,n_y}= \mP_{n_x} \otimes (\Phi_{n_y} \oplus \Phi_{n_y})$, where $\mP_{n_x}$ denotes the span of the first $n_x$ Legendre polynomials in $x$, and $\Phi_{n_y}$ denotes the span of the first $n_y$ Hermite functions in $y.$ In the sequel we use $\mV$ for short. Empirically, we observe that in this basis the solution of the projected system converges spectrally (super-algebraically) in $n_x$ and $n_y$ to the solution of the original system, given sufficient regularity of the potential $V.$ Interested readers can refer to \cite{kress1989linear} for applicable theorems on convergence. 

Clearly, there are two linear operators in \eqref{IntegralFormulation} which must be truncated and whose action must be computed numerically: integration with $G$ and multiplication of $V$. 
\begin{itemize}
\item \textbf{ Kernel integration with $G$.} The integration with respect to $y_0$ is analytically available, given by equations \eqref{Gphi1}
 and \eqref{Gphi2}. In the $x$ direction, the integrands have a singularity in their derivatives when $x_0=x$. Thus, using the $n_x$ point quadrature rule for Legendre polynomials (i.e. the $n_x$ Gauss-Legendre quadrature rule) leads to slow convergence of the integral over the support of $V$ in $x$, which we denote by $[x_L, x_R].$ To circumvent this issue, for any fixed $x\in[x_L, x_R]$, we divide the interval $[x_l,x_R]$ into two subintervals $[x_L, x]$ and $[x,x_R]$. On both intervals, the integrand is smooth and we can apply standard smooth quadrature rules to compute the integral of the Green's function (after integrating over $y_0$ analytically) against the basis functions (i.e. suitably rescaled, shifted, and dilated Legendre polynomials). Furthermore, we observe that by choosing the values of $x$ at which to evaluate the integrals as the nodes of the $n_x$-point Gauss-Legendre quadrature rule, the map from values at these nodes to the coefficients in an $n_x$-th order Legendre expansion is invertible and stable. We let $\hat{G}$ denote the $2n_x n_y\times2n_xn_y$ square matrix obtained by discretizing the application of the Green's function in this way.

    \item {\bf Multiplication by $V$.} We recall that by assumption, $V$ is smooth and compactly supported in both the $x$ and $y$ directions. Evidently, pointwise multiplication by $V$ is a linear operator, and in the orthonormal basis $\mV,$ has matrix entries given by  $(\mV_i,V\mV_j)_{x,y}$, where $\mV_i$ and $\mV_j$ are two basis functions from $\mV$. To compute it, we first evaluate $\mV_j$ on a tensor product of Legendre and Hermite quadrature nodes. The orders of these quadrature rules should be large than $n_x$ and $n_y$, respectively. Indeed, using standard estimates from approximation theory it is possible to derive explicit upper bounds for the oversampling required as a function of the smoothness and support of $V.$ In practice, we observe that for most of the examples considered here, increasing the integration orders by no more than 20 suffices, though obviously this would need to be increased for rougher $V$'s. One then computes the product of $\mV_i,$ $\mV_j,$ and $V,$ at these nodes and sums against the quadrature weights. This process generates a $2n_x n_y\times2n_xn_y$ square matrix $\hat{V}$ which encodes this transformation.
\end{itemize}
In addition to the previously defined $\hat{G}$ and $\hat{V}$, we denote $\Pi_\mV$ as the quadrature projection to the product basis $\mV$. Using these matrices, the discretization of our integral formulation is given by
\begin{align} \nonumber
(I+\hat{V}\hat{G})\hat{\rho}=-\hat{V}\Pi_\mV\psin.
\end{align}

The last step in computing the generalized eigenfunctions is to recover $\psiout$ from the integral representation \eqref{representation_psiout}. The eigenfunction $\psi$ is given by 
\begin{align}
\hat{\psi}(x,y)=\psin(x,y)+\int G(x,y;x_0,y_0)\hat{\rho}(x_0,y_0)dx_0dy_0.
    \label{eqn:psinumerical}
\end{align}
It is worth emphasizing that after the computation of $\rho$ we are not restricted to evaluate $\psi_{out}$ only within the support of $V.$ Indeed, again using the integral representation \eqref{representation_psiout}, $\psiout$ can be computed directly by numerically evaluating $\int G(x,y;x_0,y_0)\rho(x_0,y_0)dx_0dy_0$  at any point $(x,y)$. The discretization of this integral is similar to the one employed in the construction of $\hat{G}$, i.e., in the $x$ direction, we can split the domain of integration, if necessary, at $x_0=x,$ while in the $y$ direction, \eqref{Gphi1}, \eqref{Gphi2} give analytic expressions for the integrals.

\subsection{Computing eigenfunctions by merging TR matrices}
\label{sec:meging_alg}
In this section, we use the \textbf{transmission reflection (TR) matrix} formalism (see \cite{kexiang1999modal} and the references therein, for example) to solve the integral equation (\eqref{IntegralFormulation}) on large domains.

We begin by defining the TR matrices. To that end, first suppose that $V$ is compactly supported in the x direction on the interval $[x_L,x_R]$ and $\psi$ is the eigenfunction of $H-E+V$ given some incoming condition. We decompose $\psi$ as
\begin{align}\label{psi_decomposition}
	\psi(x,y) = \dsum_{m\in M}  \alpha_m(x)  \phi_m(y),
\end{align}
with $\alpha_m(x)=(\phi_m(y),\psi(x,y))_y$ proportional to $e^{i\xi_m x}$ outside of the support of $V$ and $\phi_m$ defined in \eqref{def:phi}. In this way, instead of mapping the incoming field to the outgoing field in the far field, we consider the linear map right at the boundary of $[x_L, x_R]$.  The incoming condition is then the coefficients of the right traveling modes $\phi_m$ ($\epsilon_m>0$) at left boundary and the left traveling modes $\phi_m$ ($\epsilon_m<0$) at the right boundary, namely  $\alpha_{+}(x_L)$ and $\alpha_{-}(x_R)$. The outgoing solution consists of the coefficients of the left traveling modes $\phi_m$,($\epsilon_m<0$), at left boundary and the right traveling modes $\phi_m$ ($\epsilon_m>0$) at right boundary, namely  $\alpha_{-}(x_L)$ and $\alpha_{+}(x_R)$. In summary, the TR matrix for $V$ restricted to the interval $[x_L,x_R]$, denoted by $M$, can be defined as,
\begin{align}\label{def:TRMatrix}
    \begin{pmatrix}
    \alpha_{-}(x_L)\\
    \alpha_{+}(x_R)
    \end{pmatrix}=
    \begin{pmatrix}
    M_{11} & M_{12}\\
    M_{21} & M_{22}
    \end{pmatrix}
    \begin{pmatrix}
    \alpha_{-}(x_R)\\
    \alpha_{+}(x_L)
    \end{pmatrix}.
\end{align}
In numerical computation, if we restrict the density $\rho$ to $\Phi_{n_y}$ in the $y$ direction, in which case the TR matrix is a $(2n_y-1)\times(2n_y-1)$ matrix. $M_{11}$ ($M_{22}$) denotes the left (right) transmission coefficients, and $M_{21}$ ($M_{12}$) denotes the left (right) reflection coefficients. 

\paragraph{Merging Two TR Matrices}
It is of particular interest in physics to consider the case when there is an elongated perturbation near the interface. The TR matrix in this case will depend on the length of interface, $l$, in the $x$ direction. Obtaining the TR matrix requires solving for the total field $\psi$ for each of the $2n_y-1$ independent modes as incoming conditions. When solving $\psi$, if the support of $V$ in the $x$ direction is large, accurate computation of $\rho$ requires very high order quadrature rules in the $x$ direction in order to resolve the local structure of $V$ and $\rho$ in a relatively long interval. Clearly, a naive implementation then leads to a significant computational cost (both in time and memory), as well as possible numerical instabilities. To avoid this, we break the problem up into small intervals or {\it leaves} in the $x$ direction and compute the TR matrices of these sub-problems. It is then possible to {\it merge} two TR matrices for adjacent intervals to obtain a TR matrix for their union. We can continue merging in this way until the TR matrix for the entire domain is obtained.

To that end, let $L$ and $R$ denote the TR matrices of two adjacent interval in the $x$ direction, $I_L$ and $I_R$, respectively. Then the TR matrix for $I_L\cup I_R$ is 
\begin{align}
    \begin{pmatrix}
    L_{11}(I-R_{12}L_{21})^{-1}R_{11} & L_{11}(I-R_{12}L_{21})^{-1}R_{12}L_{22} +L_{12}\\
    R_{22}(I-L_{21}R_{12})^{-1}L_{21}R_{11}+R_{21} & R_{22}(I-L_{21}R_{12})^{-1}L_{22}
    \end{pmatrix}\label{scattermatrix_formula}
\end{align}
Furthermore, given incoming waves at the left and right boundary, we can also calculate the Fourier coefficient at the intersection of two merging intervals. Specifically, the \textbf{intersection coefficient matrix} $\mM$ is defined by,
\begin{align}\label{eqn:midmatrix}
    \begin{pmatrix}
    \alpha_{-,\mM}\\
    \alpha_{+,\mM}
    \end{pmatrix}=
    \begin{pmatrix}
    \mM_{11} & \mM_{12}\\
    \mM_{21} & \mM_{22}
    \end{pmatrix}
    \begin{pmatrix}
    \alpha_{-,R}\\
    \alpha_{+,L}
    \end{pmatrix},
\end{align}
where $\begin{pmatrix}
    \alpha_{-,\mM}\\
    \alpha_{+,\mM}
    \end{pmatrix}$ denotes the Fourier coefficient projected to unperturbed eigenfunction basis $\{\phi_m\}$ at the intersection. $\mM$ can be directly calculate by,
\begin{align}
    \mM=\begin{pmatrix}
    (I-R_{12}L_{21})^{-1}R_{11} & (I-R_{12}L_{21})^{-1}R_{12}L_{22}\\
    (I-L_{21}R_{12})^{-1}L_{21}R_{11} & (I-L_{21}R_{12})^{-1}L_{22}
    \end{pmatrix}.\label{intermatrix_formula}
\end{align}

\begin{figure}[ht!]
	\centering
		\includegraphics[width = 1 \linewidth]{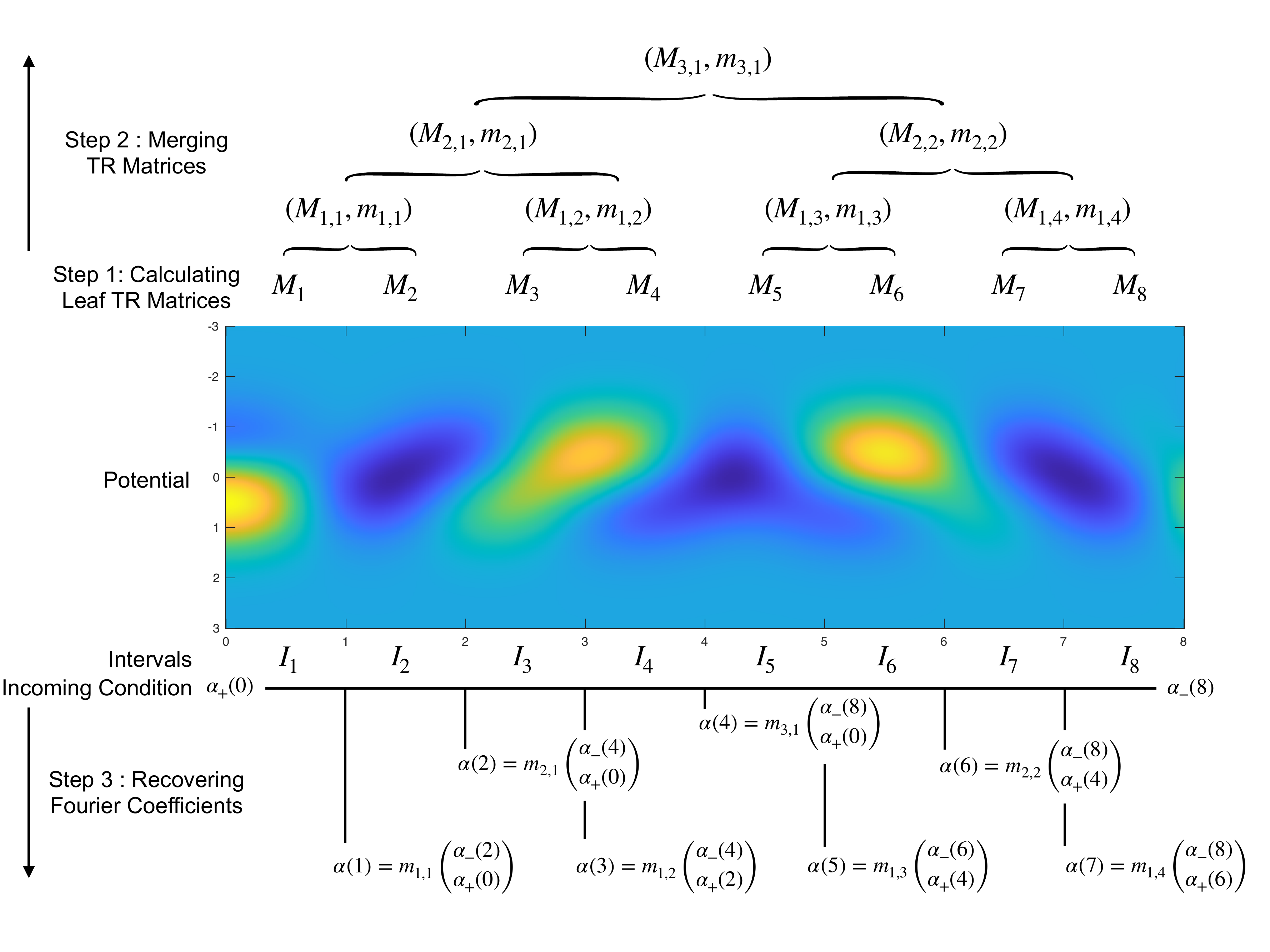}
		\caption{An Illustration of Merging/Recovering Algorithm}
    \label{fig:illustration}
	\end{figure}

\paragraph{Binary merging and recovering the eigenfunctions inside the material}
 Merging TR matriices is independent of interval position and length, so for a general partition of a single interval we can compute the TR matrix first for each leaf by Alg.\ref{alg_rho} and merge them. Here we apply a binary merging strategy. To be more precise, we first partition the whole interval into $2^L$ leaves $\{I_i\}_{i=1:2^L}$ of equal length. Here $L$ denotes the total number of levels. We compute the TR matrices on each interval $I_i$ and merge two adjacent intervals $I_{2k-1}$ and $I_{2k}$ where $k=1\cdots2^{L-1}$. We continue this merging $L$ times (i.e. at each level) until we have the TR matrix for the whole interval. 
 When merging two adjacent intervals, the TR matrix reflects the outgoing waves as a linear transform of incoming waves. The intersection coefficient matrix then reveals the Fourier coefficient ($\alpha$) at intersection is a linear function of incoming waves at the boundaries. In this way, after merging all the intervals, we can recover $\alpha$ from the last level of merging back to the first level. The full algorithm is summarized in Alg.\ref{alg}. An illustration of the algorithm is given by Fig.\ref{fig:illustration}, where, for concreteness, we compute the TR matrix of $[0,8]$.
\section{The algorithm}
\label{sec:alg}
In this section we summarize our algorithms for computing generalized eigenfunctions for compact smooth perturbations of Dirac equations with linear domain walls. The first algorithm (Alg.\ref{alg_rho}) computes the density and eigenfunction in a single slab without merging. As input, it takes in a perturbation $V$ (a compactly supported smooth, $2\times 2$ Hermitian matrix-valued function), an interval $[x_l,x_R]$ containing the support of $V$ in the $x$-direction, an integer $n_y$ giving the order of the Hermite expansion to be used, an integer $n_x$ giving the order of the Legendre expansion to be used  It also requires as input two vectors $(\alpha_{+},\alpha_{-1}) \in \mathbb{C}^{n_y-1}\times\mathbb{C}^{n_y}$ corresponding to the amplitudes of the incoming waves. As output it returns a tensor-product (piecewise) Legendre - Hermite approximation to the density $\rho$ and outgoing amplitudes $\alpha_-(x_L)$ and $\alpha_+(x_R)$.
\begin{minipage}{\textwidth}
% switch off footnote line locally
\renewcommand*\footnoterule{}
\begin{savenotes}
\begin{algorithm}[H]
	\caption{Computing density and eigenfunction in a single slab (leaf)}  \label{alg_rho}
	\begin{algorithmic}[1]
	\Require Potential Field $V$; Interval of $V$ that is compacted supported $I=[x_L,x_R]$; Level of binary merging $L$;  Incoming wave condition $\alpha_+(x_L)$ and $\alpha_-(x_R)$; discretization configuration $(n_x,n_y)$.
	\Ensure Density $\hat{\rho}=\hat{\rho}^{n_x,n_y}$ and eigenfunction $\hat{\psi}$ given incoming condition; outgoing wave $\alpha_-(x_L)$ and $\alpha_+(x_R)$.
	\State Construct $\hat{V}$ and $\hat{G}$ projected to $n_x$ Legendre polynomials and $n_y$ Hermite functions. 
	\State Compute $\Pi_{\mV}\psin$ with $\alpha_-(x_L)$ and $\alpha_+(x_R)$.
	\State Solve $\hat{\rho}$ by $\hat{\rho}=-(I+\hat{V}\hat{G})^{-1}\hat{V}\Pi_\mV\psin$.\footnote{$(I+\hat{V}\hat{G})^{-1}$ can be re-used when constructing TR matrix.\\ \mbox{$\quad$} * optional step}
	
	\State Recover $\psi$ by $\hat{\psi}(x,y)=\psin(x,y)+\int G(x,y;x_0,y_0)\hat{\rho}(x_0,y_0)dx_0dy_0$.*
    \State Extract $\alpha_-(x_L)$  by $\alpha_{n,-}(x_L)=\int \phi_{n,-}(y)G(x_L,y;x_0,y_0)\hat{\rho}(x_0,y_0)dx_0dy_0dy$.*
    \State Extract $\alpha_+(x_R)$ by $\alpha_{n,+}(x_R)=\int \phi_{n,+}(y)G(x_R,y;x_0,y_0)\hat{\rho}(x_0,y_0)dx_0dy_0dy$.*
	\end{algorithmic}
	\end{algorithm}	
\end{savenotes}
% footnotes are displayed at the end of the savenotes environment
\end{minipage}

The second algorithm (Alg.\ref{alg}) describe the hierarchical merging that compute generalized eigenfunctions efficiently. In addition to inputs that are required by Alg.\ref{alg_rho}, we used $L$ to specified total level of merging.

\begin{minipage}{\textwidth}
% switch off footnote line locally
\renewcommand*\footnoterule{}
\begin{savenotes}
\begin{algorithm}[H]
	\caption{Computing eigenfunctions in a slab with potential $V$}  \label{alg}
	\begin{algorithmic}[1]
	\Require Potential Field $V$; Interval of $V$ that is compacted supported $I=[x_L,x_R]$; Level of binary merging $L$;  Incoming wave condition $\alpha_+(x_L)$ and $\alpha_-(x_R)$.
	\Ensure TR matrix of interval $I$, $M_I$;  eigenfunctions given incoming wave, $\alpha$.
	\State Partition $I$ to be $2^L$ intervals. Denote the intervals as $\{I_k\}$, $k=1,2,\cdots, 2^L$ and the grid points as $x_k$, $k=0,1,\cdots, 2^L$.
	\For{$k$ in $1,\cdots ,2^L$}
	\State Compute TR matrix $M_{0,k}$ with potential field $V$ limited on interval $I_k$ by Alg.\ref{alg_rho}.
	\EndFor
	\For{$l$ in $1\to L$}
	\For{$k$ in $1\to2^{(L-l)}$}
	    \State Merge $M_{l,2k-1}$ with $M_{l,2k}$:
	    \State \quad Calculate the TR matrix $M_{l+1,k}$ for $[x_{(k-1)2^l},x_{k2^l}]$ from \eqref{scattermatrix_formula} 
	    \State \quad Calculate intersection coefficient matrix $\mM_{l+1,k}$ at $x_{k2^l-2^{l-1}}$ from \eqref{intermatrix_formula}.
	\EndFor 
	\EndFor
	$M_I=M_{L+1,1}$
	\State Assign $\alpha_+(x_0)=\alpha_+(x_L)$, $\alpha_-(x_0)=0$,$\alpha_+(x_{2^L})=0$,$\alpha_-(x_{2^L})=\alpha_-(x_R)$.
	\For{$l$ in $L\to 1$}
	\For{$k$ in $1,\cdots ,2^{(L-l)}$}
	\State $\begin{pmatrix}
 \alpha_-(x_{(2k-1)2^{l-1}})\\
 \alpha_+(x_{(2k-1)2^{l-1}})
 \end{pmatrix}=\mM_{l+1,k}\begin{pmatrix}
 \alpha_-(x_{k2^{l}})\\
 \alpha_+(x_{(k-1)2^{l}})
 \end{pmatrix}$
	\EndFor
	\EndFor
\For{$k$ in $1,\cdots,2^L$}
\State Recover $\rho|_{[x_{k-1},x_k]}$ by $\alpha_+(x_{k-1})$ and $\alpha_-(x_k)$ via Alg.\ref{alg_rho}.\footnote{optional step, we can then compute values of $\psiout$ at any point by \eqref{representation_psiout}. }
	\EndFor
	\end{algorithmic}
\end{algorithm}	
\end{savenotes}
% footnotes are displayed at the end of the savenotes environment
\end{minipage}
	
\begin{remark}
For ease of exposition, here we have restricted our attention to uniform leaf size. The method described above generalizes naturally to the non-uniform case, in which the leaf size is chosen to depend on the local smoothness of $V.$ Moreover, it can also be easily extended to the fully adaptive case in which, after a coarse solve, leaves are refined adaptively until the solution $\rho$ is fully resolved. See \cite{lee1997} for a description of this approach in the context of ODE solvers.  
\end{remark}	
\iffalse
\begin{remark}
The merging procedure used here is a minor modification of  \cite{greengard1991numerical}, where it was used to solve two point boundary value problems for second order ODEs. Similar merging procedures arise in a number of contexts in the integral equations community, see \cite{} for example. 
Additionally, when solving the PDE directly, there are similar constructions which appear in scattering problems electromagnetic and acoustic scattering problems \cite{}, as well as in the study of  electromagnetic waveguides, see \cite{} for example. Recently, there have been several papers which have applied these techniques to related problems involving the Dirac equation \cite{}.
\end{remark}
\fi

\section{Numerical results}
\label{sec:num}
\begin{figure}[htbp]
	\centering
	\begin{subfigure}{0.95\textwidth}
		\includegraphics[width = \linewidth]{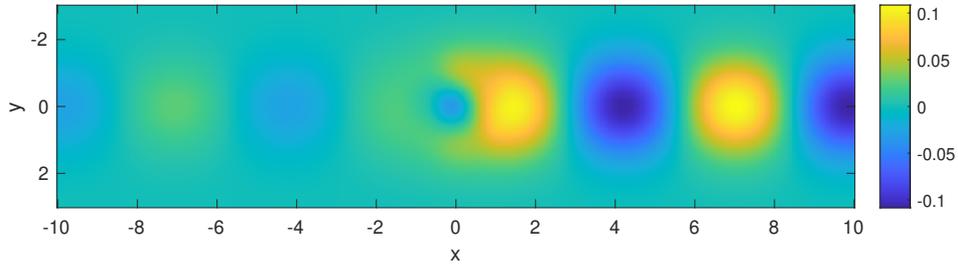}
		\caption{Real part of $\psi_1$}
	\end{subfigure}
		\begin{subfigure}{0.95\textwidth}
		\includegraphics[width = \linewidth]{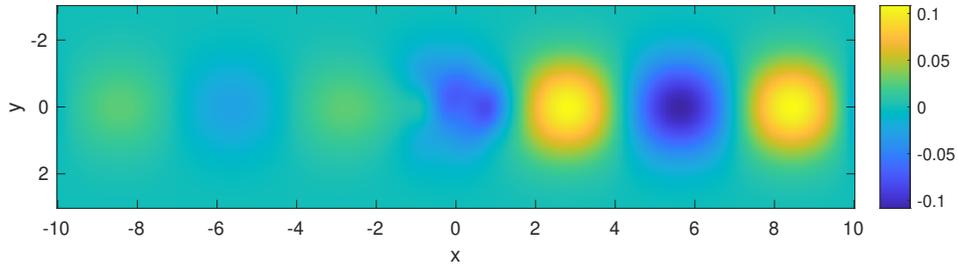}
		\caption{Imaginary part of $\psi_1$}
	\end{subfigure}
		\begin{subfigure}{0.95\textwidth}
		\includegraphics[width = \linewidth]{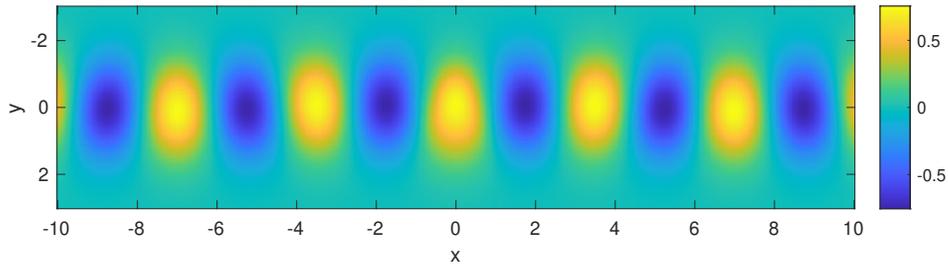}
		\caption{Real part of $\psi_2$}
	\end{subfigure}
		\begin{subfigure}{0.95\textwidth}
		\includegraphics[width = \linewidth]{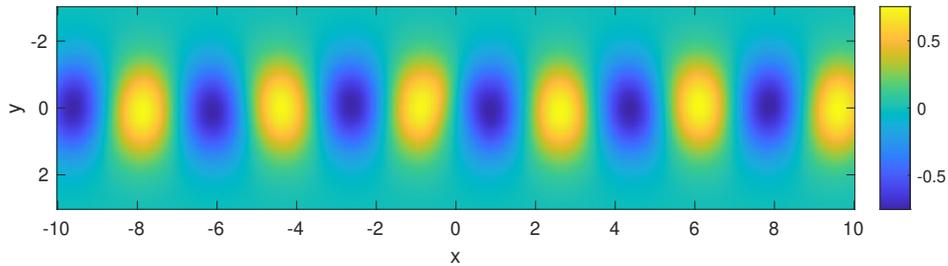}
		\caption{Imaginary part of $\psi_2$}
	\end{subfigure}
		\begin{subfigure}{0.95\textwidth}
		\includegraphics[width = \linewidth]{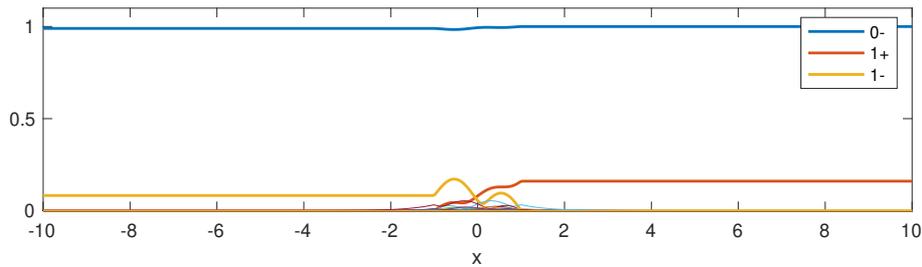}
		\caption{Amplitude of Fourier coefficients, $|\alpha|$; Slender lines denotes evanescent modes.}
	\end{subfigure}
	\caption{Solution $\psi=\psi_{in}+\psi_{out}$}\label{fig:solution}
	\end{figure}

\paragraph{Numerical solution in one interval.}
As a first example, we consider a single interval with potential $V=V_1\chi_I$ where 
\begin{align}\label{V1}
    V_1=\exp \left(-y^{2}\right) y \cos \left(\left(\xi_{(0,-1)}-\xi_{(1,1)}\right)x\right), 
\end{align}
%where we recall that $\xi_{0-}$ and $\xi_{(1,1)}$ are shorthand notation for $\xi_{(0,-1)}$ and $\xi_{(1,1)}$. 
and it is supported on $I=[-1,1]$. For the incoming field we choose $\psi_{in}=e^{i\xi_{(0,-1)}x}\phi_{(0,-1)}(y)$. It is a left traveling plane wave in $x$ direction. %+e^{i\xi_{(1,-1)}x}\phi_{1-}(y)+e^{i\xi_{(1,1)}x}\phi_{1+}(y)
We compute $\rho$ via Alg.\ref{alg_rho} with $n_x=20$ and $n_y=108$ (we will justify the accuracy in the following paragraph). In Fig.\ref{fig:solution}, we recover the generalized eigenfunction obtained via the formula \eqref{eqn:psinumerical}.

\begin{figure}[htbp]
	\centering
	\begin{subfigure}{0.45\textwidth}
		\includegraphics[width = \linewidth]{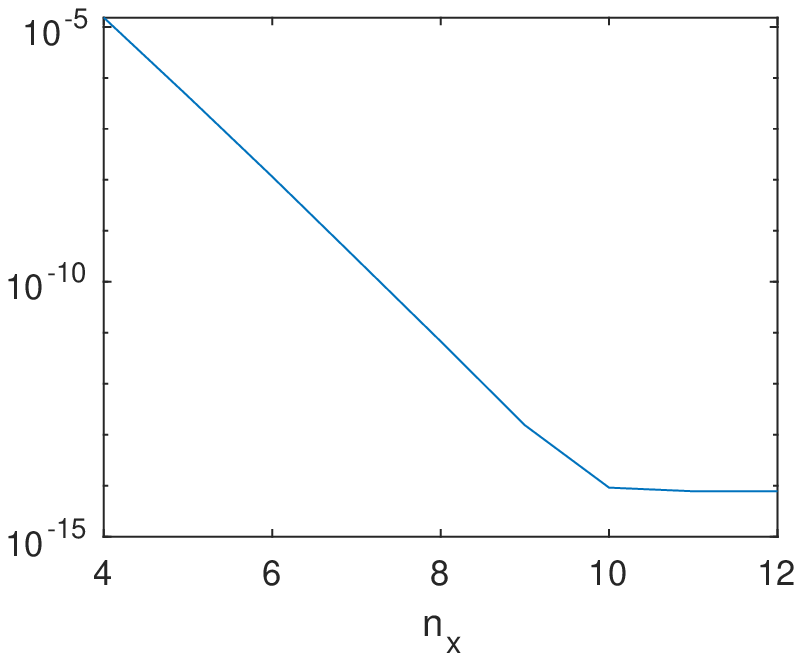}
		\caption{Convergence with respect to $n_x$, i.e. $e_{n_x,n_y'}$ versus $n_x$}
	\end{subfigure}
		\begin{subfigure}{0.45\textwidth}
		\includegraphics[width = \linewidth]{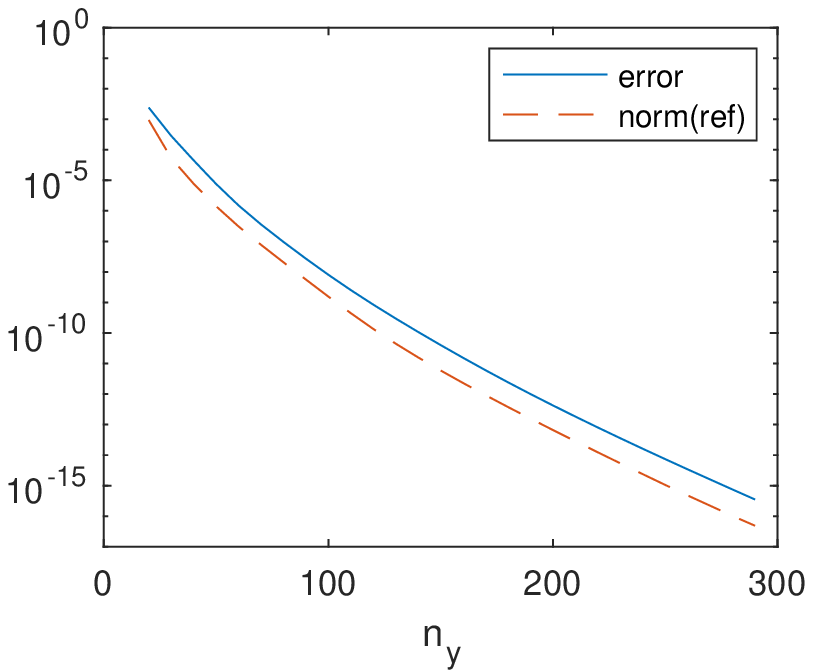}
		\caption{Convergence with respect to $n_y$, i.e. $e_{n_x',n_y}$ versus $n_y$}
	\end{subfigure}
	\caption{Convergence of $\rho$}\label{fig:rho_convergence}
	\end{figure}

Looking at Fig.\ref{fig:solution}, we note that our integral formulation preserves the plane wave structure outside the support of $V$, i.e. the interval $I$. Moreover, in comparing the plot of $\psi_1$ with $\psi_2$, one can clearly observe the difference in wavelength. This is expected, since the  $(0,-1)$ mode vanishes in the first entry of $\psi$ and hence the dominant wavelength in $\psi_2$ is $\frac{2\pi}{\sqrt{E^2-2}}\approx 5.64$ and the dominant wavelength in $\psi_1$ is $\frac{2\pi}{E}\approx 3.49$.
\iffalse
From the amplitude plot of the Fourier coefficients $|\alpha|$ of $\psi$ projected onto the $\{\phi_m\}$ basis, we can see the amplitude of $(1,1)$ and $(1,-1)$ grows to roughly $0.1$. This comes from the scattering by the potential $V$.  There is also a amplitude difference between left and right side of $\psi_1$ which represents the amplitude difference between outgoing $(1,-1)$ and $(1,1)$ mode. The solution has some oscillation inside the interval $I$ (eg. see $\psi_1$ near $x=0$). They represent the evanescent modes which is developed by potential.
\fi
\medskip
\paragraph{Verification of spectral convergence.}
Next, we discuss the results of a self-convergence study to verify the spectral convergence in computing $\rho$ by Alg.\ref{alg_rho}. In Fig.\ref{fig:rho_convergence} we show the error between the reference solution and the solutions with different discretization orders ( $n_x$ or $n_y$). The error in our experiment is an approximated $L_2$ relative error. Recalling definition of $\rho_{i,n}$ in \eqref{rho_coeff}, we denote $\hat{\rho}_{in}^{n_x,n_y}=(P_i\phi_n,\hat{\rho}^{n_x,n_y})_{x,y}$, where $\hat{\rho}^{n_x,n_y}$ is the numerical solution computed in Alg.\ref{alg_rho} on the product space of $n_x$ Legendre polynomial and $n_y$ Hermite basis. Then the error in the convergence study is defined as,
\begin{align}
    e_{n_x,n_y}=\sqrt{\frac{\dsum_{i,n}(\hat{\rho}_{in}^{n_x,n_y}-\rho_{in}^{n_x',n_y'})^2}{\dsum_{i,n}(\rho_{in}^{n_x',n_y'})^2}},
\end{align}
where $\rho^{n_x',n_y'}$ is the solution computed from finer discretization, which is considered as ground truth. In our study, we take $n_x'=16$, $n_y'=300$ on the interval $I=[0,100/1024]$.

Clearly, $\hat{\rho}$ converges to around machine precision ($10^{-15}$) when $n_x\geq 10$. We remark that the convergence with respect to $n_x$ is significantly faster than the convergence with respect to $n_y$, since we are considering a relatively thin leaf in the $x$ direction, so the decay of coefficients is faster in the $x$ direction. In Fig.\ref{fig:rho_convergence}(b), in addition to the error, we also plot the norm of the Fourier coefficients of the corresponding Hermite function basis. Clearly the error decays at a similar rate to the ground truth coefficients.

\begin{figure}[htbp]
	\centering
	\begin{subfigure}{0.32\textwidth}
		\includegraphics[width = \linewidth]{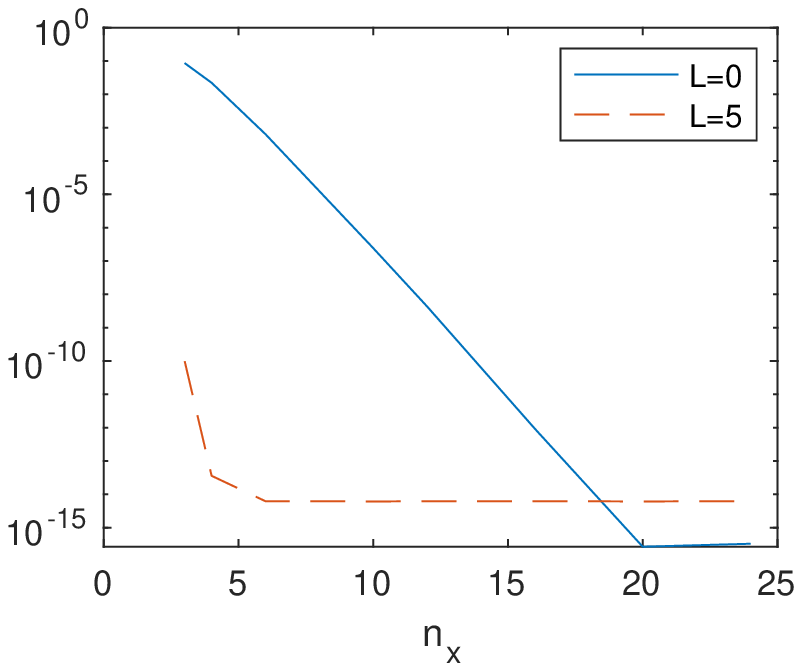}
		\caption{Convergence with respect to $n_x$}
	\end{subfigure}
		\begin{subfigure}{0.32\textwidth}
		\includegraphics[width = \linewidth]{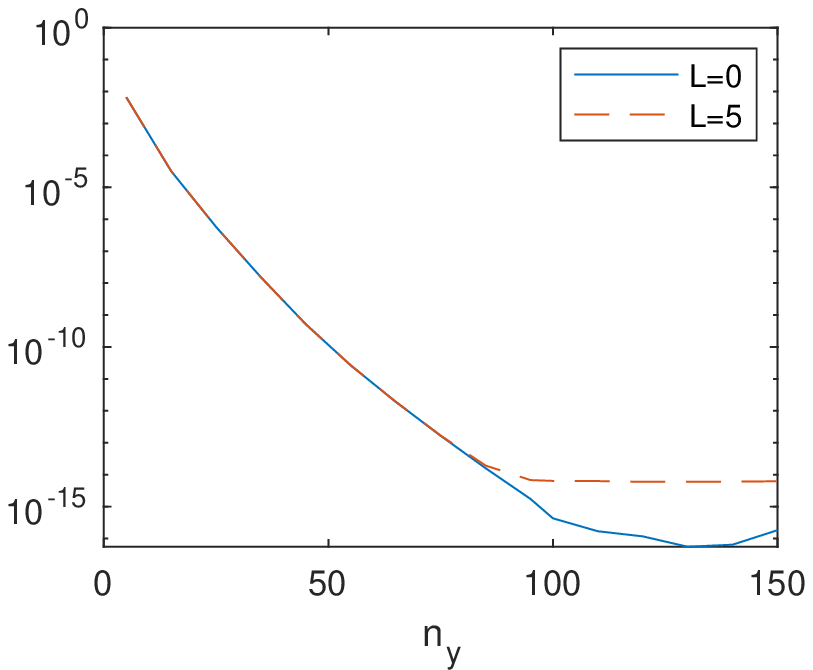}
		\caption{Convergence with respect to $n_y$}
	\end{subfigure}
	\begin{subfigure}{0.32\textwidth}
		\includegraphics[width = \linewidth]{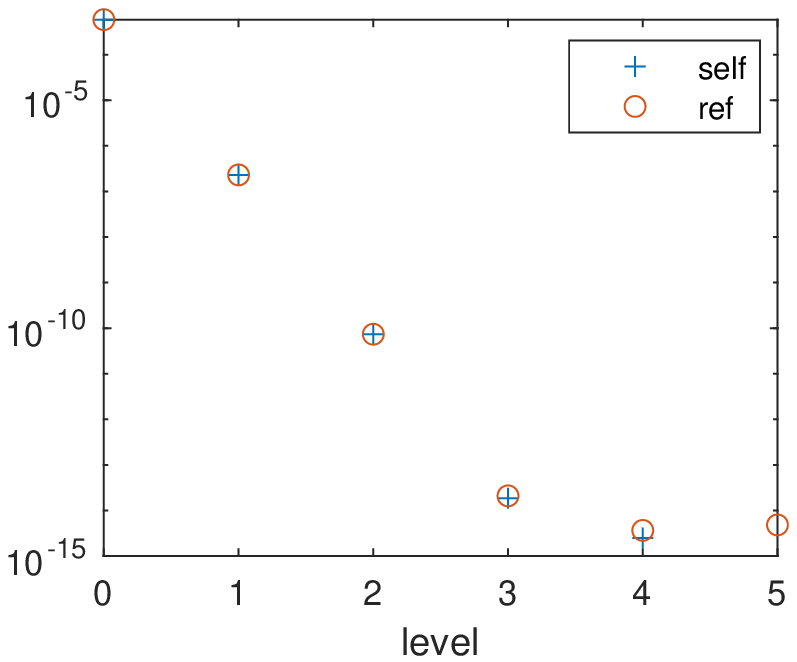}
		\caption{Convergence with respect to level of merging $L$ }
	\end{subfigure}
	\caption{Convergence of scattering matrix with Alg.\ref{alg}, error measured by Frobenius norm}
	\label{fig:TRmatrixConvergence}
	\end{figure}

\paragraph{Convergence of scattering matrices by Alg.\ref{alg}}
	 Now we are going to investigate the error of computing scattering matrix by by Alg.\ref{alg}. We consider $V=V_1\chi_I$, see \eqref{V1}, on the interval $I=[-1,1]$ and $E=1.8$. The the scattering matrix under consideration is a $3\times 3$ square matrix representing the incoming/outgoing coefficients between $(0,-1)$, $(1,1)$ and $(1,-1)$ mode. We apply Alg.\ref{alg} with $n_x'=30$, $n_y'=200$ and $L=0$ to generate ground truth scattering matrix matrix. In Fig.\ref{fig:TRmatrixConvergence} we show the Frobenius norm of error in computing scattering matrix with various $n_x$ and $n_y$ configuration and different levels of merging. 
	
 In Fig.\ref{fig:TRmatrixConvergence}, we compare the performance of our merging algorithm, Alg.\ref{alg} with direct solving on the whole interval, Alg.\ref{alg_rho}. We the claim that the leaf-and-merge algorithm is very efficient and accurate in computing eigenfunctions in a long slab. This is because, we can divide the slab to arbitrary small pieces. For each piece, the discretization in $x$ direction only takes few modes ($6$, see dashed line in Fig.\ref{fig:TRmatrixConvergence}(a)). Meanwhile the overall accuracy is high as the merge/recover procedure is purely algebraic and its error comes from roundoff in the computation procedure (see difference of dashed and solid line when $n_x$ or $n_y$ is sufficiently large in Fig.\ref{fig:TRmatrixConvergence}(a) and Fig.\ref{fig:TRmatrixConvergence}(b). In Fig.\ref{fig:TRmatrixConvergence}(c), compare the outcome of the algorithm with different overall level configuration while keeping $n_x=6$ to be the same in each leaf.      

\section{Applications to conductivity and scattering matrix computations}
\label{sec:appli}
This section presents three applications for the above algorithm. Firstly, in section \ref{sec:conductivity}, we derive an expression for the conductivity $\sigma_I$ in \eqref{eq:sigmaI} in terms of generalized eigenfunctions of the perturbed system $(H+V-E)\psi=0$ and show that it is indeed quantized and equal to $-1$ with high accuracy independent of the energy range described by $\varphi'(E)$ and of the perturbation $V$. 

In a second application, described in section \ref{sec:scattering}, we study the far-field scattering matrix associated with the generalized eigenvectors of $(H+V-E)\psi=0$ in the presence of a compactly supported perturbation $V$. We derive the relationship between scattering matrix and conductivity in \eqref{jm}. We also investigate how the scattering matrix is related to length of the slab.

Lastly, in section \ref{sec:localizedmodes}, we present a special case that $V\mathcal{G}$ has an eigenvalue equal to $-1$. In such situation, the corresponding eigenfunction $\rho_{loc}$ recovers an eigenfunction of $H+V-E$, namely $\psi_{loc}$, that vanishes outside the support of $V$ in $x$ direction. 
\subsection{Asymmetric transport conductivity}
\label{sec:conductivity}

\paragraph{Current conservation.}
Before presenting and computing the line conductivity modeling asymmetric transport, we consider the related notion of current conservation. 

For $\psi(x,y)$ a generalized eigenfunction of the problem $(H+V-E)\psi=0$ (including both incoming and outgoing components), we define the current as
\begin{align}\label{current}
   \tilde j_\psi(x) =  (\psi(x,\cdot),\sigma_3\psi(x,\cdot))_y,
\end{align}
where we recall that $(\cdot,\cdot)_y$ is the standard inner product on $L^2(\Rm_y;\Cm^2)$. 

For a (non-local) Hermitian perturbation $V$ that only couples propagating modes, it was shown in \cite{topological} that the current $\tilde j_\psi(x)$ was conserved, i.e., independent of $x\in\Rm$. We now generalize this current conservation to local (pointwise-multiplication) Hermitian perturbations $V$.

First we decompose $\psi$ as,
\begin{align}\label{psi_decomposition_2}
	\psi(x,y) = \dsum_{m\in M}  \beta_m(x)  \psi_m(x,y),
\end{align}
with $\alpha_m(x)=\beta_m(x)e^{i\xi_m x}$ in \eqref{psi_decomposition} so that $\beta_m(x)$ is constant when $V=0$.
For $M\ni m=(n,\epm)$ and $M\ni q=(p,\epsilon_q)$, we observe that
\begin{align}
   (\psi_m,\sigma_3\psi_q)_y = \widebar{e^{i\xi_mx}}e^{i\xi_qx}(\phi_m,\sigma_3\phi_q)_y.
\end{align}
When $n\not= p$, then $(\psi_m,\sigma_3\psi_q)_y=0$
by orthogonality of the Hermite functions $\varphi_n$. When $n=p$, then
\begin{align}\label{psi_sigma3}
   (\psi_m,\sigma_3\psi_q)_y =\widebar{e^{i\xi_mx}}e^{i\xi_qx} c_mc_q (2n-(E-\bar\xi_m)(E-\xi_q)).
\end{align}
For propagating modes, $\xi_m\in \Rm$ so that \eqref{psi_sigma3} vanishes when $\xi_m=-\xi_q$ while for $\xi_m=\xi_q$, 
\begin{align} \nonumber
    (\psi_m,\sigma_3\psi_m)_y = \frac{2n-(E-\xi_m)^2}{2n+(E-\xi_m)^2} = \frac{\xi_m}{E}\not=0 . 
\end{align}
For evanescent mode $i\xi_m\in \Rm$, then \eqref{psi_sigma3} vanishes when $\xi_m=\xi_q$. When $\xi_m=-\xi_q$, we find that $\widebar{e^{i\xi_mx}}e^{i\xi_qx}=1$, and hence,
\begin{align}\nonumber
 (\psi_m,\sigma_3\psi_q)_y   = \frac{2n-E^2-i\epm E \sqrt{2n-E^2}}{2n} \not=0.
\end{align} 
Here we used
\begin{align*}
  2n-(E+\xi_m)^2 = 2(2n-E^2-\xi_m E),\quad c_m^{-2}= c_q^{-2} = 2n+|E-\xi_m|^2=4n= (c_mc_q)^{-1}.
\end{align*}
Define $\widehat{m}=\begin{cases}m \quad \xi_m\in \Rm \\ -m \quad i\xi_m \in \Rm\end{cases}$. Then from the above, $(\psi_m,\sigma_3\psi_{m'})_y$ is a nonzero constant independent of $x$ when $m'=\widehat{m}$ while it vanishes otherwise.
We observe that
\begin{align}\label{dxpsi}
    D_x(\psi,\sigma_3\psi)_y&=D_x(\dsum_{m,m'} (\beta_{m'}\psi_{m'},\sigma_3\beta_m\psi_{m})_y)=\dsum_m D_x(\widebar{\beta_{\widehat{m}} }\beta_m)(\psi_{\widehat{m}} ,\sigma_3\psi_m)_y.
\end{align}
Substituting \eqref{psi_decomposition_2} in $(H+V-E)\psi=0$ yields
\begin{align}\label{eigenproblem}
   \dsum_{m'} (D_x\beta_{m'})\sigma_3\psi_{m'}+ \beta_{m'} V\psi_{m'}=0.
\end{align}
Multiplying the above by $\widebar{\beta_{\widehat{m}}\psi_{\widehat{m}}}$ and integrating over $y$ yields
\begin{align}\nonumber
    \widebar{\beta_{\widehat{m}} } D_x \beta_m = - (\beta_{\widehat{m}}\psi_{\widehat{m}}, V\dsum_{m'}\beta_{m'} \psi_{m'})_y.
\end{align}
Similarly, taking the conjugate of \eqref{eigenproblem} and multiplying by $\beta_{m}\psi_{m}$, we observe
\begin{align}\nonumber
    \beta_{m} \widebar{D_x \beta_{\widehat{m}} } = - (V\dsum_{m'}\beta_{m'} \psi_{m'},\beta_{m}\psi_{m})_y.
\end{align}
Using \eqref{dxpsi}, we finally obtain
\begin{align*}
 D_x(\psi,\sigma_3\psi)_y&=(V\dsum_{m'}\beta_{m'} \psi_{m'},\dsum_m\beta_{m}\psi_{m})_y-(\dsum_m\beta_{\widehat{m}}\psi_{\widehat{m}}, V\dsum_{m'}\beta_{m'} \psi_{m'})_y\\
 &=(V\dsum_{m'}\beta_{m'} \psi_{m'},\dsum_m\beta_{m}\psi_{m})_y-(\dsum_m\beta_{m}\psi_{m}, V\dsum_{m'}\beta_{m'} \psi_{m'})_y=0.
\end{align*}
The last identity holds since $V$ is Hermitian. This concludes our derivation that $\tilde j_\psi(x)$ is independent of $x$ and hence current is conserved.
\paragraph{Numerical verification of current conservation.}
As an example, we consider the potential $V=V_1\chi_I$ in \eqref{V1}, with support in $I=[0,100]$. We use 10 levels in Alg.\ref{alg} ($L=10$) to compute the eigenfunctions of the slab with $n_x=12$ and $n_y=100$ in each leaf. In the convergence study in (Fig.\ref{fig:TRmatrixConvergence}), empirically we saw that for these parameters, the error in a leaf of a corresponding size (length $\approx 0.1$) was almost machine precision. Since the merging algorithm is purely algebraic, we expect similar errors (up to accumulated rounding errors) in the computation of the final TR matrix and the subsequent solution of $\rho$. %% $(n_x,n_y)=(12,100)$. 

%\gb{Mode below do not seem defined anywhere? What's the energy? Here is an attempt. }

\begin{figure}[htbp]
	\centering
	\begin{subfigure}{0.32\textwidth}
		\includegraphics[width = \linewidth]{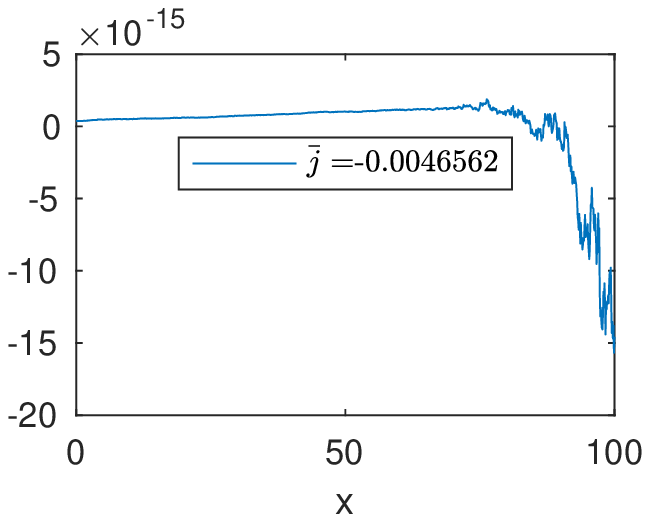}
		\caption{$\psin=e^{i\xi_{(0,-1)}x}\varphi_{(0,-1)}$}
	\end{subfigure}
		\begin{subfigure}{0.32\textwidth}
		\includegraphics[width = \linewidth]{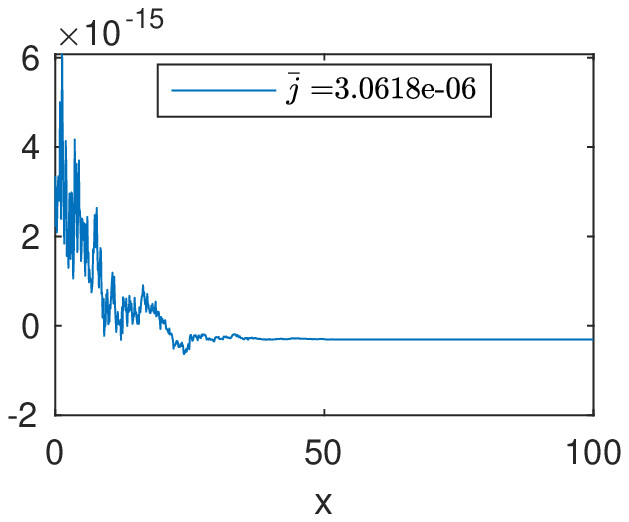}
		\caption{$\psin=e^{i\xi_{(1,1)}x}\varphi_{(1,1)}$}
	\end{subfigure}
	\begin{subfigure}{0.32\textwidth}
		\includegraphics[width = \linewidth]{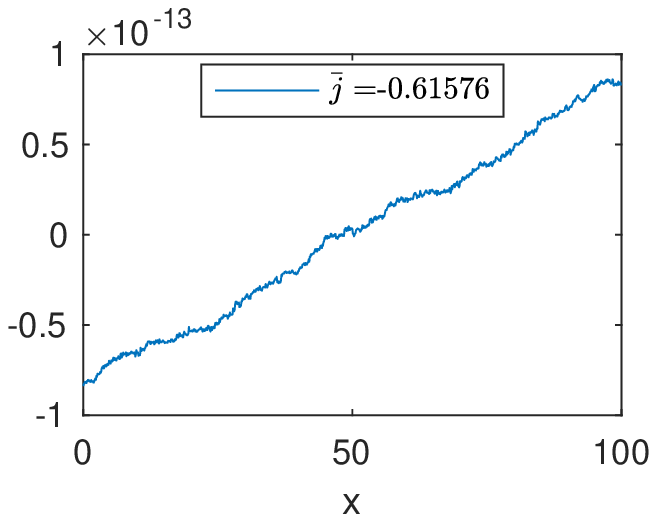}
		\caption{$\psin=e^{i\xi_{(1,-1)}x}\varphi_{(1,-1)}$}
	\end{subfigure}
	\caption{Spatial fluctuations of the currents $\tilde j_\psi(x)-\overline{\tilde j_\psi(\cdot)}$ for different incoming condition, mean current $\overline{\tilde j_\psi(\cdot)}$ is shown in the legend. } %\gb{This is $\varphi$, right?}}yes
	\label{fig:current}
\end{figure}

As an example, we consider the case where the energy $E$ is $1.8.$ Since for this value $2<E^2<4$, $\xi_m$ is real-valued for three modes $\psi\equiv\psi_m^V=\psi_{m,{\rm in}}+\psi_{m,{\rm out}}$ solution of $(H+V-E)\psi^V_m=0$ corresponding to $ m=(0,-1)$, $ m=(1,1)$, and $ m=(1,-1)$. Fig.\ref{fig:current} shows the variations in the current about their average $\tilde j_\psi(x)-\overline{\tilde j_\psi(\cdot)}$ for each of these propagating modes. 
%\jh{(i.e. taking them to be the incoming field with unit amplitude)}. 
We observe that current is indeed preserved up to numerical errors of order $10^{-13}$. 

%\gb{The values of the mean currents are not readable in the figures. }

The average values of the currents corresponding to (a), (b), and (c) in the figure are given approximately by $4.66\,10^{-3}$, $3.06\,10^{-6}$, and $-0.62$, respectively. These numbers may be explained as follows. The current of mode $(1,-1)$ is close to the value it would have when $V_1=0$ since this mode is only moderately affected by the potential $V_1$. Since the latter has oscillations in the $x$ variable that concentrate near $\xi_{(0,-1)}-\xi_{(1,1)}$, it effectively couples the modes $(0,-1)$ and $(1,1)$. As a result, significant backscattering occurs and the resulting currents associated to modes $(0,-1)$ and $(1,1)$ are correspondingly smaller. See discussion in the subsequent section \ref{sec:scattering} for additional detail on transmission and reflection coefficients.

\paragraph{Computation of line conductivity.}
The interface conductivity $\sigma_I$ introduced in \eqref{eq:sigmaI} is a physical observable that models asymmetric transport along the interface $y\sim0$. Observing current for modes at energy $E$ and at the interface point $x_0$, we consider the specific conductivity
\begin{align}\label{eq:sigmaIE0}
    2\pi\sigma_I (E,x_0)= {\rm Tr}\ 2\pi i[H_V,P] \chi'(H_V)
\end{align}
with $P'=\delta(x-x_0)$ and $\chi'(H)=\delta(H-E)$ formally. Computations of general conductivities as in \eqref{eq:sigmaI} may be obtained as a superposition of such conductivities $2\pi\sigma_I(E,x_0)$.

In the unperturbed case with $V=0$, the generalized eigenfunctions of $H$ are given by $\psi_m(x;\xi)=e^{ix\xi}\phi_m(\xi)$ in \eqref{def:phi}. Then $\chi'(H)$ can be represented spectrally by
\begin{align} \nonumber
    \chi'(H) = \mF^{-1} \chi'(\hat H(\xi)) \mF = \dsum_m \mF^{-1} \chi'(E_m(\xi)) \Pi_m(\xi) \mF,
\end{align}
with $\mF$ Fourier transform in the first variable and $\Pi_m(\xi)=\phi_m(\xi)\otimes\phi_m(\xi)$ the rank-one projector for the $m$th branch at wavenumber $\xi$ corresponding to an energy $E_m(\xi)=\epm\sqrt{\xi^2+2n}$. With this, the Schwartz kernel of the operator $\chi'(H)$ is given by
\begin{align} \nonumber
    K_{\chi'(H)}(x,x') =  \dsum_m \dint_{\Rm} \dfrac{d\xi}{2\pi} \chi'(E_m(\xi)) \psi_m(x;\xi) \otimes \psi_m(x';\xi) .
\end{align}
For the Dirac equation, $i[H,P]=P'(x)\sigma_3$, so that for $\chi'(H)=\delta(H-E)$,
\begin{align} \nonumber
    {\rm Tr} \ 2\pi i[H,P] \chi'(H) &=\dsum_m \dint_{\Rm} d\xi \chi'(E_m(\xi)) \big(\psi_m(x_0;\xi),\sigma_3\psi_m(x_0;\xi)\big)_{y} 
    \\\nonumber
    & = \dsum_m   \Big|\pdr{\xi_m}E\Big|   \big(\psi_m(x_0;\xi_m),\sigma_3\psi_m(x_0;\xi_m)\big)_{y} ,
\end{align}
where we have defined the points $\xi_m$ such that $E_m(\xi_m)=E$. They are given by $\xi_m=\epsilon_m \sqrt{E^2-2n}$ and are real-valued as the above sum over $m$ involves only propagating modes since $(\phi_m,\sigma_3\phi_m)=0$ for evanescent modes. Here we have implicitly used the results obtained in, e.g., \cite{B19b,bal2021topological} that the trace in \eqref{eq:sigmaIE0} may be computed as the integral along the diagonal of the Schwartz kernel of the operator $2\pi i[H_V,P] \chi'(H_V)$.

Now, when $V\not=0$ is compactly supported, the Fourier transform may formally be replaced by a generalized Fourier transform, where on each branch, $e^{ix\xi}\phi_m(\xi)$ is replaced by $\psi^V_m(x;\xi)$; see 
\cite{simon1982schrodinger} for similar completeness results for the Schr\"odinger equation.

Here, $\psi_m^V(x;\xi_m)$ is the solution in the kernel of $H_V-E$ with incoming condition given by $e^{i\xi_m x}\phi_m(\xi_m)$ for $\xi_m$ solving $E=E_m(\xi)$, i.e. $\xi_m=\epsilon_m \sqrt{E^2-2n}$.

We thus formally obtain the following expression for the conductivity:
\begin{align}\label{conductivity_formula}
    2\pi \sigma_I = \dsum_m   \Big|\pdr{\xi_m}E\Big|   \big(\psi_m^V(x_0;\xi_m),\sigma_3\psi^V_m(x_0;\xi_m)\big)_{y} ,\quad \Big|\pdr{\xi_m}E\Big|   = \dfrac{E}{\sqrt{E^2-2n}}.
\end{align}
The above expression relates the conductivity describing asymmetric transport to the generalized eigenvectors that our algorithm may estimate robustly and accurately.

For simplicity of presentation, we decompose the conductivity as follows
\begin{align}\label{jm}
     2\pi\sigma_I= \dsum_m j_m(x),\qquad   j_m(x):= \Big|\pdr{\xi_m}E\Big|   \big(\psi_m^V(x,\cdot;\xi_m),\sigma_3\psi^V_m(x,\cdot;\xi_m)\big)_{y}.
\end{align}
Due to the current conservation shown above, $j_m(x)$ is in fact independent of $x$.
\paragraph{Numerical Example: Conductivity dependence.}
The computation of the conductivity for a given energy $E$ involves all generalized eigenfunctions of the operator $H+V-E$. We consider a potential that couples the propagating modes of the unperturbed system given by $V=V_0\chi_{[-1,1]}$, where
\begin{align}\nonumber
    V_0&=\exp(-y^2)y\cos(x(\xi_{(0,-1)}-\xi_{(1,1)}))+\exp(-y^2)y\cos(x(\xi_{(0,-1)}-\xi_{(1,-1)}))\\[1mm]\label{eq:V0}
    &+\exp(-y^2)\cos(x(\xi_{(1,-1)}-\xi_{(1,1)})).
\end{align}
Note that in this construction, $V_0$ and hence $V$ depend on $E$ implicitly since the wavenumbers $\xi_m$ also do. We then apply Alg.\ref{alg} with four levels to calculate the generalized eigenfunctions $\psi_m^V$ in the interval $[-1,1]$ with $L=4$ level binary merging. In each leaf, we have $n_x=10$, $n_y=100$ to achieve sufficient accuracy. The conductivity is evaluated at $x_0=0$ and $2<E^2<4$ ($E>0$). Then we have two branches $m=(0,-1)$ and $m=(1,\pm 1)$ crossing $E$ with the first branch crossing at $\xi=-E$ and the second branch crossing at $\xi_{1\pm}=\pm \sqrt{E^2-2}$. 

\begin{figure}[ht!]
	\centering
	\begin{subfigure}{0.24\textwidth}
		\includegraphics[width =\linewidth]{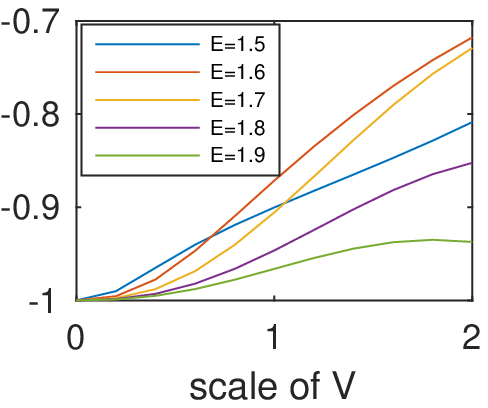}
	\caption{ $j_{(0,-1)}$}
	\end{subfigure}
		\begin{subfigure}{0.24\textwidth}
		\includegraphics[width =\linewidth]{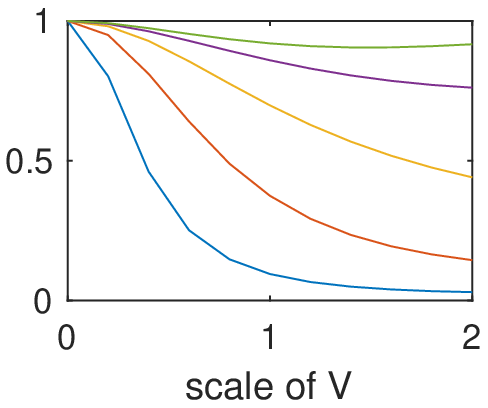}
	\caption{ $j_{(1,1)}$}
	\end{subfigure}
		\begin{subfigure}{0.24\textwidth}
		\includegraphics[width =\linewidth]{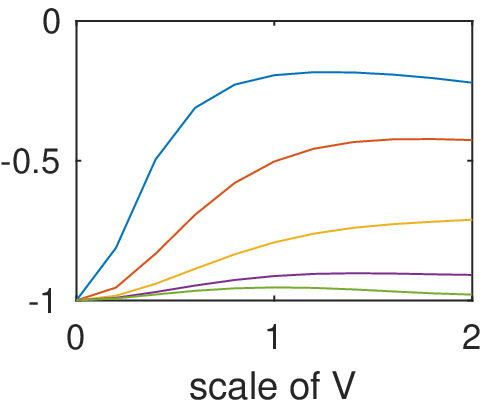}
	\caption{ $j_{(1,-1)}$}
	\end{subfigure}
		\begin{subfigure}{0.24\textwidth}
		\includegraphics[width =\linewidth]{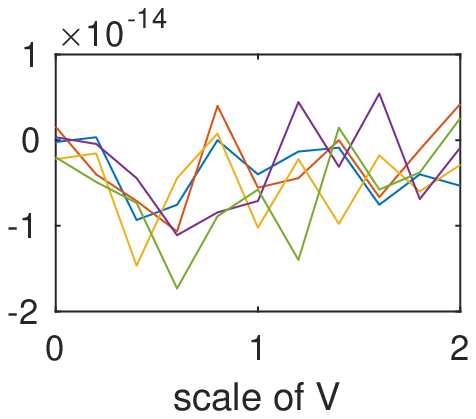}
	\caption{$\dsum_m j_m+1$}
	\end{subfigure}
	\caption{
	%\jh{legends too small, labels too small}
	Current and conductivity dependence on $E$ and scale $\lambda$ of $V$ (currents of generalized eigenvectors of $(H+\lambda V_0\chi_{[-1,1]}-E)$). } 
	\centering
	\label{fig:cond_VS_EV}
	\end{figure}

In Fig.\ref{fig:cond_VS_EV}, we show $j_m$ defined in \eqref{jm} for different energy levels and different scalings $\lambda$ of the (energy dependent) potential (where the perturbation $V$ is replaced by $\lambda V$). We observe that the individual terms in the sum \eqref{conductivity_formula} depend on $E$ and $V$. The change of the current of a particular eigenfunction given some fixed incoming waves is due to the change of scattering coefficients when varying $E$ and $V$. However, as a topological invariant, as expected $\sum_m j_m$ is close to $-1$ with high accuracy (13 correct digits in this example).

\begin{figure}[ht!]
	\centering
	\begin{subfigure}{0.24\textwidth}
		\includegraphics[width =\linewidth]{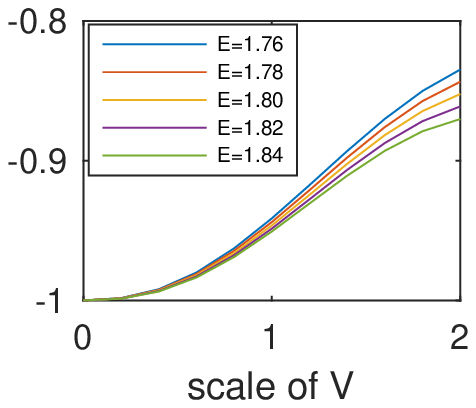}
	\caption{ $j_{(0,-1)}$}
	\end{subfigure}
		\begin{subfigure}{0.24\textwidth}
		\includegraphics[width =\linewidth]{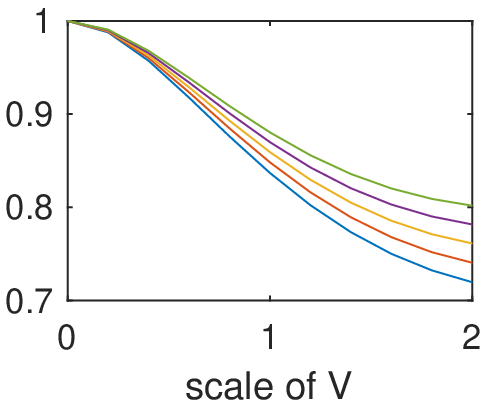}
	\caption{ $j_{(1,1)}$}
	\end{subfigure}
		\begin{subfigure}{0.24\textwidth}
		\includegraphics[width =\linewidth]{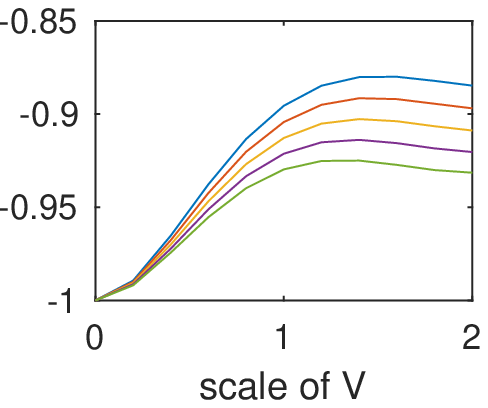}
	\caption{ $j_{(1,-1)}$}
	\end{subfigure}
		\begin{subfigure}{0.24\textwidth}
		\includegraphics[width =\linewidth]{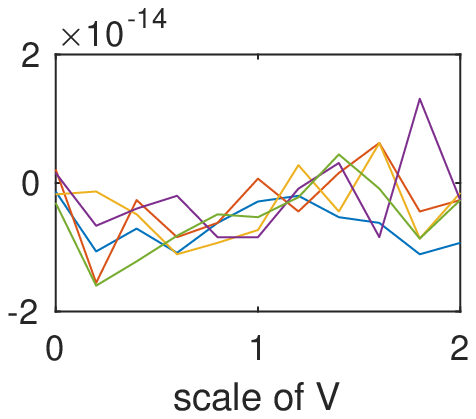}
	\caption{ $\dsum_m j_m+1$}
	\end{subfigure}
	\caption{Current and conductivity dependence on $E$ and scale $\lambda$ of $V$ (current of generalized eigenvectors of $(H+\lambda V_0(E_0=1.8)\chi_{[-1,1]}-E)$). }
	\centering
	\label{fig:cond_VS_EV_set2}
	\end{figure}

We also consider the setting of a fixed energy $E_0=1.8$ with $V_0=V_0(E_0=1.8)$ in \eqref{eq:V0} with wavenumbers $\xi_m=\xi_m(E_0)$. We present in Fig.\ref{fig:cond_VS_EV_set2} the currents as a function of $E$ close to $E_0$ and $\lambda$ associated to the generalized eigenvectors of $(H+\lambda V(E_0=1.8)-E)$. As above, we observe quantization of the line conductivity while its individual components (continuously) depend on the energy and potential fluctuations.

\subsection{Computation of far-field scattering matrices}
\label{sec:scattering}
\paragraph{Line conductivity and scattering matrix.}
Given a potential $V$ compactly supported on $I=[x_L,x_R]$ in the variable $x$, the far-field scattering matrix may be extracted from the computed TR matrix by only considering propagating modes. Note from \eqref{jm} that $\sqrt{\frac{E}{|\xi_m|}}\psi_m$ carries normalized current equal to $\pm1$.

For example, when $E=1.8$, the scattering matrix is the $3\times 3$ submatrix of the TR matrix related to the propagating modes $(0,-1)$, $(1,1)$ and $(1,-1)$.  

In general, assuming $k$ propagating modes travelling right (with positive current) and $k+1$ modes travelling left (with negative current so that the sum of all currents is $2\pi\sigma_I=-1$), the scattering matrix $S$ is defined by,
\begin{align}\nonumber
    S=\left(\begin{array}{ll}
R_{+} & T_{-} \\
T_{+} & R_{-}
\end{array}\right)
\end{align}
where $R_{+}$ is the $k \times(k+1)$ matrix of reflection of the modes from left to left, $T_{+}$ the $(k+1) \times(k+1)$ transmission matrix of the same modes to the right (positive values of $y), R_{-}$ the $(k+1) \times k$ matrix of reflection of the modes from right to right and $T_{-}$ the $k \times k$ matrix of transmission of those modes to the left. Then, from the definition of the TR matrix, we deduce that
\begin{align}\nonumber
\begin{cases}
    R_{+,pn}=e^{ix_L\xi_{(n,1)}-ix_R\xi_{(p,-1)}}\sqrt{\Big|\frac{\xi_p}{\xi_n}\Big|}M_{(p,-1),(n,1)}\\
    T_{+,pn}=e^{ix_L\xi_{(n,1)}-ix_R\xi_{p+}}\sqrt{\Big|\frac{\xi_p}{\xi_n}\Big|}M_{(p,1),(n,1)}\\
    R_{-,pn}=e^{ix_L\xi_{(n,-1)}-ix_R\xi_{(p,1)}}\sqrt{\Big|\frac{\xi_p}{\xi_n}\Big|}M_{(p,1),(n,-1)}\\
    T_{-,pn}=e^{ix_L\xi_{(n,-1)}-ix_R\xi_{(p,-1)}}\sqrt{\Big|\frac{\xi_p}{\xi_n}\Big|}M_{(p,-1),(n,-1)}.\\
    \end{cases}
\end{align}
 
The scattering matrix is constrained by the quantization of the asymmetric transport $2\pi\sigma_I=-1$. Substituting the decomposition \eqref{psi_decomposition}, $\psi=\sum_m\alpha_m\phi_m$ into the individual term of the conductivity in \eqref{jm}, we obtain
\begin{align}\label{current2}
     j_m(x_0)=\Big|\pdr{\xi_m}E\Big|\dsum_{q,q'} \widebar{\alpha_{q;m}'(x_0)}\alpha_{q';m}(x_0) (\phi_q',\sigma_3\phi_q),
\end{align}
here $q=(p,\eps_q)$, $m=(n,\eps_m)$, and $\alpha_{q;m}$ denotes the Fourier coefficients of $\psi_m^V$ projected onto the $\{\psi\}$ basis. 
Since evanescent modes vanish outside of the support of $V$, while we proved current remains constant in $x_0$, we may choose $|x_0|\to\infty$ in \eqref{current2} to limit the summation to propagating modes only. 

More precisely, when $\eps_m>0$, the incoming condition is a right travelling modes. We take $x_0\to \infty$ in \eqref{current2} and also substitute $(\phi_q',\sigma_3\phi_q)=\delta_q(q') \frac{\xi_m}{E}$ to get,
\begin{align*}
    j_m(x_0)&=\frac{E}{|\xi_m|}\dsum_{q}|\alpha_{q;m}|^2(x_0)\frac{\xi_q}{E}\ =\ \dsum_{\eps_q>0}|\alpha_{q;m}|^2(x_0)\frac{\xi_q}{|\xi_m|}.
\end{align*}
The second equation is due to incoming condition $\psi^V_{{\rm in};m}=\psi_m$ which is solely right travelling. Still for $\eps_m>0$, we have
\begin{align} \nonumber
    |\alpha_{q;m}|^2(x_0)\frac{\xi_q}{\xi_m}=|M_{q,m}|^2\Big|\frac{\xi_q}{\xi_m}\Big|=|T_{+,pn}|^2,
\end{align}
%Back to $j_m$, it turns to, (for $\eps_m>0$),
so that 
\begin{align}\label{jm_current}
    j_m=\sum_{p}|T_{+,pn}|^2.
\end{align}
For $\eps_m<0$, we consider $x_0\to-\infty$ instead since each $j_m(x)$ is independent of $x$ and obtain that
\begin{align}\label{jm_current_2}
    j_m=-\sum_{p}|T_{-,pn}|^2.
\end{align}
Summing up the above calculations, we thus deduce that the conductivity in \eqref{jm} is given by
\begin{align}\nonumber
    2\pi\sigma_I={\rm tr} \ T_+^*T_+- {\rm tr} \ T_-^*T_-.
\end{align}

\paragraph{Scattering matrix of a slab of length $l$.}
 
We now investigate the properties of the scattering matrix beyond the above quantized constraint, and in particular how it depends on the potential $V$. 
 
For computational purpose, instead of the binary merging proposed in Alg.\ref{alg}, we merge the leaves sequentially. More precisely, for $i=1\cdots N$, let $M_i$ be the sequence of leaf TR matrices on the interval $I_i$. We start from the first leaf TR matrix $M_1$ and merge it with the adjacent $M_2$ to get a scattering matrix $I_1\cup I_2$. Iterating the process until $k=N-1$ to get the TR matrix of $I_1\cup\cdots\cup I_k$, we merge it with $M_{k+1}$ to get the TR matrix for $I_1\cup\cdots\cup I_k\cup I_{k+1}$. This modified merging procedure is slightly more convenient for showing the relationship between the length of the interval and the corresponding scattering matrix.

\begin{figure}[ht!]
	\centering
	\begin{subfigure}{0.23\textwidth}
		\includegraphics[width = \linewidth]{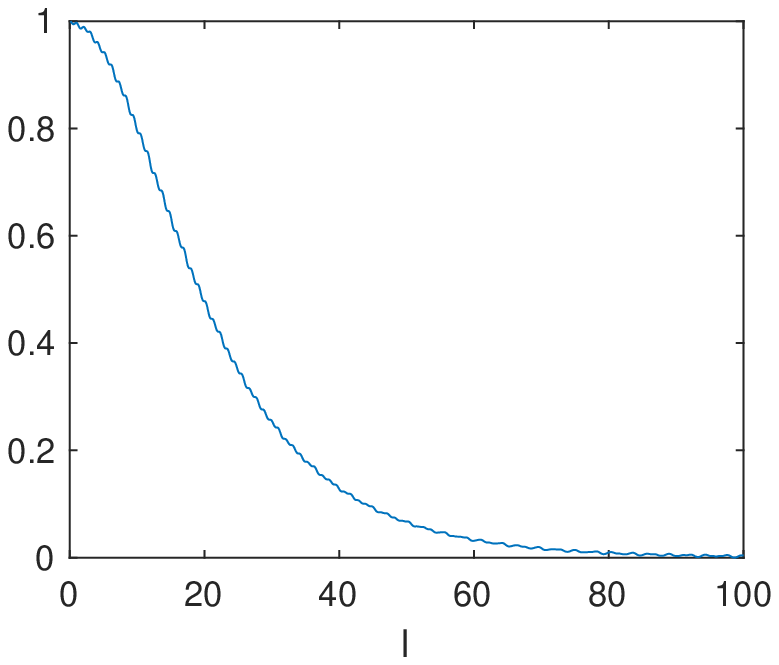}
		\caption{$(0,-1)\to(0,-1)$}
	\end{subfigure}
	\begin{subfigure}{0.23\textwidth}
		\includegraphics[width = \linewidth]{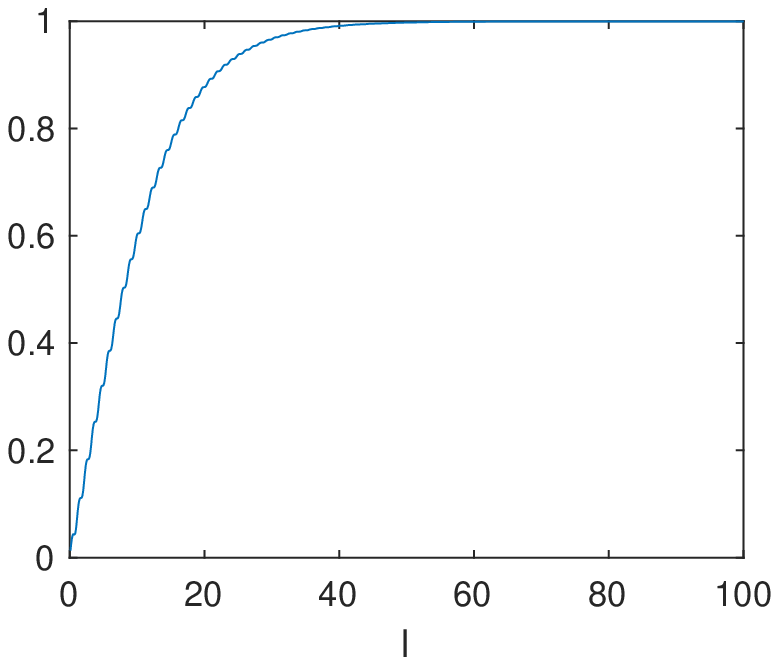}
		\caption{$(1,1)\to(0,-1)$}
	\end{subfigure}
	\begin{subfigure}{0.23\textwidth}
		\includegraphics[width = \linewidth]{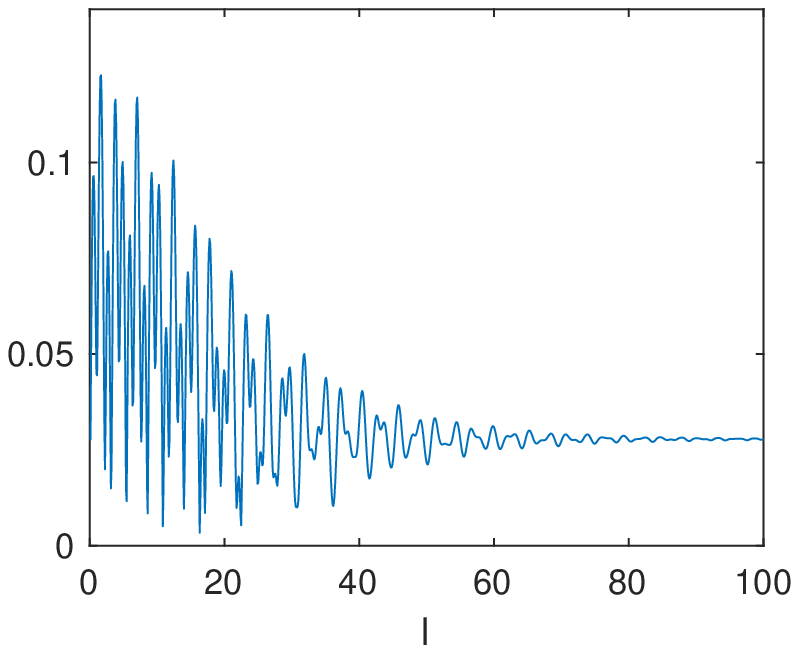}
		\caption{$(1,-1)\to(0,-1)$}
	\end{subfigure}\\
	\begin{subfigure}{0.23\textwidth}
		\includegraphics[width = \linewidth]{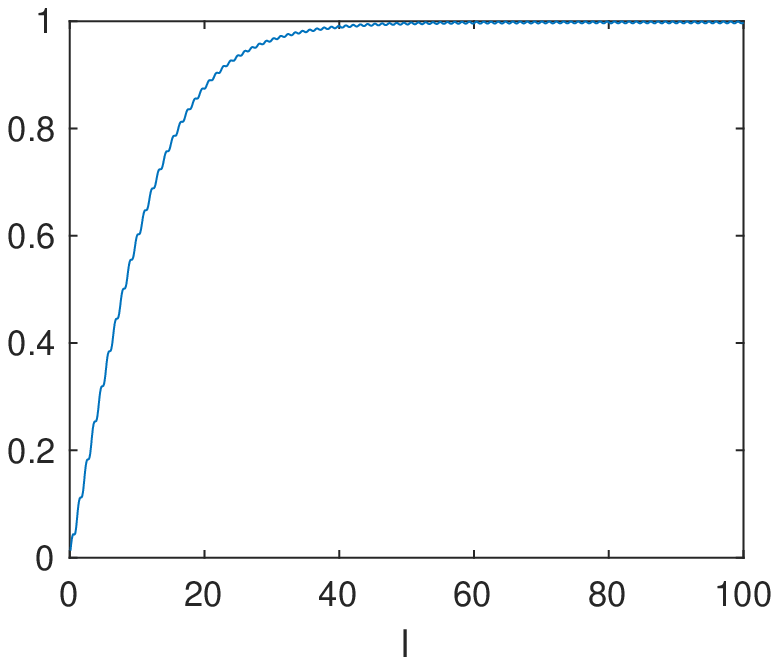}
		\caption{$(0,-1)\to(1,1)$}
	\end{subfigure}
	\begin{subfigure}{0.23\textwidth}
		\includegraphics[width = \linewidth]{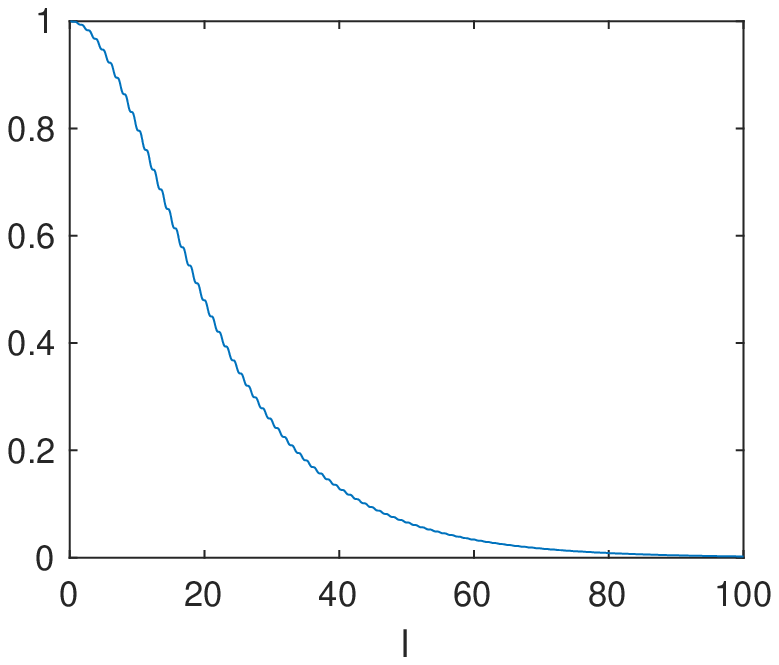}
		\caption{$(1,1)\to(1,1)$}
	\end{subfigure}
	\begin{subfigure}{0.23\textwidth}
		\includegraphics[width = \linewidth]{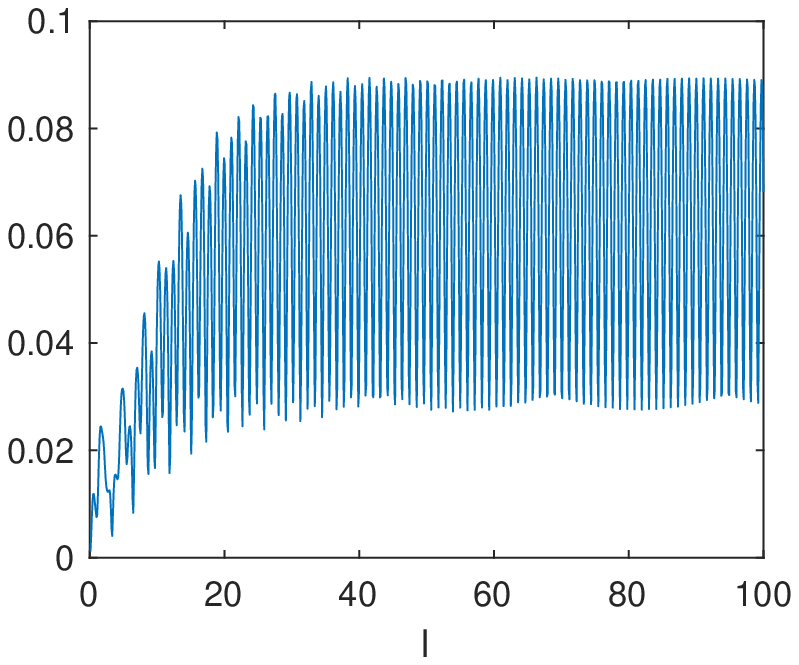}
		\caption{$(1,-1)\to(1,1)$}
	\end{subfigure}\\
	\begin{subfigure}{0.23\textwidth}
		\includegraphics[width = \linewidth]{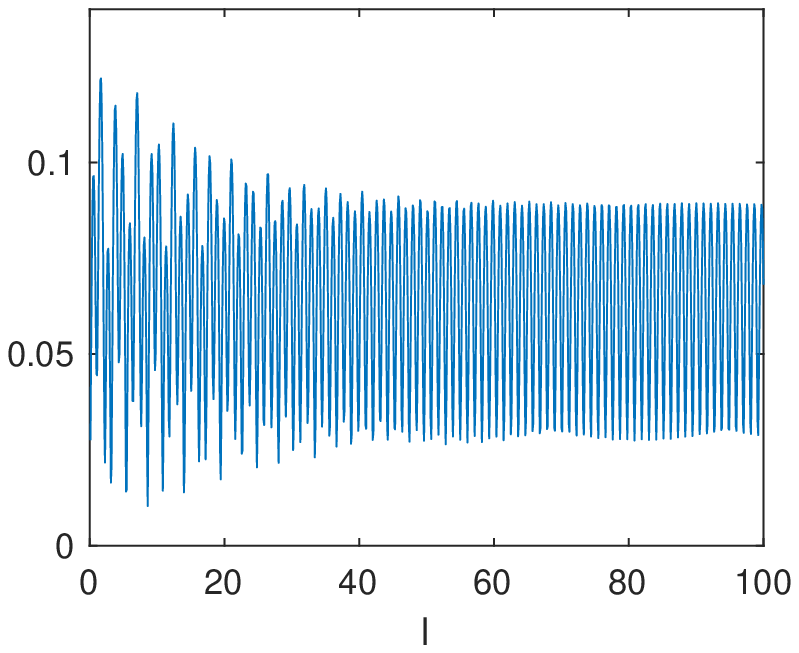}
		\caption{$(0,-1)\to(1,-1)$}
	\end{subfigure}
	\begin{subfigure}{0.23\textwidth}
		\includegraphics[width = \linewidth]{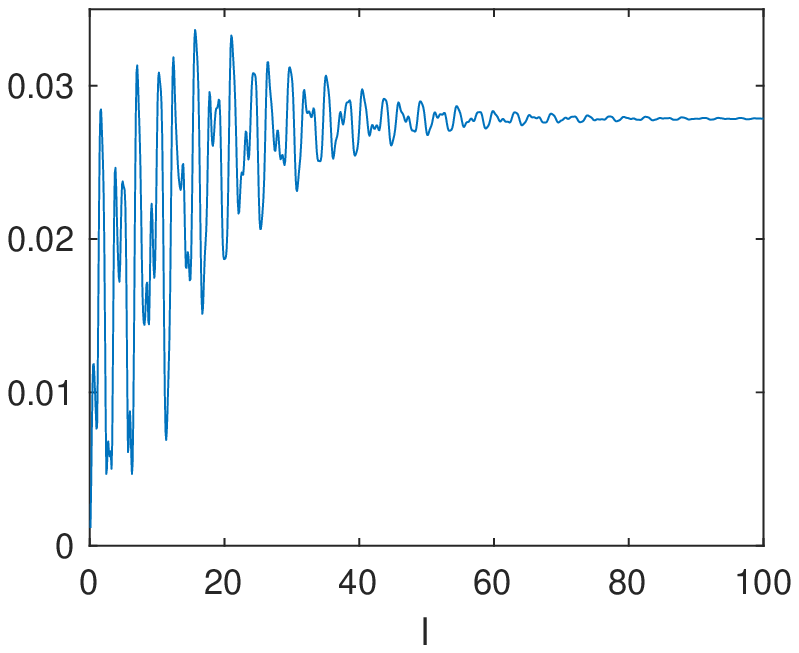}
		\caption{$(1,1)\to(1,-1)$}
	\end{subfigure}
	\begin{subfigure}{0.23\textwidth}
		\includegraphics[width = \linewidth]{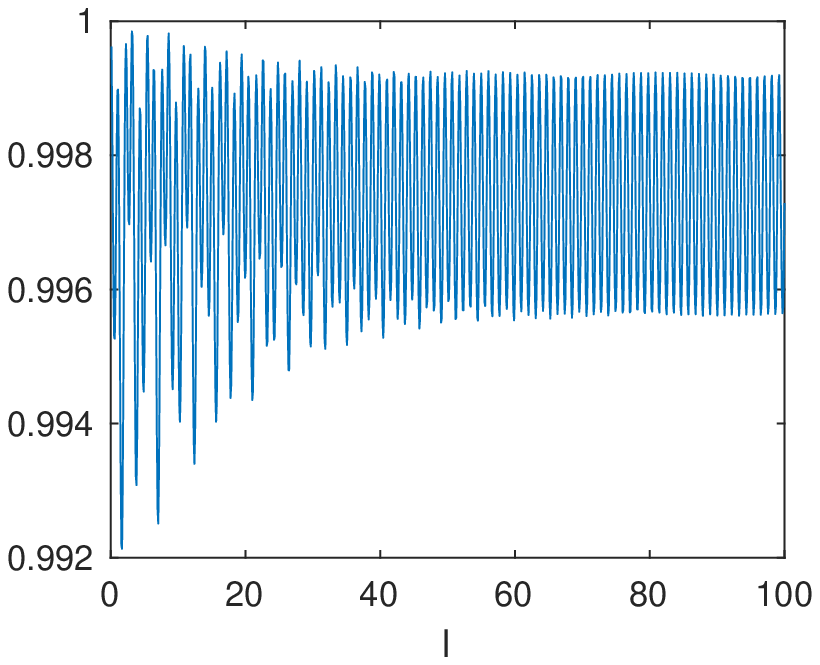}
		\caption{$(1,-1)\to(1,-1)$}
	\end{subfigure}\\
	\caption{Entries of scattering matrix against length of slab $l$, $V=V_1\chi_{[0,l]}$, $E=1.8$}\label{fig:unprotected_bs}
	\end{figure}

%%%%
\paragraph{Topologically non-trivial back-scattering.}
%%%
Fig.\ref{fig:unprotected_bs} displays the scattering matrix for energy $E=1.8$ (so that $2<E^2<4$) corresponding to the three propagating modes $(0,-1)$, $(1,-1)$, and $(1,1)$; recall Fig. \ref{fig:EvsXi}. The parameters used in the simulation are identical to those used for the current calculation, i.e. for computing the results in numerical verification of current conservation in section \ref{sec:conductivity}.

The main observation is that in the presence of sufficiently strong fluctuations $V$ (i.e., when the support of $V$ is sufficiently large), the modes $(0,-1)$ and $(1,1)$ are almost fully back-scattered while the mode $(1,-1)$ is almost fully transmitted. The absence of backscattering is therefore not topologically protected. Only the global asymmetry $2\pi\sigma_I=-1$ is, which is consistent with the analysis in \cite{topological} in the presence of random fluctuations $V$.

We note that the oscillations in the scattering coefficient of the $(1,-1)$ mode, while small, are not numerical artifacts. These oscillations, which depend on the choice of $V$, are significantly larger than accuracy of the algorithm as exemplified by current conservation in, e.g., Fig.\ref{fig:current}.
Although we chose a potential $V=V_0\chi_{[0,l]}$ to maximize coupling between the $(0,-1)$ and $(1,1)$ mode, this does not guarantee that the remaining modes are exactly protected.   
	
The above strong backscattering is in sharp contrast with the setting of energies $0<E^2<2$. In that case, the only allowed propagating mode is $(0,-1)$. Back-scattering is then obviously absent since no mode carries a positive current. We thus obtain a back-scatter-free setting for a combination of topological ($\sigma_I=1)$ and energetic ($E^2<2$) reasons.  
	
\begin{figure}[ht!]
	\centering

	\begin{subfigure}{0.45\textwidth}
		\includegraphics[width = \linewidth]{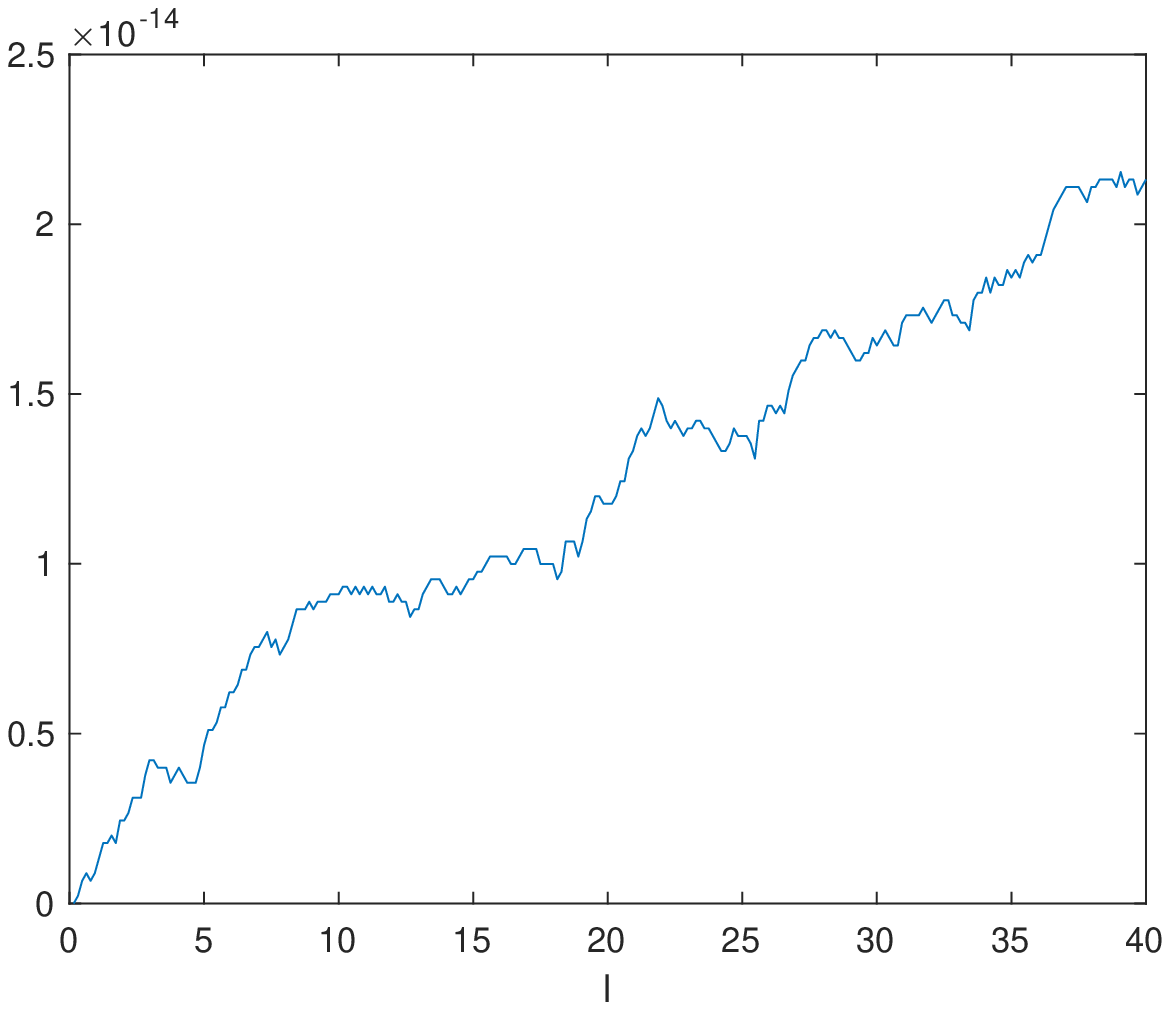}
		\caption{$|S_{(1,-1),(1,-1)}|-1$}
	\end{subfigure}
	\begin{subfigure}{0.45\textwidth}
		\includegraphics[width = \linewidth]{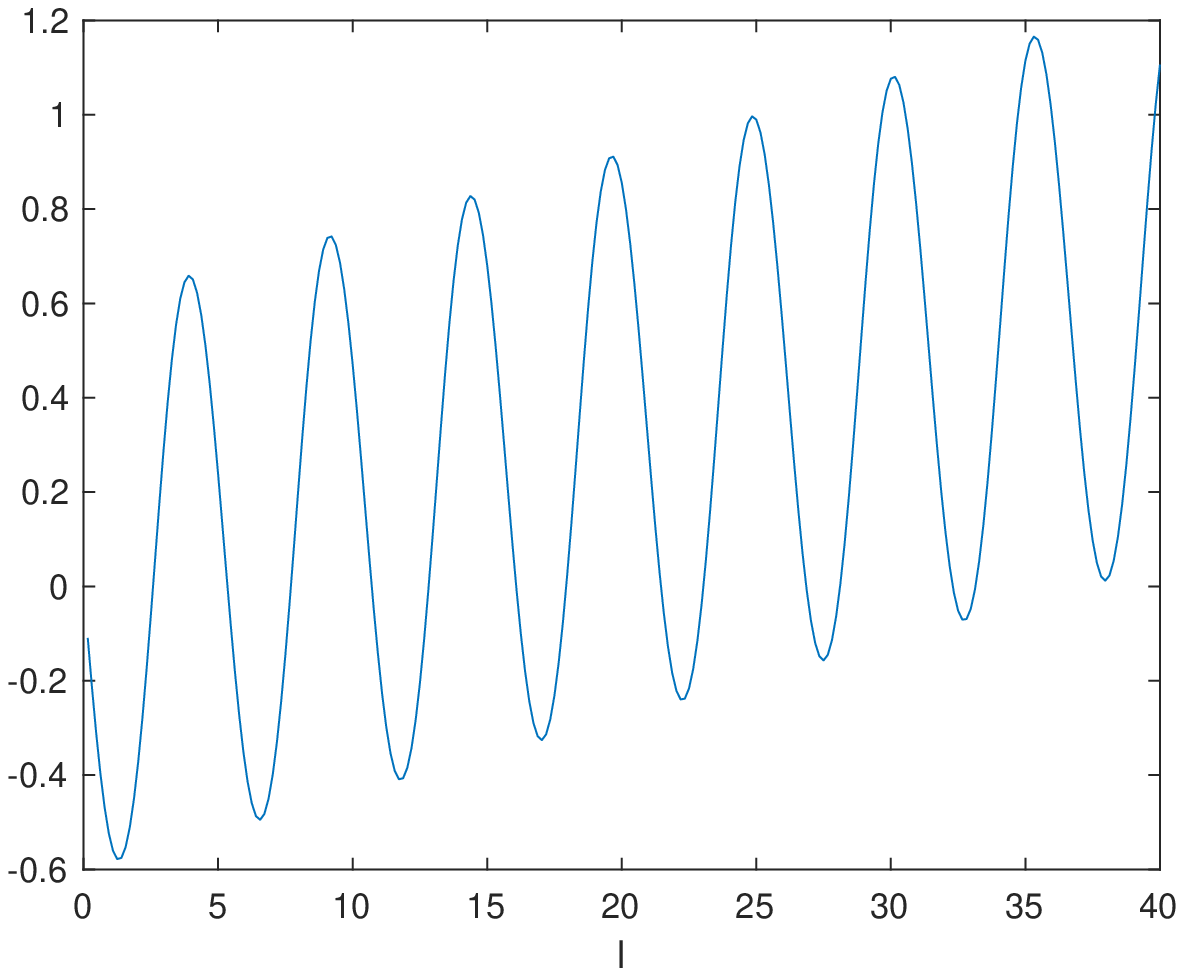}
		\caption{$arg(S_{(1,-1),(1,-1)})$}
	\end{subfigure}
\\
	
	\caption{Scattering coefficient $(1,-1)\to(1,-1)$ against length of slab $l$, $V=V_4\chi_{[0,l]}$, $E=1.2$}\label{fig:1sys}
	\end{figure}

Consider the case $E=1.2$ with a potential $V=V_4\chi_{[0,l]}$ where 
\begin{align}\nonumber
    V_4(x,y)=\exp(-y^2)\cos(Ex)
\end{align}. In Fig.\ref{fig:1sys}, we find the scattering coefficient $(0,-1)\to(0,-1)$, which is a $1\times1$ (transmission) matrix.  The coefficient oscillates as a periodic function of $l$ with absolute value necessarily equal to $1$. However, the potential does have an influence on the phase of this mode; see Fig.\ref{fig:1sys}(b)).
	
\begin{figure}[ht!]
	\centering
	\begin{subfigure}{0.23\textwidth}
		\includegraphics[width = \linewidth]{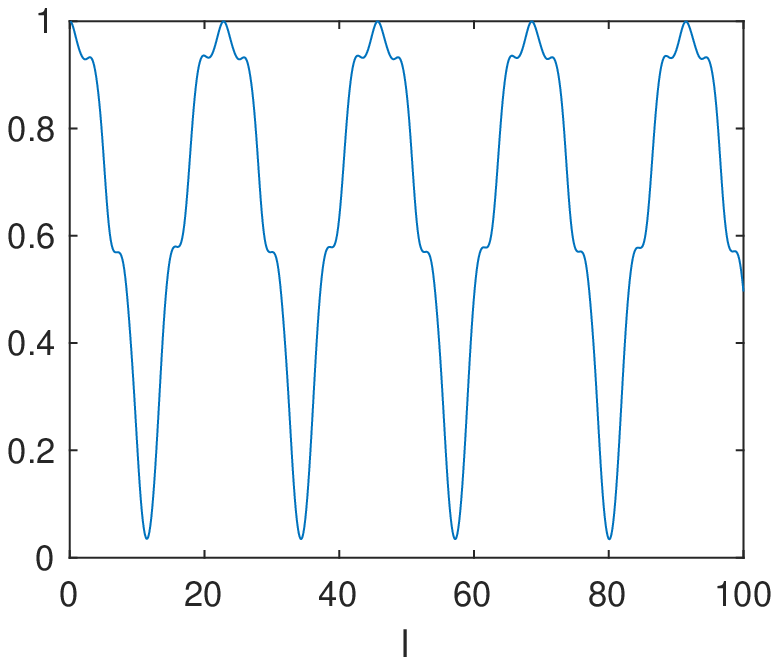}
		\caption{$(0,-1)\to(0,-1)$}
	\end{subfigure}
	\begin{subfigure}{0.23\textwidth}
		\includegraphics[width = \linewidth]{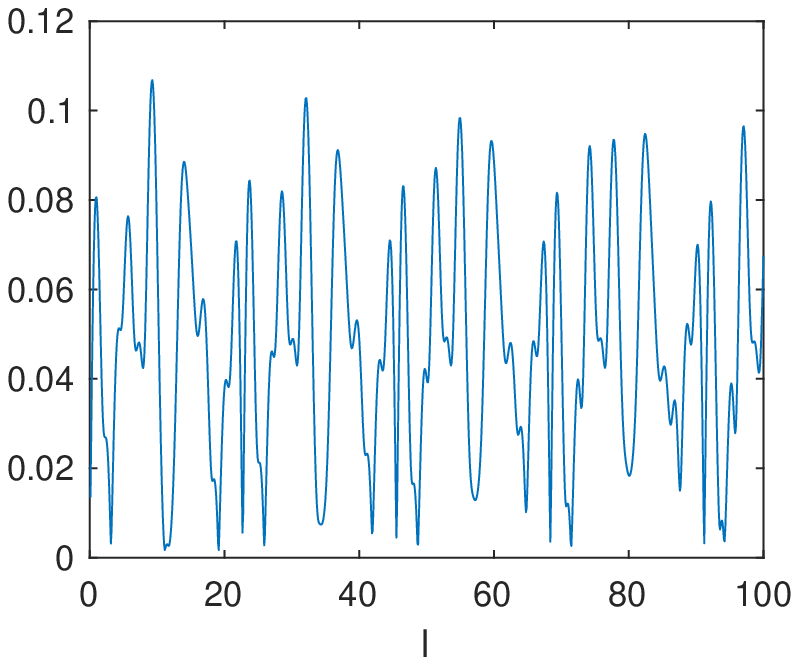}
		\caption{$(1,1)\to(0,-1)$}
	\end{subfigure}
	\begin{subfigure}{0.23\textwidth}
		\includegraphics[width = \linewidth]{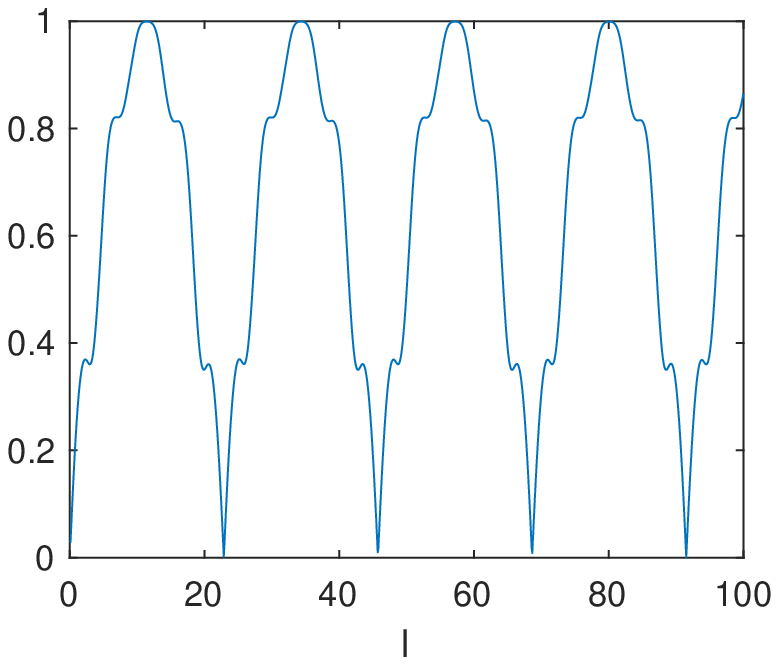}
		\caption{$(1,-1)\to(0,-1)$}
	\end{subfigure}\\
	\begin{subfigure}{0.23\textwidth}
		\includegraphics[width = \linewidth]{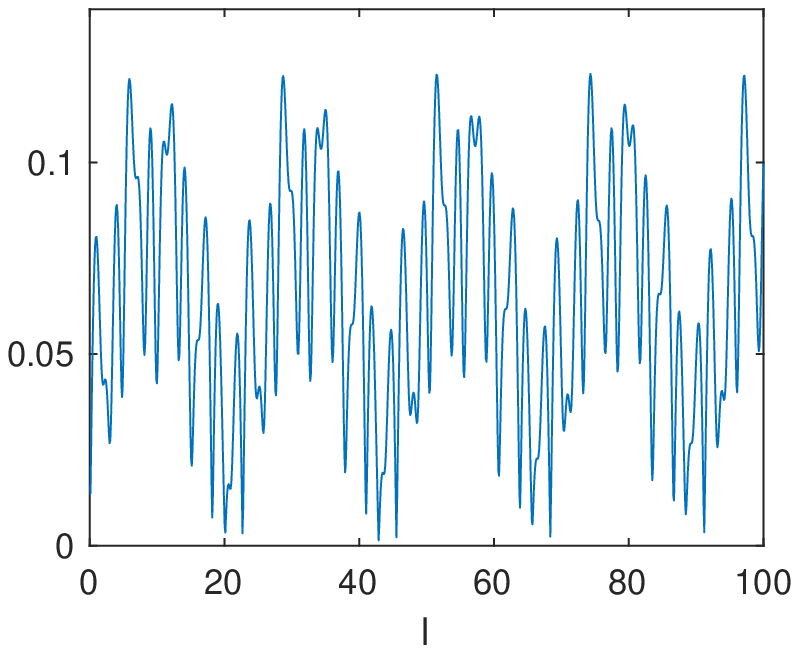}
		\caption{$(0,-1)\to(1,1)$}
	\end{subfigure}
	\begin{subfigure}{0.23\textwidth}
		\includegraphics[width = \linewidth]{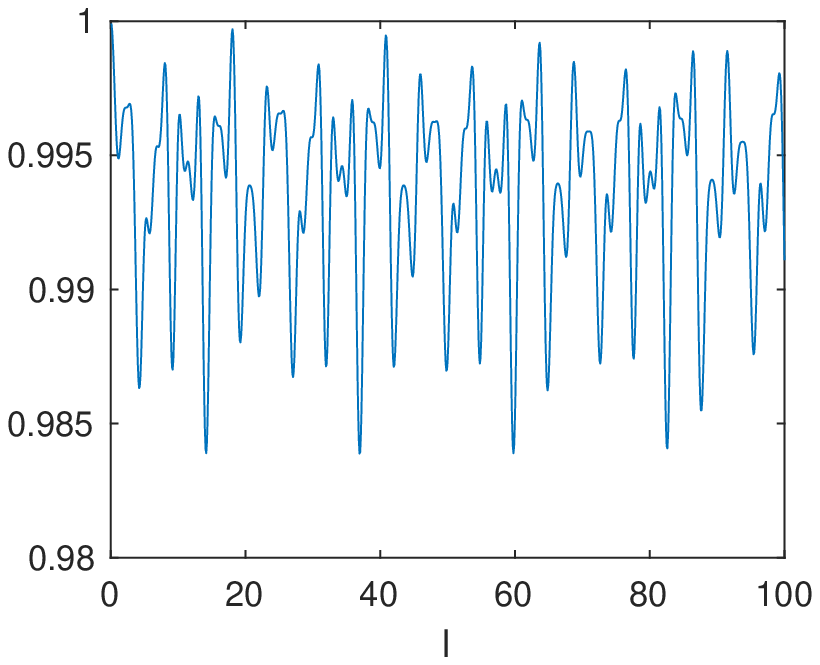}
		\caption{$(1,1)\to(1,1)$}
	\end{subfigure}
	\begin{subfigure}{0.23\textwidth}
		\includegraphics[width = \linewidth]{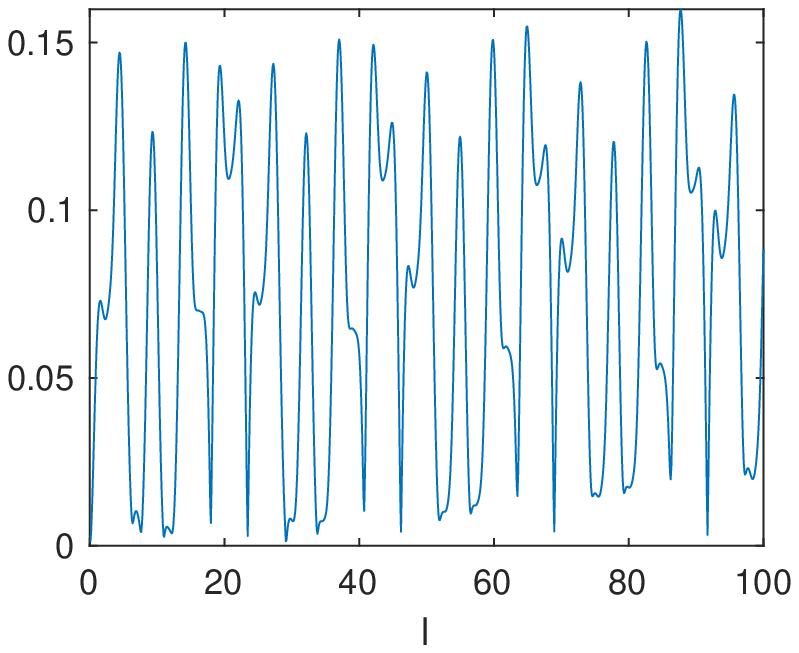}
		\caption{$(1,-1)\to(1,1)$}
	\end{subfigure}\\
	\begin{subfigure}{0.23\textwidth}
		\includegraphics[width = \linewidth]{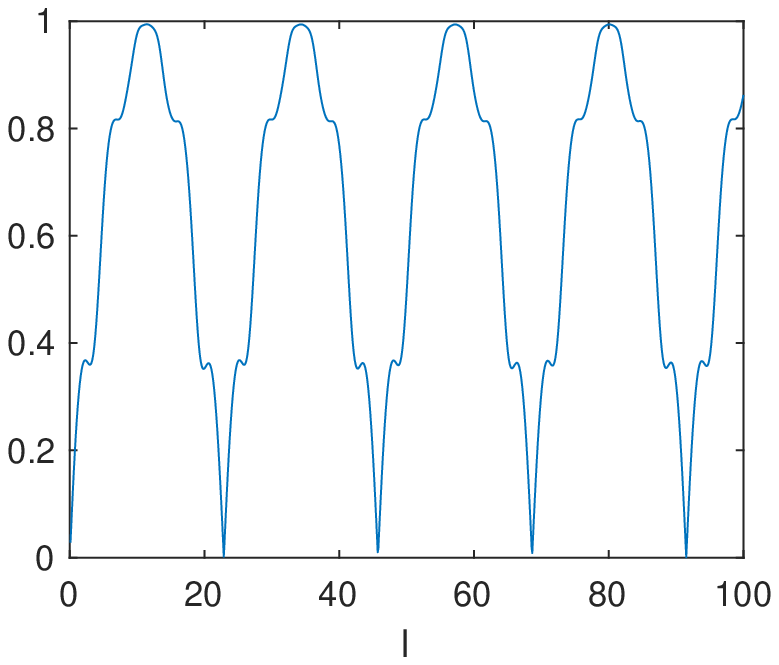}
		\caption{$(0,-1)\to(1,-1)$}
	\end{subfigure}
	\begin{subfigure}{0.23\textwidth}
		\includegraphics[width = \linewidth]{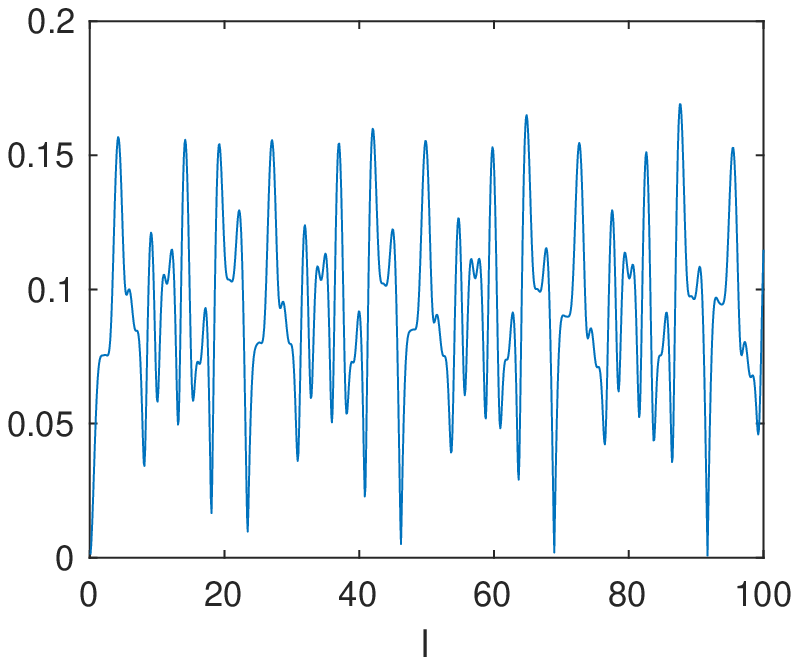}
		\caption{$(1,1)\to(1,-1)$}
	\end{subfigure}
	\begin{subfigure}{0.23\textwidth}
		\includegraphics[width = \linewidth]{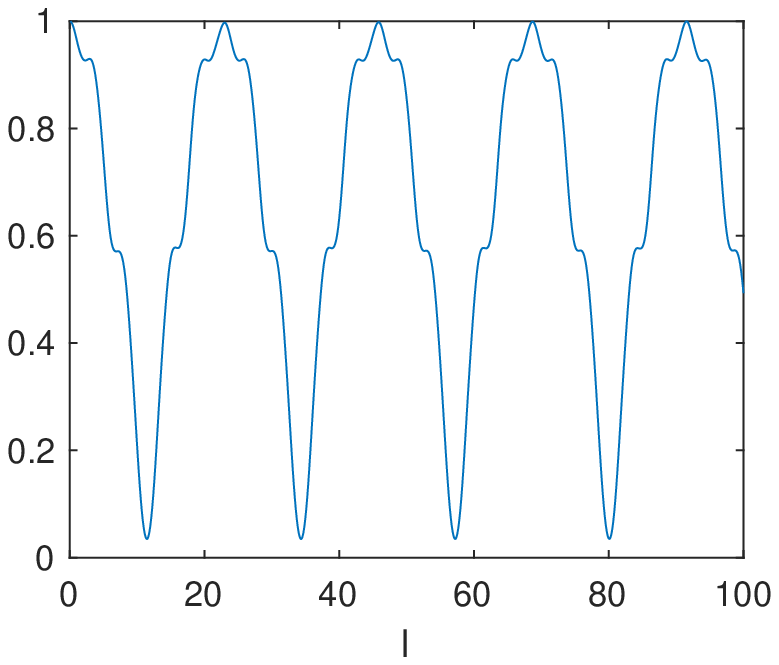}
		\caption{$(1,-1)\to(1,-1)$}
	\end{subfigure}\\
	\caption{Entries of scattering matrix against length of slab $l$, $V=V_2\chi_{[0,l]}$, $E=1.8$.}\label{fig:periodic_transmission}
	\end{figure}

\paragraph{Coupling generating oscillatory transmission.} We now consider the potential
\begin{align} \nonumber
    V=V_2\chi_{[0,l]}, \quad V2=\exp \left(-y^{2}\right) y \cos \left(\left(\xi_{(0,-1)}-\xi_{(1,-1)}\right)x\right),
\end{align}
which couples the two left travelling modes  $(0,-1)$ and $(1,-1)$. The computational configuration is the same as for $V=V_1\chi_{[0,l]}$ above. In Fig.\ref{fig:periodic_transmission}, we show the entries of the scattering matrix as a function of the length of the slab $l$. 

Recall that for $V=V_1\chi_{[0,l]}$, increasing the length of the scattering interval increases the coupling between the modes $(0,-1)$ and $(1,1)$, which are eventually fully backscattered. 

\begin{figure}[ht!]
	\centering
	\begin{subfigure}{0.19\textwidth}
		\includegraphics[width = \linewidth]{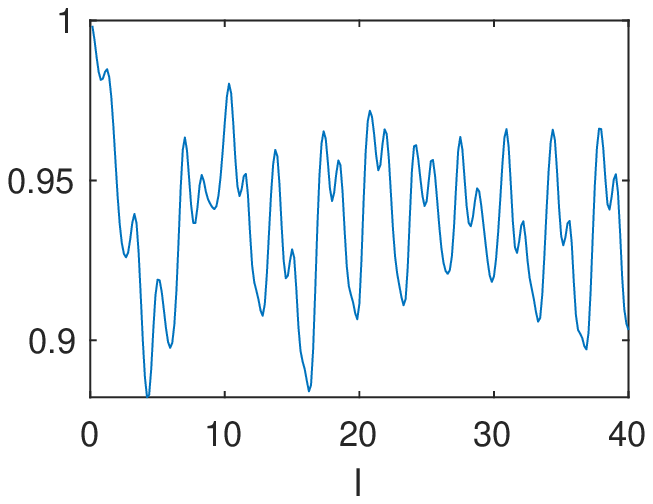}
		\caption{\scriptsize$(0,-1)\to(0,-1)$}
	\end{subfigure}
	\begin{subfigure}{0.19\textwidth}
		\includegraphics[width = \linewidth]{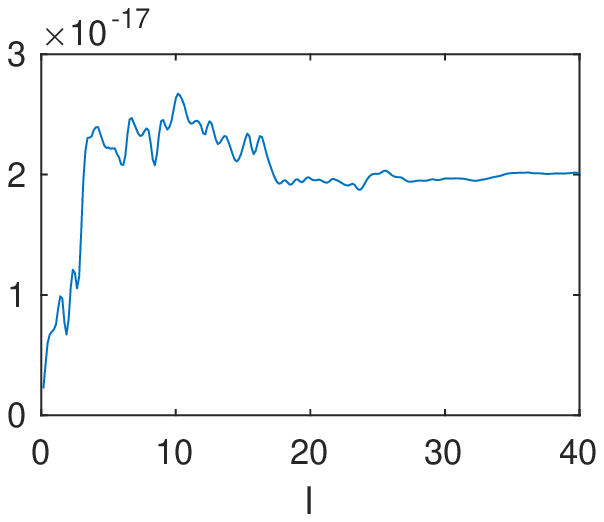}
		\caption{\scriptsize$(1,1)\to(0,-1)$}
	\end{subfigure}
	\begin{subfigure}{0.19\textwidth}
		\includegraphics[width = \linewidth]{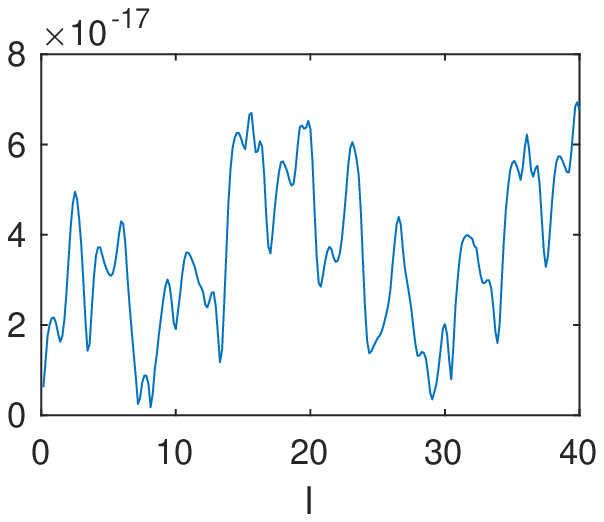}
		\caption{\scriptsize$(1,-1)\to(0,-1)$}
	\end{subfigure}
	\begin{subfigure}{0.19\textwidth}
		\includegraphics[width = \linewidth]{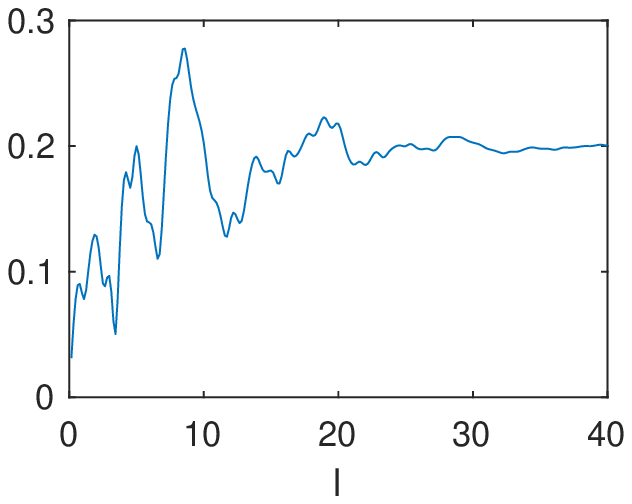}
		\caption{\scriptsize$(2,1)\to(0,-1)$}
	\end{subfigure}
	\begin{subfigure}{0.19\textwidth}
		\includegraphics[width = \linewidth]{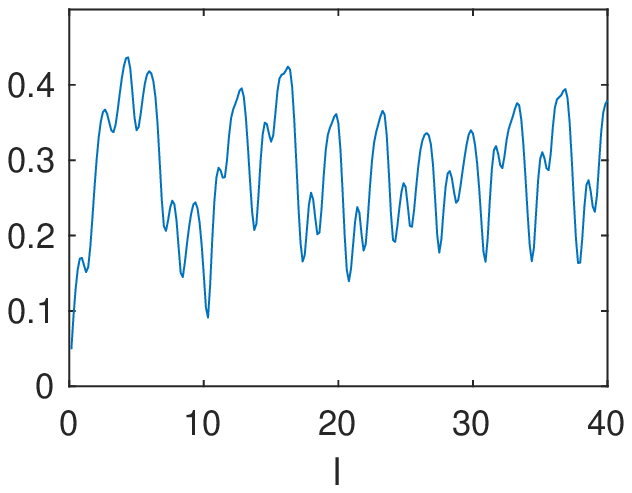}
		\caption{\scriptsize$(2,-1)\to(0,-1)$}
	\end{subfigure}\\
		\begin{subfigure}{0.19\textwidth}
		\includegraphics[width = \linewidth]{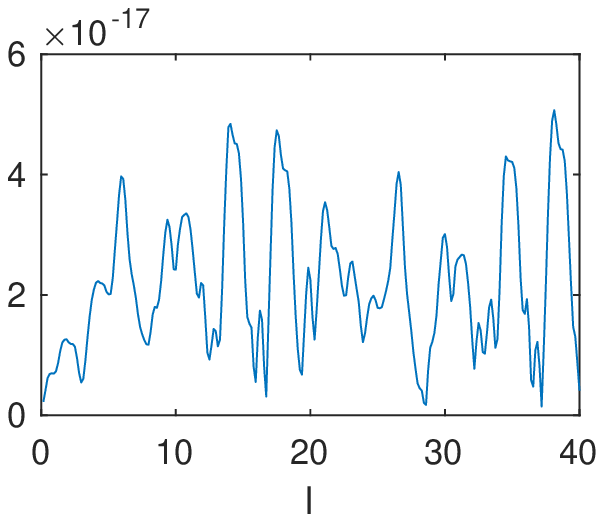}
		\caption{\scriptsize$(0,-1)\to(1,1)$}
	\end{subfigure}
	\begin{subfigure}{0.19\textwidth}
		\includegraphics[width = \linewidth]{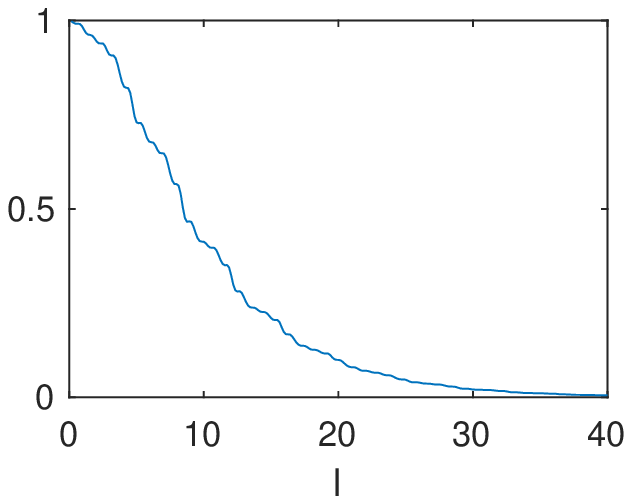}
		\caption{\scriptsize$(1,1)\to(1,1)$}
	\end{subfigure}
	\begin{subfigure}{0.19\textwidth}
		\includegraphics[width = \linewidth]{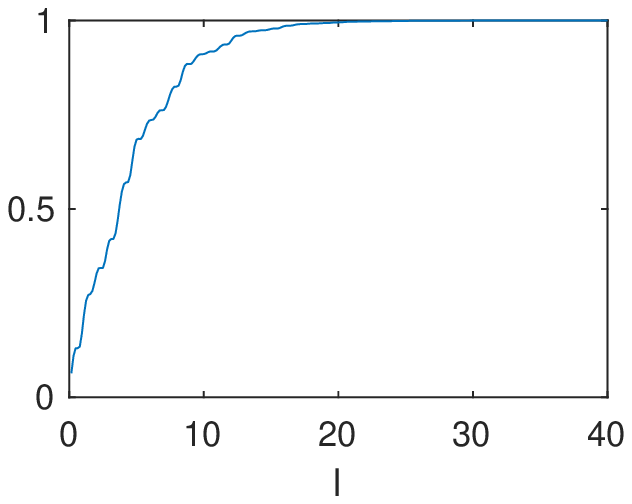}
		\caption{\scriptsize$(1,-1)\to(1,1)$}
	\end{subfigure}
	\begin{subfigure}{0.19\textwidth}
		\includegraphics[width = \linewidth]{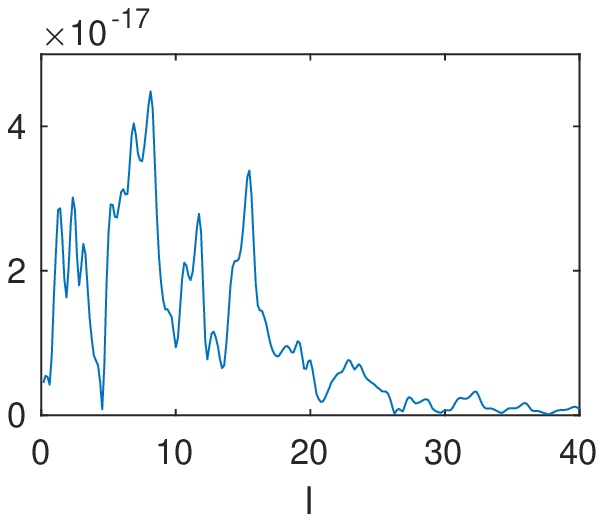}
		\caption{\scriptsize$(2,1)\to(1,1)$}
	\end{subfigure}
	\begin{subfigure}{0.19\textwidth}
		\includegraphics[width = \linewidth]{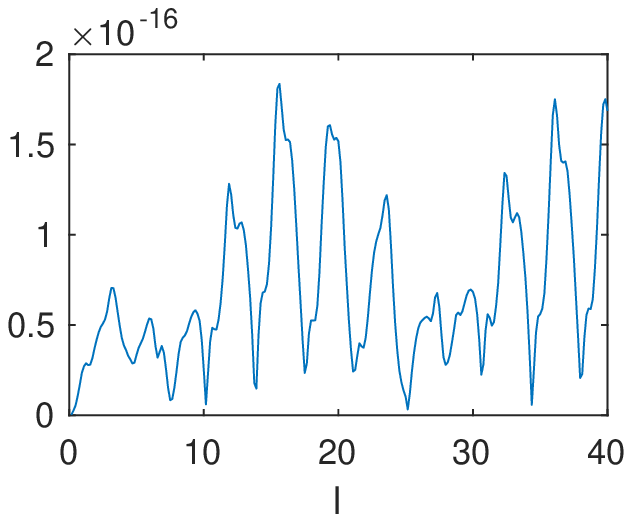}
		\caption{\scriptsize$(2,-1)\to(1,1)$}
	\end{subfigure}\\
		\begin{subfigure}{0.19\textwidth}
		\includegraphics[width = \linewidth]{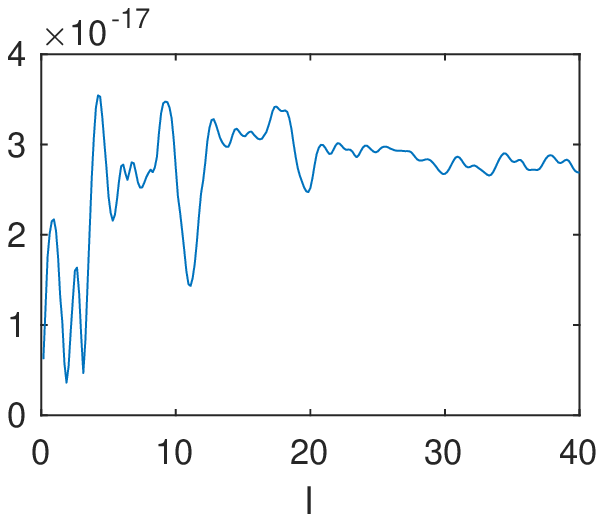}
		\caption{\scriptsize$(0,-1)\to(1,-1)$}
	\end{subfigure}
	\begin{subfigure}{0.19\textwidth}
		\includegraphics[width = \linewidth]{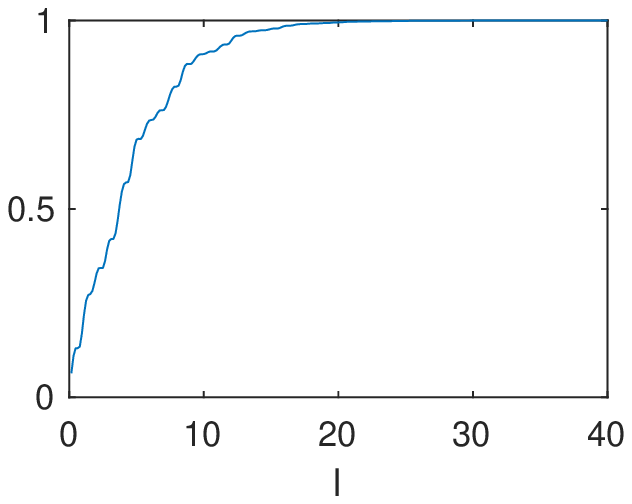}
		\caption{\scriptsize$(1,1)\to(1,-1)$}
	\end{subfigure}
	\begin{subfigure}{0.19\textwidth}
		\includegraphics[width = \linewidth]{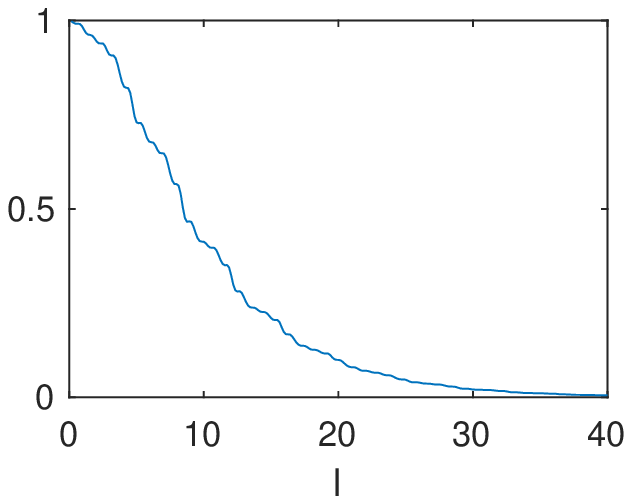}
		\caption{\scriptsize$(1,-1)\to(1,-1)$}
	\end{subfigure}
	\begin{subfigure}{0.19\textwidth}
		\includegraphics[width = \linewidth]{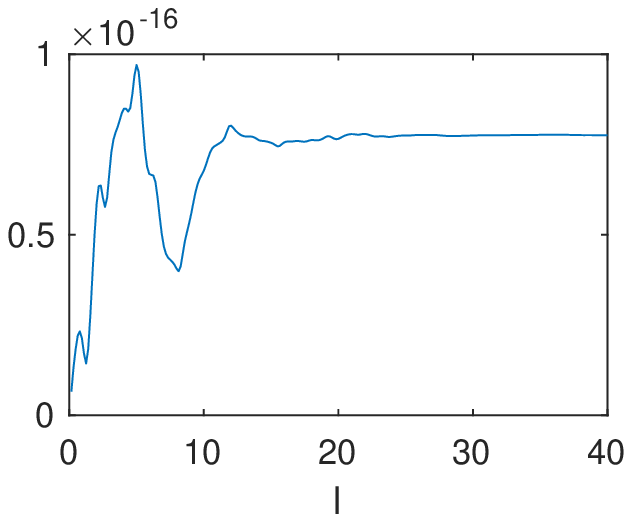}
		\caption{\scriptsize$(2,1)\to(1,-1)$}
	\end{subfigure}
	\begin{subfigure}{0.19\textwidth}
		\includegraphics[width = \linewidth]{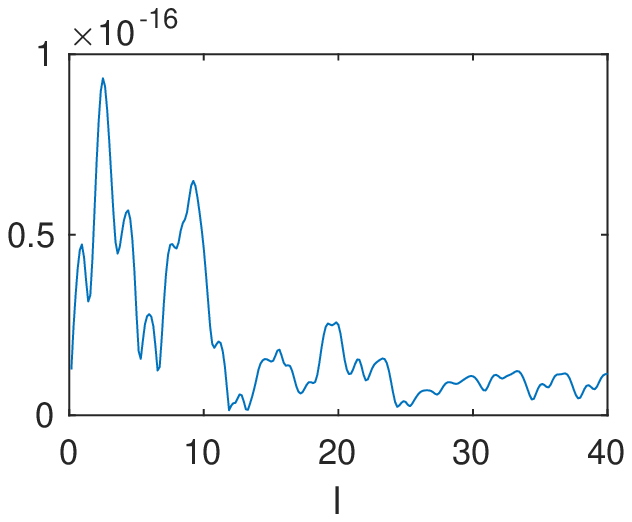}
		\caption{\scriptsize$(2,-1)\to(1,-1)$}
	\end{subfigure}\\
	
		\begin{subfigure}{0.19\textwidth}
		\includegraphics[width = \linewidth]{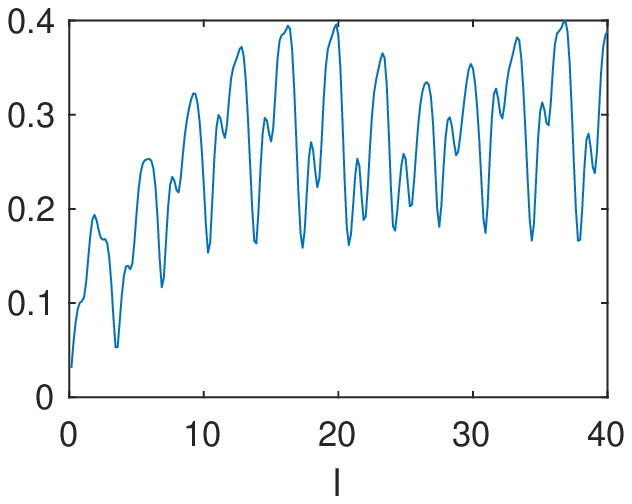}
		\caption{\scriptsize$(0,-1)\to(2,1)$}
	\end{subfigure}
	\begin{subfigure}{0.19\textwidth}
		\includegraphics[width = \linewidth]{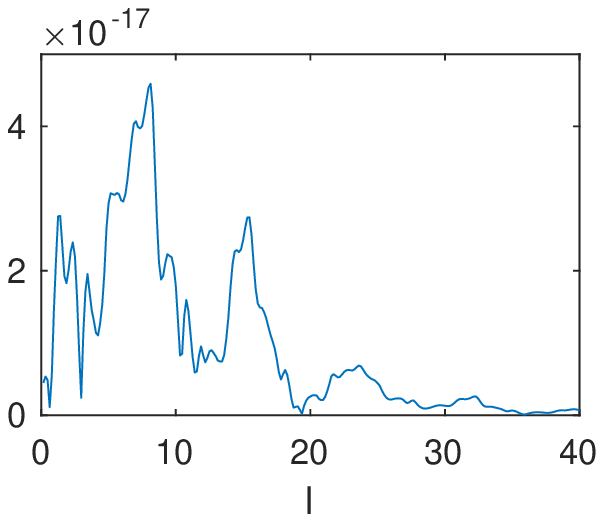}
		\caption{\scriptsize$(1,1)\to(2,1)$}
	\end{subfigure}
	\begin{subfigure}{0.19\textwidth}
		\includegraphics[width = \linewidth]{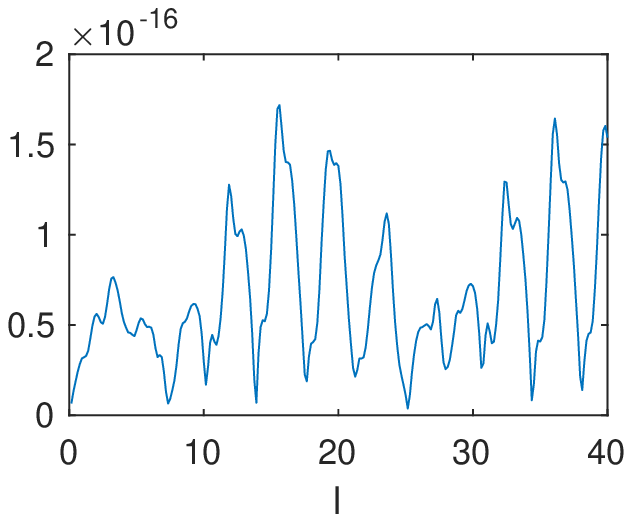}
		\caption{\scriptsize$(1,-1)\to(2,1)$}
	\end{subfigure}
	\begin{subfigure}{0.19\textwidth}
		\includegraphics[width = \linewidth]{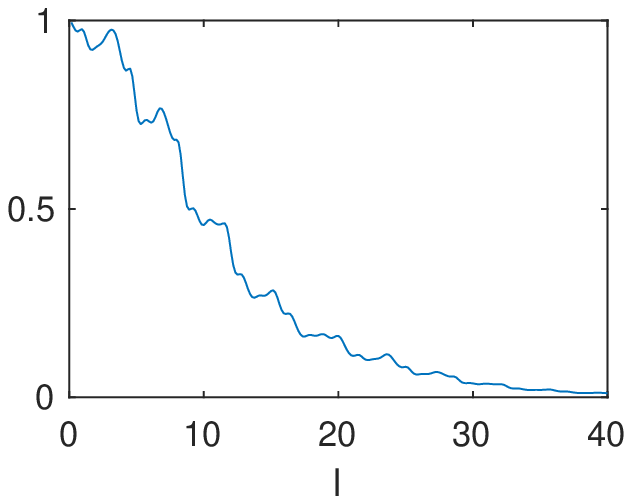}
		\caption{\scriptsize$(2,1)\to(2,1)$}
	\end{subfigure}
	\begin{subfigure}{0.19\textwidth}
		\includegraphics[width = \linewidth]{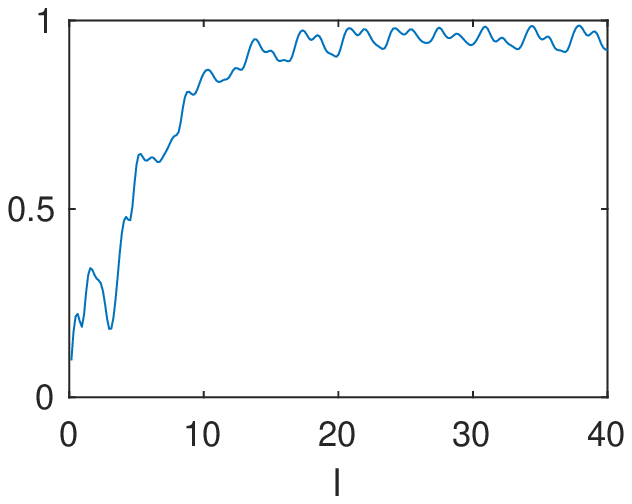}
		\caption{\scriptsize$(2,-1)\to(2,1)$}
	\end{subfigure}\\
	
		\begin{subfigure}{0.19\textwidth}
		\includegraphics[width = \linewidth]{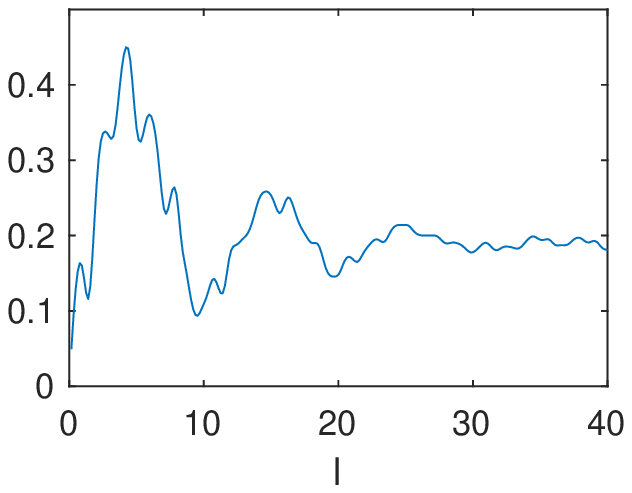}
		\caption{\scriptsize$(0,-1)\to(2,-1)$}
	\end{subfigure}
	\begin{subfigure}{0.19\textwidth}
		\includegraphics[width = \linewidth]{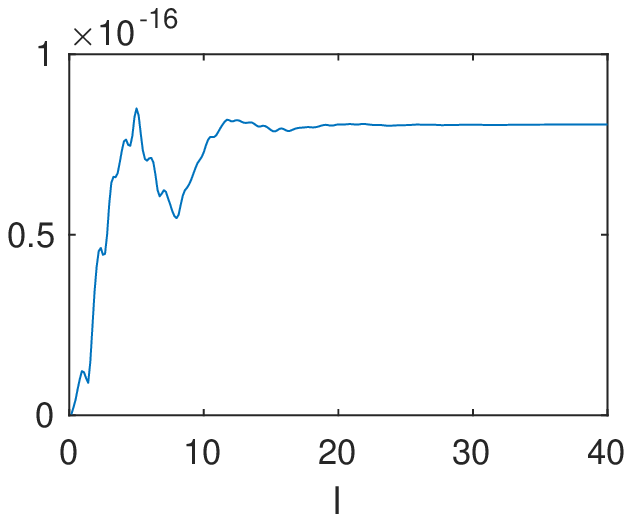}
		\caption{\scriptsize$(1,1)\to(2,-1)$}
	\end{subfigure}
	\begin{subfigure}{0.19\textwidth}
		\includegraphics[width = \linewidth]{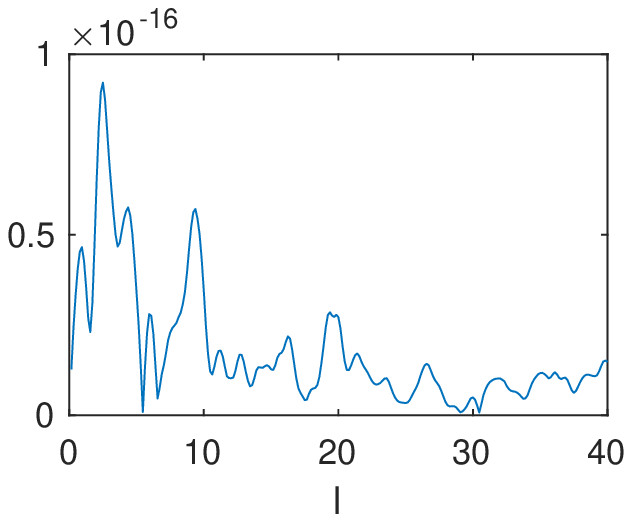}
		\caption{\scriptsize$(1,-1)\to(2,-1)$}
	\end{subfigure}
	\begin{subfigure}{0.19\textwidth}
		\includegraphics[width = \linewidth]{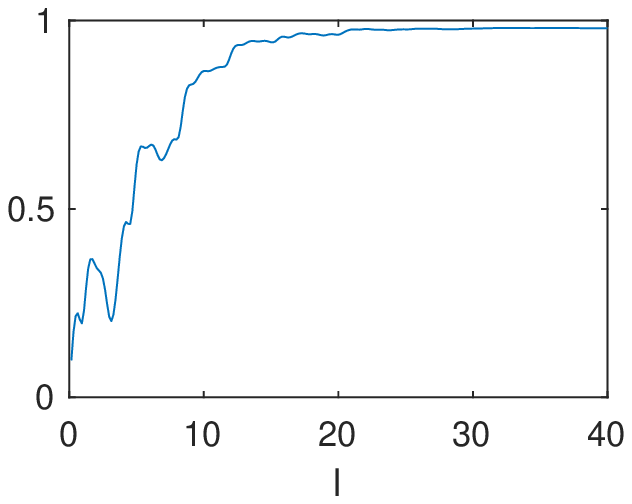}
		\caption{\scriptsize$(2,1)\to(2,-1)$}
	\end{subfigure}
	\begin{subfigure}{0.19\textwidth}
		\includegraphics[width = \linewidth]{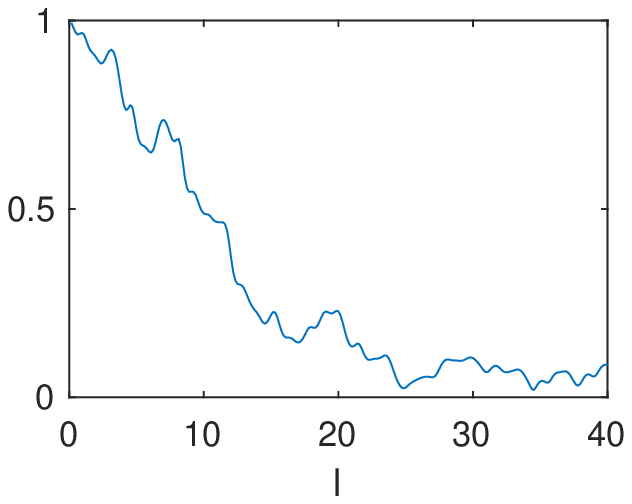}
		\caption{\scriptsize$(2,-1)\to(2,-1)$}
	\end{subfigure}\\
	\caption{Entries of scattering matrix against length of slab $l$, $V=V_3\chi_{[0,l]}$, $E=2.2$.}\label{fig:5sys}
	\end{figure}

In contrast, when $V=V_2\chi_{[0,l]}$, the modes with negative current $(0,-1)$ and $(1,-1)$ are efficiently coupled, which results in the periodic pattern observed in Fig.\ref{fig:periodic_transmission}.
\paragraph{Scattering matrix at higher energy.}
Our algorithm allows us to compute generalized eigenfunctions of arbitrary energy level and matrix-valued (Hermitian) perturbations. Consider $E=2.2$, which corresponds to $5$ propagating modes, i.e., $(0,-1)$, $(1,\pm)$, $(2,\pm)$, and a perturbation of the form $V=V_3\chi_{[-1,1]}$ with
\begin{align} \nonumber
    V_3=\begin{pmatrix} \exp(-y^2)\cos((\xi_{(1,-1)}-\xi_{(1,1)})x) & 0\\
0& \exp(-y^2)\cos((\xi_{(2,-1)}-\xi_{(2,1)})x)
\end{pmatrix},
\end{align}
which couples $(0,-1)$ with $(2,\pm)$ and $(1,-1)$ with $(1,1)$ modes. Since $V_3$ is an even function, there is no coupling between modes with adjacent energy, e.g. $0$ (or $(2,\pm1)$) with $(1,\pm1)$. In Fig.\ref{fig:5sys} we show the absolute value of the scattering matrix associated to these $5$ modes. Firstly, we see that computed coefficients that are supposed to vanish between the non-coupled modes are indeed so up to machine precision, (see (b), (c), (f), (i), (j), (k), (n), (o), (q), (r), (v), (w)). Secondly, all modes except $(0,-1)$ are significantly back-scattered when the slab is long enough, while the $(0,-1)$ mode strongly propagates, although it admits moderate coupling with $2\pm$ modes.

\begin{figure}[ht!]
    \centering
    	\begin{subfigure}{0.32\textwidth}
		\includegraphics[width = \linewidth]{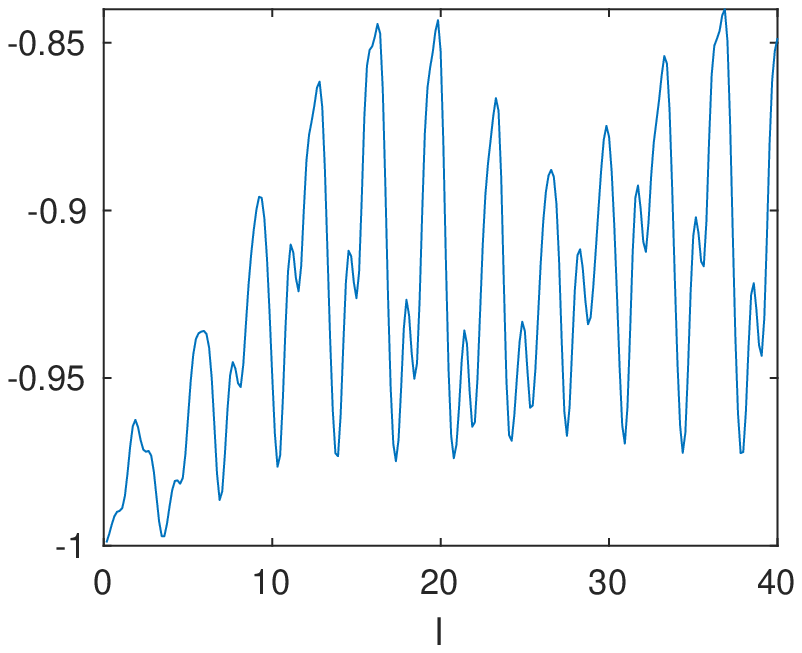}
		\caption{$j_{(0,-1)}$}
	\end{subfigure}
		\begin{subfigure}{0.32\textwidth}
		\includegraphics[width = \linewidth]{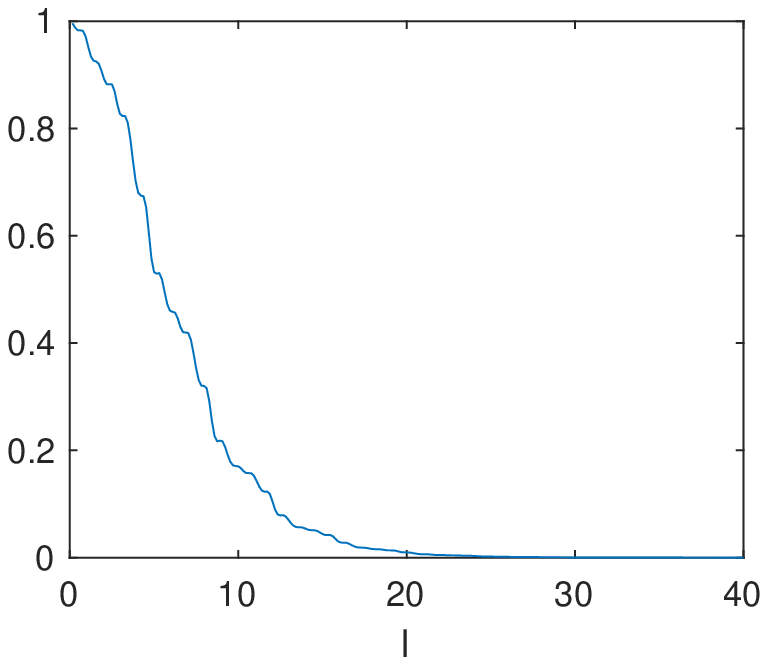}
		\caption{$j_{(1,1)}$}
	\end{subfigure}
		\begin{subfigure}{0.32\textwidth}
		\includegraphics[width = \linewidth]{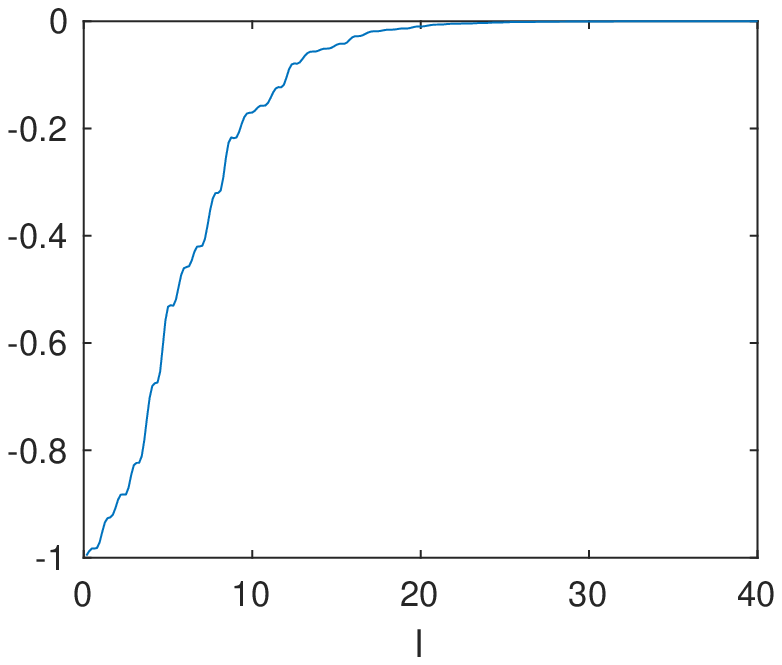}
		\caption{$j_{(1,-1)}$}
	\end{subfigure}\\
		\begin{subfigure}{0.32\textwidth}
		\includegraphics[width = \linewidth]{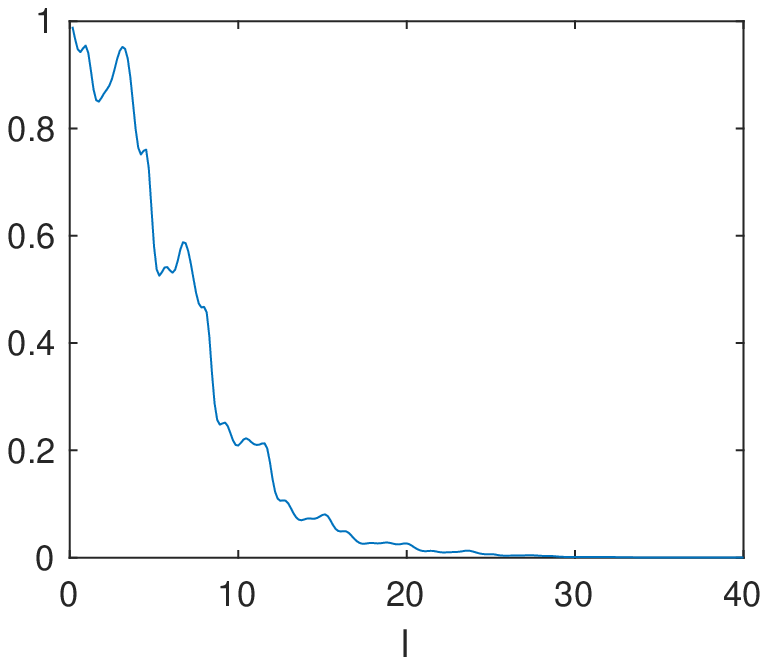}
		\caption{$j_{(2,1)}$}
	\end{subfigure}
		\begin{subfigure}{0.32\textwidth}
		\includegraphics[width = \linewidth]{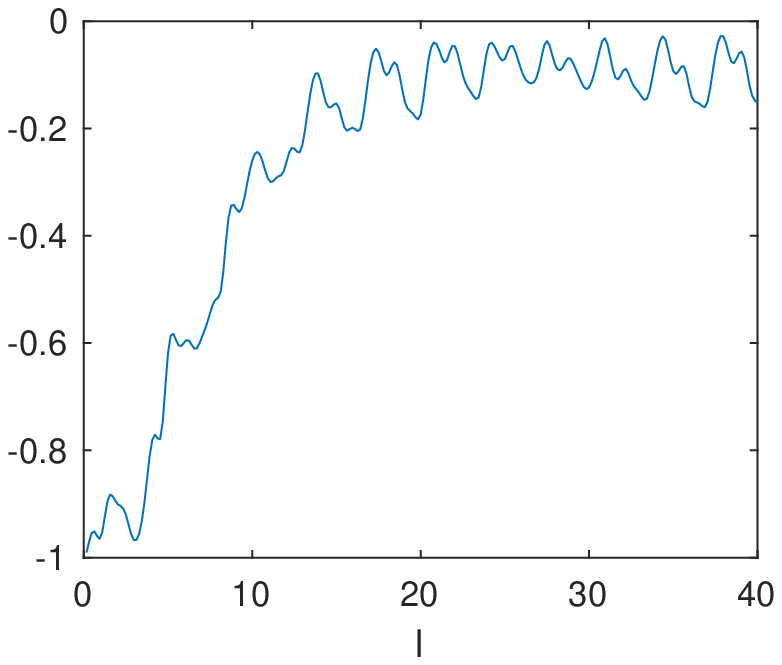}
		\caption{$j_{(2,-1)}$}
	\end{subfigure}
		\begin{subfigure}{0.32\textwidth}
		\includegraphics[width = \linewidth]{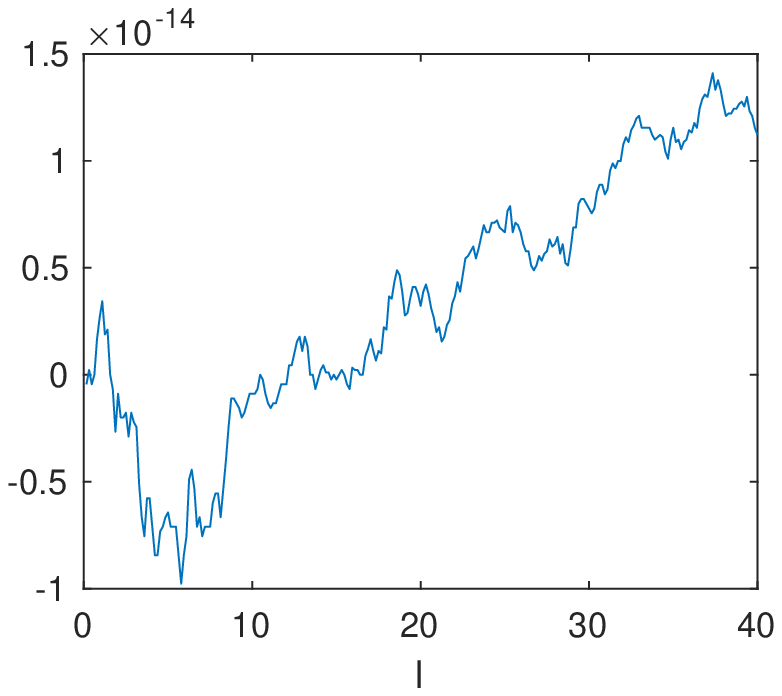}
		\caption{$\dsum_m j_m$}
	\end{subfigure}
	\caption{Currents from scattering matrix, $j_m$ against length of slab $l$, $V=V_3\chi_{[0,l]}$, $E=2.2$.}
    \label{fig:5sysjm}
\end{figure}

We next compute the currents $j_m$ from the scattering matrix using \eqref{jm_current} and \eqref{jm_current_2}. In Fig.\ref{fig:5sysjm}, we see that the currents vanish for the $(1,\pm1)$ and $(2,1)$ modes due to (essentially) complete back-scattering while for $(0,-1)$ and $(2,-1)$ mode, they oscillate as the slab length increases. These calculations also confirm the asymmetric transport properties of the topological insulator.
%%%%%
\subsection{Existence of Localized Modes}
\label{sec:localizedmodes}
%%%%%
In section \ref{sec:gal}, we argued that $I+V\mG$ with $\mG=(H-E)^{-1}$ (with outgoing conditions) was invertible as soon $-1$ was not an eigenvalue of the compact operator $V\mG$. We now show that $-1$ may indeed be in the spectrum of $\mG$ for a scalar perturbation $V(x)$ independent of $y$ and compactly supported in $x$. Specifically, we assume that $V(x,y)= V_0\chi_I(x)$, where $V_0$ is a fixed constant and $I=[0,l]$ is an interval. We now identify values of $(E,l,V_0)$ such that $-1$ is in the point spectrum of $V\mG$.

Assume the existence of $n\in\Nm$ such that $E^2-2n<0$ while $(E-V_0)^2-2n>0$. Thus, $\psi_{n,\pm 1}$ is evanescent outside of $[0,l]$ and propagating inside $[0,l]$. Denote
\begin{align}\label{def:phin}
    \phi_{\pm} = c_{(n,\pm 1)} \begin{pmatrix} \fa \varphi_n \\ (E-\xi_{(n,\pm 1)}) \varphi_n \end{pmatrix} =  c_{\pm} \begin{pmatrix} \sqrt{2n} \varphi_{n-1} \\ (E\mp\xi) \varphi_n \end{pmatrix} , 
\end{align}
where $c_{\pm}^{-2} = 2n + |E\mp\xi|^2$ and $\xi=\sqrt{E^2-2n}\in i\Rm$ and
\begin{align}\label{def:phinprime}
    \phi_{\pm}' = c_{(n,\pm 1)}' \begin{pmatrix} \fa \varphi_n \\ (E-\xi_{(n,\pm 1)}') \varphi_n \end{pmatrix} =  c_{\pm}' \begin{pmatrix} \sqrt{2n} \varphi_{n-1} \\ (E'\mp\xi') \varphi_n \end{pmatrix} , 
\end{align}
where $(c_{\pm}')^{-2} = 2n + |E'\mp\xi'|^2$, $\xi'=\sqrt{E'^2-2n}\in \Rm^+$, and $E'=E-V_0$.

We seek a solution $(H+V-E)\psi=0$ given by $\psi=e^{i\xi (x-l)}\phi_+$ for $x\geq l$, which vanishes at $+\infty$. At $x=l$, we have by continuity
\begin{align}\nonumber
    \psi(l,y)=a_+\phi'_+(y)+a_-\phi'_-(y), \qquad  \begin{pmatrix}
a_+\\a_-
\end{pmatrix}=(\phi_+',\phi_-')^{-1}\phi_+.
\end{align}
Here, $(\phi_+',\phi_-')$ is a $2\times2$ matrix with columns $\phi'_\pm$. Solving the above equation over $x\in[0,l]$ yields
\begin{align}\nonumber
    \psi(0,y)=a_+e^{-i\xi'l }\phi'_+(y)+a_-e^{i\xi'l}\phi'_-(y).
\end{align}
The function $\psi\in L^2(\Rm^2;\Cm^2)$ is normalizable only if $\psi$ is proportional to $e^{-i\xi x}\phi_-$ for $x<0$, which in turn is equivalent to
\begin{align}\nonumber
   b_+a_+e^{-i\xi'l}+b_-a_-e^{i\xi'l}=0,
\end{align}
where $(b_+,b_-)$ is the first row of $(\phi_+,\phi_-)^{-1}(\phi_+',\phi_-')$.

Simple but tedious calculations show that
\begin{align}\nonumber
    \frac{b_+a_+}{b_-a_-}=-\frac{(E-E'+\xi+\xi')(E-E'-\xi-\xi')}{(E-E'-\xi+\xi')(E-E'+\xi-\xi')}.
\end{align}
Recalling $\xi\in i\Rm$ and $\xi'\in \Rm$, we have $|\frac{b_+a_+}{b_-a_-}|=1$ and thus need:
\begin{align} \nonumber
    {\rm Arg}(\frac{(E-E'+\xi+\xi')(E-E'-\xi-\xi')}{(E-E'-\xi+\xi')(E-E'+\xi-\xi')})=2\xi'l+2k\pi, \quad k=0,\pm 1, \pm 2, \cdots.
\end{align}
Any solution $(E,V_0,l)$ of the above equation provides a localized model associated to the eigenvalue $E$ of the operator $H+V$. 

As an illustration, we consider the case with $E=1.8$, $\xi'=1$ and $n=2$ while $k=0$.
We then find the numerical approximation $l\approx 1.5422$. 

\begin{figure}[ht!]
    \centering
		\begin{subfigure}{0.24\textwidth}
		\includegraphics[width = \linewidth]{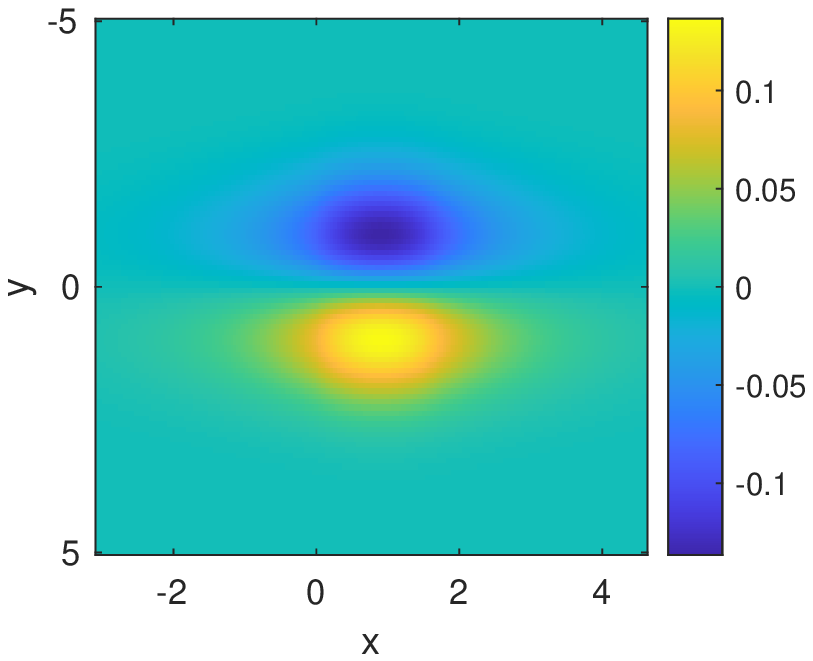}
		\caption{$Re(\psi_1)$}
	\end{subfigure}
	\begin{subfigure}{0.24\textwidth}
		\includegraphics[width = \linewidth]{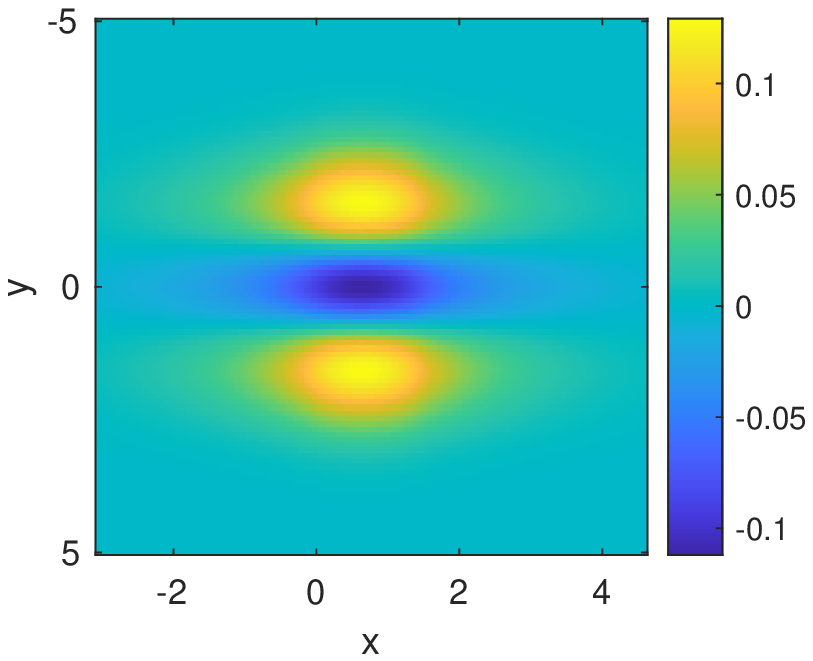}
		\caption{$Re(\psi_2)$}
	\end{subfigure}
		\begin{subfigure}{0.24\textwidth}
		\includegraphics[width = \linewidth]{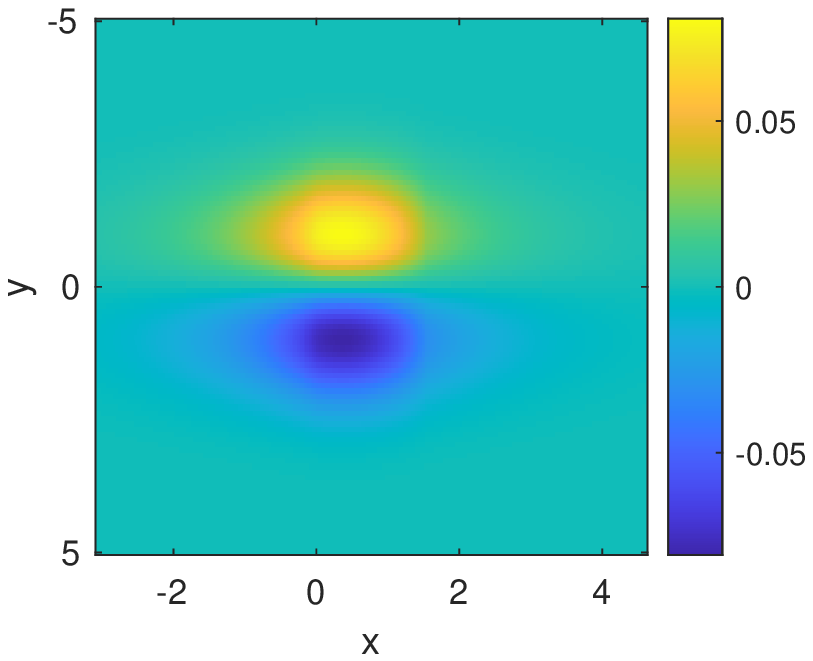}
		\caption{$Im(\psi_1)$}
	\end{subfigure}
	\begin{subfigure}{0.24\textwidth}
		\includegraphics[width = \linewidth]{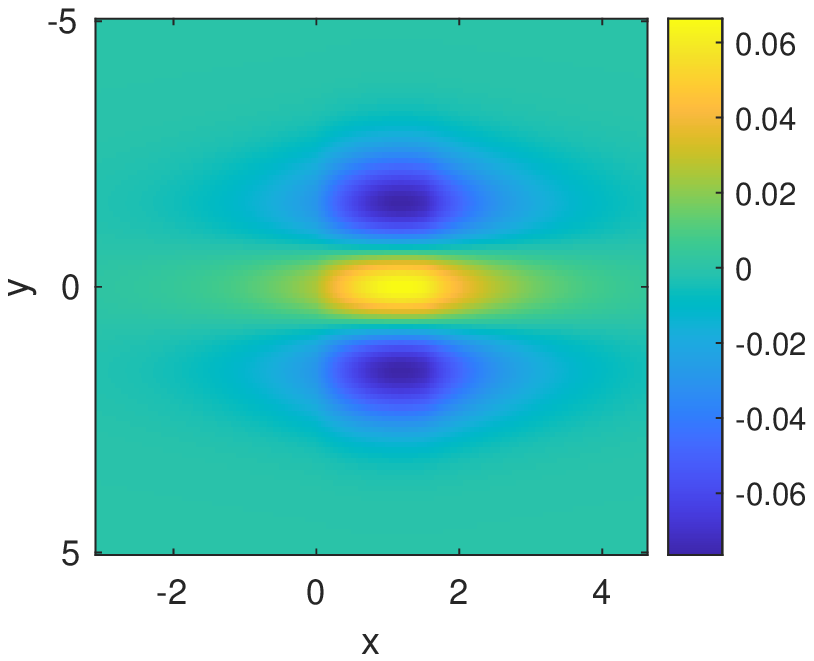}
		\caption{$Im(\psi_2)$}
	\end{subfigure}
    \caption{Localized mode computed by $\psi=\int G \hat{\rho}$.}
    \label{fig:localmodes}
\end{figure}

In the discretization of Alg.\ref{alg_rho}, we let $(n_x,n_y)=(60,20)$ to resolve local modes related to $\phi_{(2,\pm 1)}$. At step 3, we found that $I+\hat{V}\hat{G}$ was not stably invertible. Numerically, by applying singular value decomposition on $I+\hat{V}\hat{G}$, we found that the condition number, equal to the ratio of largest and smallest eigenvalues, is $3.619\times 10^{15}$. The smallest two singular values are $0.5431$ and $4.88\times 10^{-16}$. Let $\hat\rho$ be the eigenvector associated with the latter, which forms the null space of $I+\hat{V}\hat{G}$. As a verification, we compute that $|(I+\hat{V}\hat{G})\hat{\rho}|_{l_2}\approx 0$ ($2.029\times 10^{-15}$). In Fig.\ref{fig:localmodes}, we present the corresponding localized mode $\psi=\int G \hat{\rho}$ solution of $(H+V-E)\psi\approx0$.
%\gb{Looks fine, but how is $\hat\rho$ computed? }
\section{Conclusions}\label{sec:conclu}
This paper presented a fast and accurate algorithm to compute generalized eigenfunctions of two-dimensional operators displaying a domain wall that separates two insulating half-spaces. The main assumption in our construction is that the Green's function of the unperturbed operator may be described reasonably explicitly. Eigenfunctions of perturbed operators are then computed using an integral formulation for a density $\rho$ supported on the same domain as the perturbation. This allows us to construct generalized eigenfunctions with support in the whole of $\Rm^2$ from a compactly supported density.

Our main application was a Dirac operator with linear domain wall, which finds applications in the analysis of topological phases of matter. Potential generalizations of the algorithm include block-diagonal systems of Dirac operators, which also appear naturally in the study of topological insulators \cite{B19b,topological}, as well as three dimensional generalizations to high-order topological insulators \cite{bal2021topological}.

We showed a number of applications of the algorithm to quantify the transport properties of Dirac operators with domain wall. In particular, we deri=ed an expression for a quantized interface conductivity describing asymmetric transport in terms of the computed generalized eigenfunctions. We were then able to confirm numerically that the conductivity was indeed integral valued up to very high accuracy. We also computed the full far-field scattering matrix generated by the perturbations. This matrix took the form of a $2n+1$ times $2n+1$ matrix for all energies $E$ such that $2n<E^2<2n+2$. Such matrices may be computed for energies $E$ not belonging to the point spectrum of the perturbed operator $H+V$. Note that such a point spectrum is necessarily embedded in the (absolutely) continuous spectrum $\sigma_{\rm ac}(H+V)=\Rm$, a fact that would not be allowed for Schr\"odinger or Dirac operators in the absence of a domain wall. As future research, we plan to use the algorithm to analyze the structure of the point spectrum and possibly resonances for Dirac and related operators.

\section*{Acknowledgment.} This research was partially supported by the U.S. National Science Foundation, Grants DMS-1908736 and EFMA-1641100.

\bibliographystyle{unsrt}%{siam}
\bibliography{ref}
\end{document}